\documentclass[fleqn,twoside]{article}
\usepackage{amsmath,amssymb,espcrc2,color,cite}

\usepackage{graphicx,soul}
\usepackage[figuresright]{rotating}

\newcommand{\url}[1]{{\tt #1}}
\newcommand{\lsim}
{\;\raisebox{-.3em}{$\stackrel{\displaystyle <}{\sim}$}\;}

\newcommand{\gmt}{\ensuremath{(g-2)_\mu}}
\newcommand{\br}{{\rm BR}}
\newcommand{\bsg}{BR($b \to s \gamma$)}
\newcommand{\btn}{BR($B_u \to \tau \nu_\tau$)}

\newcommand{\bmm}{\ensuremath{\br(B_s \to \mu^+\mu^-)}}
\newcommand{\ssi}{\ensuremath{\sigma^{\rm SI}_p}}

\newcommand{\MW}{M_W}
\newcommand{\MZ}{M_Z}
\newcommand{\Mh}{M_h}

\newcommand{\MA}{M_A}

\newcommand{\mt}{m_t}
\newcommand{\mgl}{m_{\tilde g}}
\newcommand{\cha}[1]{\tilde \chi^\pm_{#1}}
\newcommand{\champ}[1]{\tilde \chi^\mp_{#1}}

\newcommand{\neu}[1]{\tilde \chi^0_{#1}}
\newcommand{\mneu}[1]{m_{\tilde \chi^0_{#1}}}

\newcommand{\staue}{\tilde \tau_1}
\newcommand{\astaue}{\overline{\tilde \tau_1}}
\newcommand{\sel}[1]{\tilde e_{#1}}
\newcommand{\asel}[1]{\overline{\tilde e_{#1}}}
\newcommand{\smu}[1]{\tilde \mu_{#1}}
\newcommand{\asmu}[1]{\overline{\tilde \mu_{#1}}}
\newcommand{\tb}{\tan\beta}

\newcommand{\tev}{\,\, \mathrm{TeV}}
\newcommand{\gev}{\,\, \mathrm{GeV}}
\newcommand{\mev}{\,\, \mathrm{MeV}}
\newcommand{\SM}{``SM''}

\definecolor{Orange}{named}{Orange}
\definecolor{Purple}{named}{Purple}

\newcommand{\ETslash}{/ \hspace{-.7em} E_T}

\graphicspath{{figs/}}

\title{\bf Supersymmetry in Light of 1/fb of LHC Data
 \\ \vspace{0.5em}}

\author{
{\bf O.~Buchmueller}\address[Imperial]
   {High\,Energy\,Physics\,Group,\,Blackett\,Laboratory,\,Imperial\,College,\,Prince\,Consort\,Road,\,London\,SW7\,2AZ,\,UK},
{\bf R.~Cavanaugh}\address[FNAL]
   {Fermi National Accelerator Laboratory, P.O. Box 500, 
    Batavia, Illinois 60510, USA}\hbox{$^{\rm ,}$}\address[UIC]
   {Physics Department, University of Illinois at Chicago, Chicago, 
    Illinois 60607-7059, USA},
{\bf A.~De Roeck}\address[CERN]
   {CERN, CH--1211 Gen\`eve 23, Switzerland}\hbox{$^{\rm ,}$}\address[Antwerpen]
   {Antwerp University, B--2610 Wilrijk, Belgium},
 {\bf M.J.~Dolan}\address[IPPP]
{Institute for Particle Physics
     Phenomenology,\,University\,of\,Durham,\,South 
     Road,\,Durham\,DH1\,3LE,\,UK},
{\bf J.R.~Ellis}\address[KCL]{Theoretical Particle Physics
  and Cosmology Group, Department of Physics, King's College London, London~WC2R~2LS, UK}\hbox{$^{\rm ,}$}\addressmark[CERN], 
{\bf H.~Fl\"acher}\address[Rochester]
   {H.H.~Wills Physics Laboratory, University of Bristol, Tyndall Avenue, Bristol BS8 1TL, UK},
{\bf S.~Heinemeyer}\address[Santander]
   {Instituto de F\'{\i}sica de Cantabria (CSIC-UC), 
    E--39005 Santander, Spain},
{\bf G.~Isidori}\address[Frascati]
   {INFN, Laboratori Nazionali di Frascati, Via E. Fermi 40, 
    I--00044 Frascati, Italy},
{\bf D.~Mart\'inez~Santos}\addressmark[CERN],
{\bf K.A.~Olive}\address[Minnesota] 
   {William\,I.\,Fine\,Theoretical\,Physics\,Institute,\,School of Physics\,and\,Astronomy,\,University\,of\,Minnesota,\\Minneapolis,\,Minnesota\,55455,\,USA},
{\bf S.~Rogerson}\addressmark[Imperial],
{\bf F.J.~Ronga}\address[ETHZ]
   {Institute for Particle Physics, ETH Z\"urich, CH--8093 Z\"urich, 
   Switzerland},
{\bf G.~Weiglein}\address[DESY]
   {DESY, Notkestrasse 85, D--22607 Hamburg, Germany}
}

\begin{document}

\begin{abstract}
We update previous frequentist analyses of the CMSSM and NUHM1
parameter spaces to include the public results of searches for
supersymmetric signals using $\sim 1$/fb of LHC data
recorded by ATLAS and CMS and $\sim 0.3$/fb of data recorded by LHCb
in addition to electroweak precision and B-physics observables.
We also include the constraints imposed by the cosmological dark matter density
and the XENON100 search for spin-independent dark matter scattering.
The LHC data set includes ATLAS and CMS searches for
jets + $\ETslash$ events and for the heavier
MSSM Higgs bosons, and the upper limits on \bmm\ from LHCb and CMS.
The absences of jets + $\ETslash$ signals in the LHC data favour heavier mass
spectra than in our previous analyses of the CMSSM and NUHM1,
which may be reconciled with \gmt\ if $\tan \beta \sim 40$, a possibility that is however under pressure from
heavy Higgs searches and the upper limits on \bmm.
As a result, the $p$-value for the CMSSM fit is reduced to $\sim 15 (38)$\%, and that for the NUHM1
to $\sim 16 (38)$\%, to be compared with $\sim 9 (49)$\% for the Standard Model 
limit of the CMSSM for the same set of observables (dropping \gmt), ignoring the dark
matter relic density. 
We discuss the sensitivities of the fits to the \gmt\ and \bsg\ constraints,
contrasting fits with and without the \gmt\ constraint, and
combining the theoretical and experimental errors for \bsg\ linearly or in quadrature.
We present predictions for $\mgl$, \bmm, $\Mh$ and $\MA$, and
update predictions for spin-independent dark matter
scattering, 
incorporating the uncertainty in the 
$\pi$-nucleon $\sigma$ term $\Sigma_{\pi N}$.
Finally, we present predictions based on our fits for the likely thresholds for sparticle pair
production in $e^+e^-$ collisions in the CMSSM and NUHM1.

\bigskip
\begin{center}
{\tt KCL-PH-TH/2011-28, LCTS/2011-14, CERN-PH-TH/2011-220, \\
DCPT/11/108, DESY 11-161, IPPP/11/54, FTPI-MINN-11/23, UMN-TH-3013/11}
\end{center}
\vspace{2.0cm}
\end{abstract}

\maketitle

\section{Introduction}
\label{sec:intro}

In a series of papers, we and others have reported the results of global fits to pre-LHC~\cite{mc1,mc2,mc3,mc35,mc4,pre-LHC} and
LHC 2010 data~\cite{mc5,mc6,mcweb,post-LHC} in the frameworks of simplified variants of the minimal supersymmetric
extension of the Standard Model (MSSM)~\cite{HK}
with universal supersymmetry-breaking mass parameters at the GUT scale.
We consider a class of models in which R-parity is conserved and the lightest supersymmetric
particle (LSP), assumed to be the lightest neutralino $\neu{1}$~\cite{EHNOS}, 
provides the cosmological cold dark matter~\cite{Komatsu:2010fb}.
The specific models studied have included the constrained MSSM
(CMSSM)~\cite{cmssm1},
with parameters $m_0$, $m_{1/2}$ and $A_0$ denoting common scalar, fermionic and
trilinear soft supersymmetry-breaking parameters at the GUT scale, and $\tb$ denoting the ratio of the
two vacuum expectation values of the two Higgs fields. Other models studied include a model in which 
common supersymmetry-breaking contributions to the Higgs masses are allowed to be
non-universal (the NUHM1)~\cite{nuhm1}, a very constrained model in which trilinear and bilinear
soft supersymmetry-breaking parameters are related (the VCMSSM)~\cite{vcmssm}, and minimal
supergravity (mSUGRA)~\cite{vcmssm,mSUGRA} in which the gravitino mass is required to be the same as the
universal soft supersymmetry-breaking scalar mass before
renormalization.

The impressive increase in the accumulated LHC luminosity combined with the rapid
analyses of LHC data by the ATLAS~\cite{ATLASsusy,ATLASHA}, CMS~\cite{CMSsusy,CMSHA,CMSbmm} 
and LHCb Collaborations~\cite{LHCbbmm} is putting
increasing pressure on these and other supersymmetric models, in the continuing
absence of any signal for supersymmetry. In this paper we update our previous
frequentist fits~\cite{mc6,mcweb} to include the analyses of $\sim 1$/fb of
LHC data made public in July and August 2011 at the EPS HEP
and Lepton-Photon Conferences, 
termed here LHC$_{\rm 1/fb}$, and also discuss the impact of the
result on \bmm\ by the CDF Collaboration~\cite{CDFbmm}.
As in~\cite{mc6}, we  also incorporate the results of 
the direct search for dark matter scattering by the
XENON100 Collaboration~\cite{XE100}~%
\footnote{Preliminary versions of these results were posted on~\cite{mcweb} on July 25th, 2011.}.%

The approach we use has been documented in our previous papers~\cite{mc1,mc2,mc3,mc35,mc4,mc5,mc6}, so we do
not describe it in detail here, concentrating on relevant new aspects. We construct
a global likelihood function that receives contributions from the standard portfolio of
electroweak precision observables, as well as B-decay measurements such as
\bsg\ and \btn. The
contributions to the likelihood function from \bmm, the XENON100 direct search
for dark matter scattering and the LHC searches for supersymmetric
signals are calculated within the 
{\tt MasterCode} framework~\cite{mcweb}. 
Concerning the theoretical predictions for the different
observables, the {\tt MasterCode} framework incorporates a code for the electroweak
observables based on~\cite{Svenetal} as well as the {\tt SoftSUSY}~\cite{Allanach:2001kg}, {\tt FeynHiggs}~\cite{FeynHiggs}, 
{\tt SuFla}~\cite{SuFla}, {\tt SuperIso}~\cite{SuperIso}, {\tt MicrOMEGAs}~\cite{MicroMegas} 
and {\tt SSARD}~\cite{SSARD} codes, using the SUSY Les Houches
Accord~\cite{SLHA}. The {\tt MasterCode}
framework is such that new observables can easily be incorporated via new
`afterburners', as we discuss below for the LHC$_{\rm 1/fb}$ constraints.
We use a Markov Chain Monte Carlo (MCMC) approach to sample the parameter 
spaces of supersymmetric models, and the results of this paper are based on a basic
resampling of the CMSSM with 70M points and a resampling of the NUHM1 with 70M additional points, both
extending up to $m_0, m_{1/2} = 4000 \gev$.

Our results are based on the public results of searches for supersymmetric 
signals using $\sim 1$/fb of LHC data analyzed by the ATLAS and CMS 
Collaborations and $\sim 0.3$/fb of data analyzed by the LHCb Collaboration.
For our purposes, some of the most important constraints are provided by the 
ATLAS~\cite{ATLASsusy} and CMS~\cite{CMSsusy} searches for jets +
$\ETslash$ events without leptons, as well as
searches for the heavier MSSM Higgs bosons, $H/A$~\cite{ATLASHA,CMSHA}. Also important
are the new upper limits on \bmm\ from the CMS~\cite{CMSbmm}, LHCb~\cite{LHCbbmm} and CDF Collaborations~\cite{CDFbmm},
which we incorporate in this paper as described below~\footnote{For other studies of recent data on
\bmm\ in supersymmetric frameworks, see~\cite{Bothers}.}. 
In this paper we focus on the analysis of the effects in the CMSSM
and the NUHM1. As discussed briefly below, the VCMSSM and mSUGRA models are 
further disfavoured by the LHC$_{\rm 1/fb}$ data.

The absences of signals
in the jets + $\ETslash$ searches disfavour the ranges of the model mass 
parameters $(m_0, m_{1/2})$ that had been favoured in our previous analyses 
of the CMSSM and NUHM1~\cite{mc5,mc6}, and our current best fits have $m_0 \sim 150$ to 450~GeV
and $m_{1/2} \sim 750 \gev$. Reconciling these larger values of $(m_0, m_{1/2})$
with \gmt\ favours values of $\tan \beta \sim$ 40, though with a large uncertainty.
The regions of parameter space with large $\tan \beta$ are constrained also by
the new upper limits on \bmm, as well as the LHC
$H/A$ searches. Using our standard implementation of the \gmt\ constraint based
on a Standard Model (SM) calculation~\cite{newDavier}, and
combining the theoretical and experimental errors in \bsg\ in quadrature,
we find that the $p$-value for the CMSSM best-fit point is now
$\sim 15$\%, and that for the NUHM1 
is $\sim 16$\%. On the other hand, we show that if the \gmt\
constraint is dropped much larger regions of the
$(m_0, m_{1/2})$ and other parameter planes are allowed at the 68 and 95\% CL,
and these $p$-values increase to 38\% in both models.
In contrast, changing the treatment
of \bsg\ by adding linearly the theoretical and experimental errors has relatively little
impact on the fits and their $p$-values, increasing them both to 18\%.

On the basis of these results, we present updated predictions for the gluino mass
$\mgl$, \bmm, and the light and heavy Higgs masses $\Mh$ and 
$\MA$. We also
present updated predictions for the spin-independent 
dark matter scattering cross section, \ssi, stressing the importance of the uncertainty in the 
$\pi$-nucleon $\sigma$ term $\Sigma_{\pi N}$~\cite{sigma}. 

In addition, we use our results to present 
likelihood functions for the thresholds for sparticle pair production in $e^+e^-$ 
collisions. These results indicate that, within the CMSSM and NUHM1, the
best-fit values for the
sparticle thresholds lie above 
$E_{\rm CM} = 500 \gev$. However, we 
emphasize that these results are derived in the context of specific models with
specific universal soft supersymmetry-breaking masses at the GUT
scale, and do not apply to other classes of supersymmetric models.


\section{Implementations of the New LHC Constraints}

\subsection*{\it Jets + $\ETslash$ searches}

The CMS and ATLAS Collaborations have both announced new exclusions in the
$(m_0, m_{1/2})$ plane of the CMSSM based on searches for events with jets + $\ETslash$
unaccompanied by charged leptons, assuming $\tan \beta = 10$, $A_0 = 0$ and $\mu > 0$.
The updated CMS $\alpha_T$ analysis is based on 1.1/fb of data~\cite{CMSsusy}, and the updated ATLAS
0-lepton analysis is based on 1.04/fb of data~\cite{ATLASsusy}.
It is known that 0-lepton analyses are in general relatively insensitive to the $\tan \beta$ and $A_0$
parameters of the CMSSM, as has been confirmed specifically for the CMS $\alpha_T$ analysis,
and they are also insensitive to the amount of Higgs non-universality in the NUHM1.
Therefore, we treat these analyses as constraints in the $(m_0, m_{1/2})$ planes of the CMSSM
and NUHM1 that are independent of the other model parameters. The ATLAS~\cite{ATLASlepton} and CMS
Collaborations~\cite{CMSlepton} have also announced new exclusions for searches for jets + $\ETslash$ 
events with one or more charged leptons with $\sim 1$/fb of data, 
but these have in general less expected sensitivity, and are more dependent 
on the other model parameters, so we do not include them in our analysis. A similar remark
applies to the new ATLAS limits on events with $b$ jets + $\ETslash$ unaccompanied by charged leptons
using 0.83/fb of data~\cite{ATLASb} 
and on events with $b$ jets + $\ETslash$ + 1 lepton using 1.03/fb
of data~\cite{ATLASb2}~\footnote{It would facilitate the modelling of 
LHC constraints on SUSY
if each Collaboration could combine the results from its different missing-energy searches,
as is already done for Higgs searches.}.

The CMS and ATLAS 0-lepton searches are more powerful in complementary regions of the
$(m_0, m_{1/2})$ plane. Along each ray in this plane, we compare the expected CMS and ATLAS
sensitivities, select the search that has the stronger expected 95\% CL limit, and apply the
constraint imposed by that search~\footnote{It would also facilitate the modelling of LHC constraints on
supersymmetry if the results from different Collaborations were combined officially,
as was done at LEP, is already done for \bmm\ searches, and is planned for Higgs searches.}. We assign $\Delta \chi^2 = 5.99$, corresponding to 1.96 effective standard deviations,
along the CMS and ATLAS 95\% 0-lepton exclusion contours in the $(m_0, m_{1/2})$ plane.
In the absence of more complete experimental information, we approximate the impact of
these constraints by assuming that event numbers scale along rays in this plane
$\propto {\cal M}^{-4}$ where ${\cal M} \equiv \sqrt{m_0^2 + m_{1/2}^2}$
(more details can be found in \cite{mc6}).
We then use these
numbers to calculate the effective numbers of standard deviations 
and corresponding values of
$\Delta \chi^2$ at each point in the plane.

\subsection*{\it Searches for heavy MSSM Higgs bosons}

The CMS Collaboration has announced a new constraint on the heavy MSSM Higgs bosons
from a search for the neutral bosons $H/A \to \tau^+ \tau^-$ using 1.6/fb of data~\cite{CMSHA} and ATLAS
has presented a similar constraint using 1.06/fb of data~\cite{ATLASHA}, the results
being presented as 95\% CL upper limits on the product of the production cross section
times $\tau^+ \tau^-$ branching ratio as a function of the common mass,
$M_{H/A}$, for masses smaller than about $500 \gev$. In our analysis we use the CMS constraint,
which has the greater expected sensitivity. The
CMS Collaboration has also announced a constraint on the decay chain 
$t \to H^+ \to \tau^+ \nu$ using 1.09/fb of data~\cite{CMSHpm}, but this yields a constraint in a generic 
$(\MA, \tan \beta)$ plane that is much weaker than the above searches for $H/A$, so we do not implement it
in our analysis.

We assign $\Delta \chi^2 = 3.84$, corresponding to 1.96 effective standard deviations, to model
parameter sets predicting an $H/A$ signal at the 95\% CL given by the 
CMS constraint, for each fixed value of $\MA$. Other model parameter sets are assigned
values of $\Delta \chi^2$ according to the numbers of effective standard deviations
corresponding to the numbers of events they predict. For any fixed value of $M_{H/A}$, these
event numbers scale approximately as $(\tan \beta)^2$.

\subsection*{\it Constraints on \bmm}

Three new results on \bmm\ have been announced recently. One is an excess of
candidate $B_s \to \mu^+ \mu^-$ events reported by the CDF Collaboration~\cite{CDFbmm},
which corresponds to \bmm\ $= (1.8^{+1.1}_{-0.9})\times 10^{-8}$ or \bmm\ $< 4.0 \times 10^{-8}$
at the 95\% CL. The other two new results are upper limits from the CMS Collaboration using
1.14/fb of data~\cite{CMSbmm}: \bmm\ $< 1.9 \times 10^{-8}$ at the 95\% CL, and from the LHCb Collaboration using 0.34/pb
of data~\cite{LHCbbmm}: \bmm\ $< 1.5 \times 10^{-8}$ at the 95\% CL. 

These three results are reasonably compatible, though there
is some tension between the CDF and CMS/LHCb results. The two latter collaborations have
released an official combination of their results~\cite{CMSLHCb}, which yields \bmm\ $< 1.08 \times 10^{-8}$ at the 95\% CL.
In our implementation of the \bmm\ constraint, we use $\Delta \chi^2$ corresponding to the full likelihood 
function provided by this combination, which has a global minimum close to the SM
prediction. We also comment on the changes in our results that would follow from an (unofficial)
combination with the CDF result~\cite{CDFbmm}, which would yield a $\Delta \chi^2$ function with a minimum at
\bmm\ at about twice the SM value.

\subsection*{\it Constraints on dark matter scattering}

We incorporate the upper limit on the spin-independent dark matter
scattering cross section \ssi\ provided by the XENON100 Collaboration~\cite{XE100}
in a similar manner to~\cite{mc6}. In that paper we discussed extensively the
uncertainty in the spin-independent scattering matrix element induced by the relatively ill-determined
value of the $\pi$-nucleon $\sigma$ term, $\Sigma_{\pi N}$. In this paper we use 
$\Sigma_{\pi N} = 50 \pm 14 \mev$, and neglect other uncertainties, e.g., in modelling the dark
matter distribution. We also do not consider here other experiments reporting signatures
that would require \ssi\ above the XENON100 limit. Nor do we consider limits on
spin-dependent dark matter scattering and astrophysical signatures of dark matter annihilations,
which currently do not constrain the CMSSM and NUHM1~\cite{mc6}.


\section{Results}

\subsection*{\it The $(m_0, m_{1/2})$ planes in the CMSSM and NUHM1}

\begin{figure*}[htb!]
\resizebox{8.7cm}{!}{\includegraphics{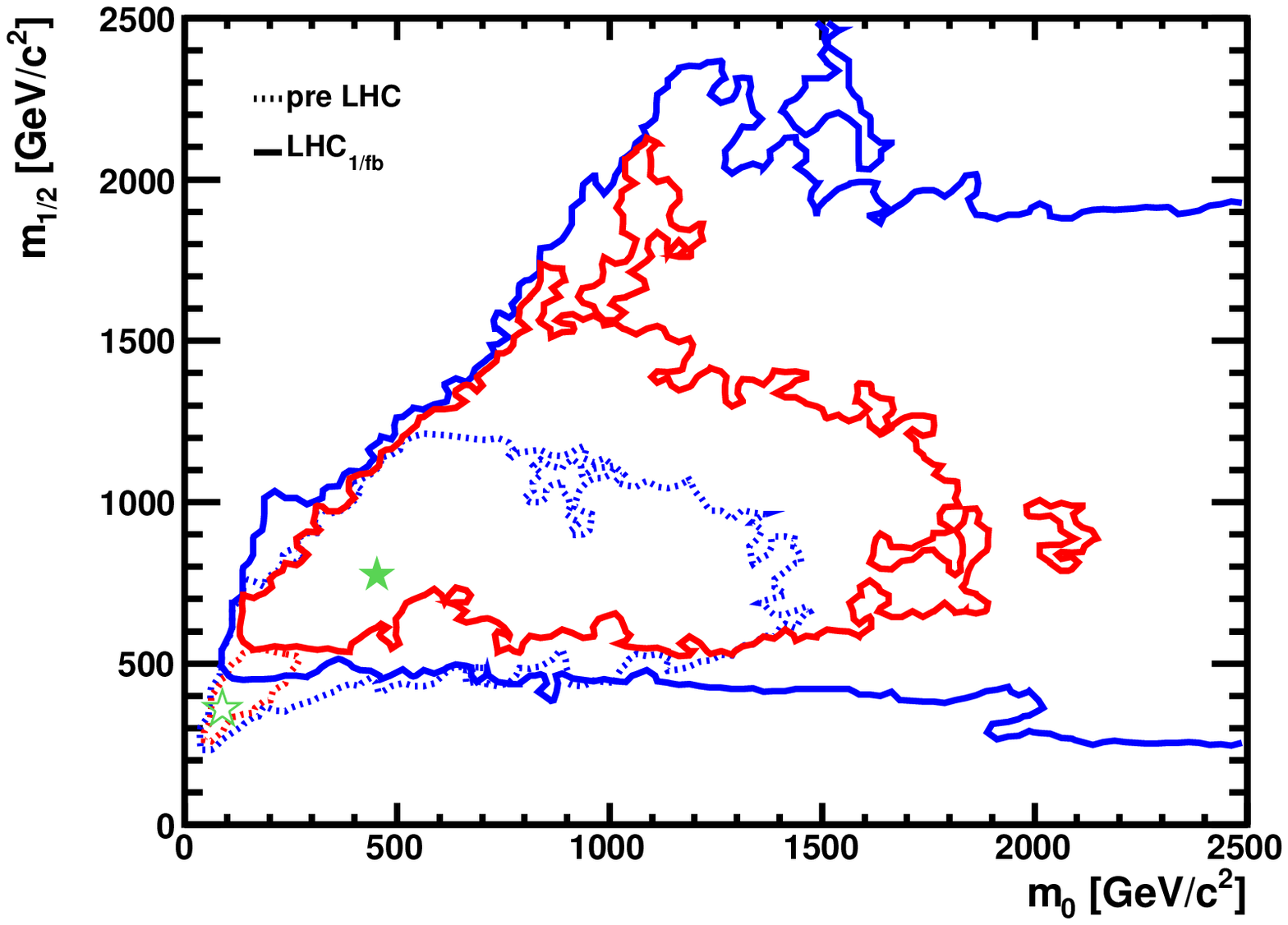}}
\resizebox{8.7cm}{!}{\includegraphics{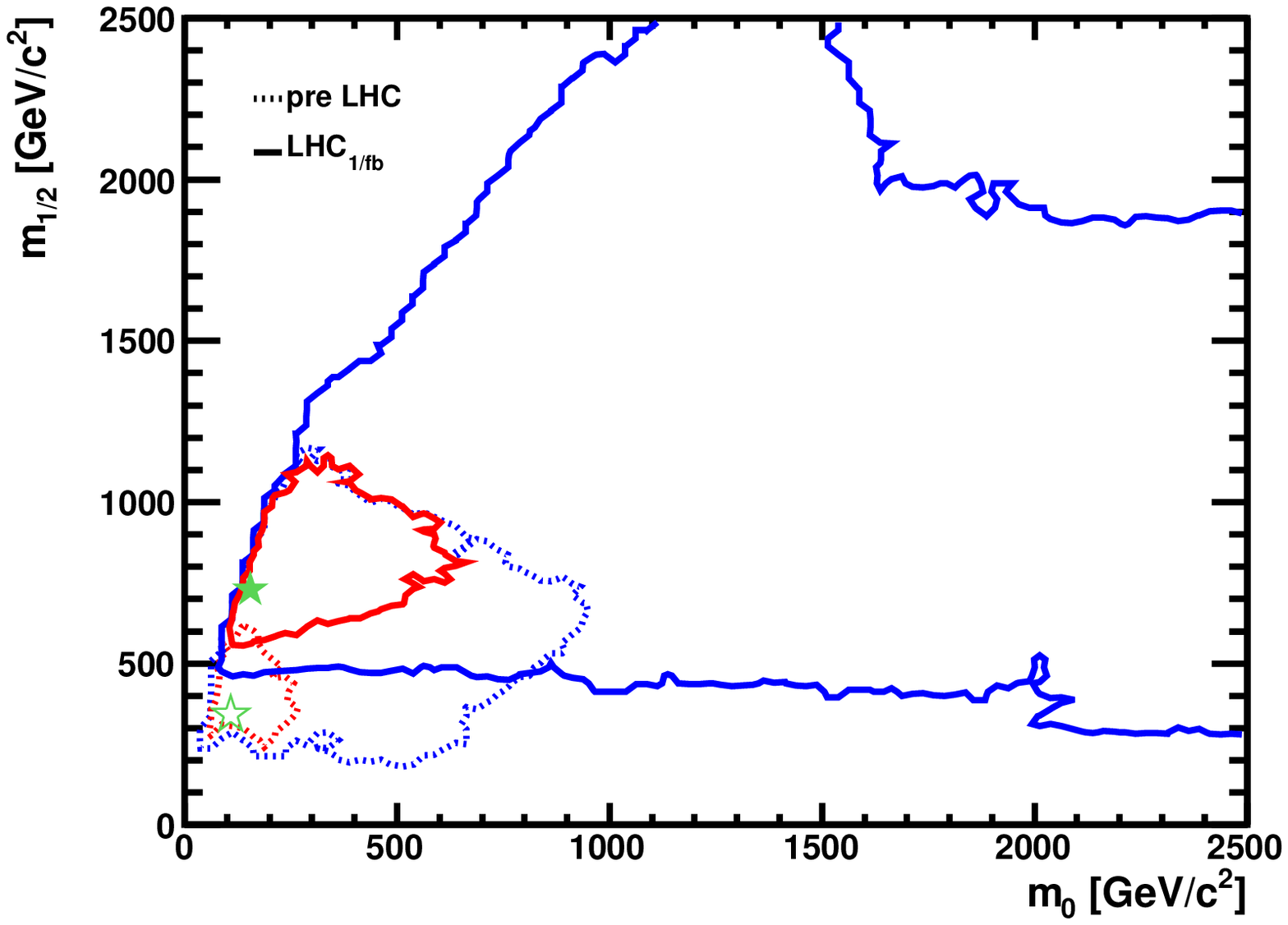}}
\vspace{-1cm}
\caption{\it The $(m_0, m_{1/2})$ planes in the CMSSM (left) and the
  NUHM1 (right). In each plane, the best-fit point after incorporation of the
  LHC$_{\rm 1/fb}$ constraints is indicated by a filled green star, and the
  pre-LHC fit~\protect\cite{mc4} by an open star. The $\Delta \chi^2= 2.30$ and $5.99$
  contours, commonly interpreted as the boundaries of the 68 and 95\% CL regions, are indicated
  in red and blue, respectively, the solid lines including the
  LHC$_{\rm 1/fb}$ data and the dotted lines showing the pre-LHC fits.
}
\label{fig:6895}
\end{figure*}

Fig.~\ref{fig:6895} displays contours with $\Delta \chi^2= 2.30$ (red) and 5.99 (blue)
relative to the minimum values of $\chi^2$ at the best-fit points (denoted by green stars) in the
$(m_0, m_{1/2})$ planes for the CMSSM and NUHM1~\footnote{The NUHM1 analysis includes both
the dedicated NUHM1 sample, which is efficient for smaller values of $(m_0, m_{1/2})$,
and the basic set of CMSSM points, which provide extra NUHM1 sampling at larger values of $(m_0, m_{1/2})$.}. 
Such contours are
commonly interpreted as 68 and 95\% CL contours.
The solid lines are the contours after incorporation of the LHC$_{\rm 1/fb}$ results,
and the dotted lines are the CL contours obtained from an analysis of the pre-LHC and pre-XENON100
data~\cite{mc4}. Consequently, the differences in the contours 
show the {\em full} impact of the $\sim 1$/fb data set
of LHC data.
The crinkles in these contours give an indication
of the sampling uncertainties in our analysis.

We see that 
the new best-fit points with $(m_0, m_{1/2}) = (450, 780) \gev$ in the CMSSM and
$(150, 730) \gev$ in the NUHM1 (denoted by solid green stars) 
lie well within the previous 95\% CL region.
On the other hand, the pre-LHC best-fit points with
$(m_0, m_{1/2}) = (90, 360) \gev$ in the CMSSM  and $(110, 340) \gev$ in
the NUHM1 (denoted by open
stars), lie far outside the regions allowed by the LHC$_{\rm 1/fb}$ data.
Thus, we see that there is now significant tension between the LHC$_{\rm 1/fb}$ and pre-LHC data sets.
The full set of parameters of the post-LHC$_{\rm 1/fb}$ and pre-LHC 
best-fit points are shown in Table~\ref{tab:bestfits}~\footnote{The parameters of the pre-LHC best-fit CMSSM and NUHM1 points given in Table~\ref{tab:bestfits} differ by up to 1 $\sigma$ from those given in~\cite{mc6}. These differences are caused primarily by changes in the data inputs.}.

In both the CMSSM and the NUHM1, as we discuss later, the 68\% CL upper limits on
$m_{1/2}$ in Fig.~\ref{fig:6895} are largely driven by \gmt, and the 95\% CL upper limits
are largely driven by the relic density constraint, with large-$m_{1/2}$ points lying in the
heavy-Higgs rapid-annihilation funnel at large $\tan \beta$.
The fact that much larger regions in the $(m_0, m_{1/2})$ are now
allowed at the 95\% CL, as compared to the pre-LHC fit, indicates that the
tension between \gmt, favouring relatively low SUSY scales, and 
the direct search limits, favouring larger SUSY scales, has
significantly reduced the sensitivity of the fits within the CMSSM and
the NUHM1 for constraining the SUSY parameters.
Since the $\chi^2$ values of
the best-fit points are significantly higher (see Table~\ref{tab:bestfits}), 
and consequently the $\chi^2$
distribution towards higher values of $m_0$ and $m_{1/2}$ is much 
flatter than in the pre-LHC case, the precise location of the 
68\% and 95\% CL contours is less precisely determined than
before.
The narrower range of $m_0$ allowed in the NUHM1 at the 68\% CL, 
compared to the CMSSM, is due to the
appearance of a `pit' with a
lower absolute value of $\chi^2$ that is attainable in this model thanks to its flexibility in
reconciling the \gmt\ with the $\neu{1}$ LSP and other constraints by deviating from Higgs mass universality. As in our
previous analysis including 2010 LHC and XENON100 data~\cite{mc6}, we find no distinct 
enhancement of the likelihood in the focus-point region at large $m_0$.

\begin{table*}[!tbh!]
\renewcommand{\arraystretch}{1.5}
\begin{center}
\begin{tabular}{|c||c|c|c|c|c|c||c|c|} \hline
Model & Minimum & Prob- & $m_{1/2}$ & $m_0$ & $A_0$ & $\tb$ & $\Mh$ (GeV) \\
      & $\chi^2$/d.o.f.& ability & (GeV) & (GeV) & (GeV) & & (no LEP)\\ 
\hline \hline
CMSSM  pre-LHC 
    & 21.5/20 & {37\%} & $360_{-100}^{+180}$ & $90_{-50}^{+220}$ 
    & $-400^{+730}_{-970}$ & $15_{-9}^{+15}$ & {$111.5_{-1.2}^{+3.5}$}\\
CMSSM  LHC$_{\rm 1/fb}$     
    & 28.8/22 & {15\%} & $780_{-270}^{+1350}$ & $450_{-320}^{+1700}$ 
    & $-1100^{+3070}_{-3680}$ & $41_{-32}^{+16}$ & {$119.1_{-2.9}^{+3.4}$}\\
Linear $\Delta$ \bsg\ 
    &  28.0/22  &   18\%    &  $720^{+1170}_{-230}$  & $420^{+1270}_{-270}$  
    &  $-1100^{+2180}_{-2750}$  &  $39^{+18}_{-22}$  &   {$118.6_{-1.9}^{+3.9}$} \\
\gmt\ neglected &  
    21.3/{20}  &   {38}\%    &  $2000^{+}_{-}$  & $1050^{+}_{-}$  
    & $430^{+}_{-}$ &  $22_-^+$  &   {$124.8_{-10.5}^{+3.4}$} \\
Both  
    &  20.5/{20}  &  {43}\%    &  $1880^{+}_{-}$  & $1340^{+}_{-}$       
    &   1890$^{+}_{-}$  &  47$^{+}_{-}$  &   {$126.1_{-6.3}^{+2.1}$} \\
\hline
NUHM1 pre-LHC    
    & 20.8/18 & 29\% & $340_{-110}^{+280}$ & $110_{-30}^{+160}$ 
    & $520^{+750}_{-1730}$ & $13_{-6}^{+27}$ & {$118.9_{-11.4}^{+1.1}$} \\
NUHM1 LHC$_{\rm 1/fb}$     
    & 27.3/21 & 16\% & $730_{-170}^{+630}$ & $150_{-50}^{+450}$ 
    & $-910^{+2990}_{-1170}$ & $41_{-24}^{+16}$ & {$118.8_{-1.1}^{+2.7}$} \\
Linear $\Delta$ \bsg\ 
    &  26.6/21  &  18\%    &  $730_{-90}^{+220}$  &  $150_{-20}^{+80}$  
    &  $ -910^{+2990}_{-1060}$  & $41^{+16}_{-22}$  &  {$118.8_{-1.3}^{+3.1}$} \\
\gmt\  neglected 
    &  20.3/{19}  &  {38}\%    &  $2020^{+}_{-}$  &  $1410^{+}_{-}$  
    &  $ 2580^{+}_{-}$  & $48^{+}_{-}$  &  {$126.6_{-1.9}^{+0.7}$} \\
Both  
    &  19.5/{19}  & {43}\%    &  $2020^{+}_{-}$  &  $1410^{+}_{-}$  
    &   2580$^{+}_{-}$  & 48$^{+}_{-}$  &  {$126.6_{-1.9}^{+0.7}$} \\
\hline
\end{tabular}
\caption{\it Comparison of the best-fit points found in the CMSSM and NUHM1 pre-LHC (including the 
upper limit on \bmm\ available then), and with the LHC$_{\rm 1/fb}$ data set (also including the XENON100 constraint)
using the standard implementations of the \gmt and \bsg\ constraints, followed by variants first
adding linearly the theoretical and experimental errors in \bsg\ and then dropping \gmt, and finally combining both variants.
The errors for fits dropping \gmt\ are large and asymmetric, and are not indicated. The
predictions for $\Mh$ do not include the constraint from the direct LEP Higgs search,
and have an estimated theoretical error of $\pm 1.5 \gev$.
}
\label{tab:bestfits}
\end{center}
\end{table*}

The absolute values of $\chi^2$ at the best-fit points for the 
pre-LHC case
and for the LHC$_{\rm 1/fb}$ data set using our standard implementations of
the \gmt\ and \bsg\ constraints are given in Table~\ref{tab:bestfits}.
Our updated analysis of the pre-LHC data set yields $\chi^2$/d.o.f. $= 21.5/20 (20.8/18)$ in the
CMSSM and NUHM1, respectively, corresponding to $p$-values of 37\% and 29\%.
On the other hand, using the LHC$_{\rm 1/fb}$ data set, we find that the minimum values of $\chi^2$
are significantly larger than the numbers of effective degrees of freedom
in the fits, which are also shown in Table~\ref{tab:bestfits}: $\chi^2$/d.o.f. $= 28.8/22 \, (27.3/21)$
for the CMSSM and NUHM1, respectively~\footnote{For technical reasons, the $\Gamma_Z$ constraint was
not included in our previous fits, leading to changes of one unit in the numbers of effective degrees of freedom
in the fits.}. Correspondingly, the best fits
have significantly reduced probability values, $\sim 15$\% in the CMSSM and
$\sim 16$\% in the NUHM1~\footnote{The $p$-values for the VCMSSM and 
mSUGRA are somewhat smaller, as was found previously~\cite{mc6}:
$\chi^2$/d.o.f. $= 31.2/{23} (32.5/{23})$, respectively, corresponding
to $p$-values of {12\%} and {9\%}. We do 
not discuss these models further, except to comment that the light-Higgs funnel
region found previously in these models is now excluded by ATLAS
data~\cite{ATLASsusy}, in particular.}. 
If we combine the \bsg\ errors linearly instead of quadratically, the values of $\chi^2$ decrease by $0.8(0.7)$
in the CMSSM and NUHM1, respectively, and the $p$-values increase modestly to 18\% in both cases. On the
other hand, if we drop the \gmt\ constraint, we find $\chi^2$/d.o.f $= 21.3/20
(20.3/19)$, respectively,
corresponding to $p$-values of 38\% in both cases. Thus, the qualities of our best fits are not very sensitive
to the treatment of \bsg, but are much more sensitive to the inclusion of \gmt, as we discuss later in
more detail.

The degrees of non-universality, $r \equiv m_H^2/m_0^2$, for the best-fit NUHM1 points in Table~\ref{tab:bestfits} are as follows:
$r = -57$ (pre-LHC), $r = -54$ (LHC$_{\rm 1/fb}$), $r = -54$ (linear \bsg\ error combination), $r = -0.39$ (dropping \gmt\ constraint),
$r = -0.39$ (including both variants). Since $r$ is quite poorly constrained, we do not quote its uncertainties.

We note that the best-fit values of $\Mh$ are significantly higher in the CMSSM and NUHM1 fits dropping \gmt, 
with their large values of $(m_0, m_{1/2})$, than in the fits that include the \gmt\ constraint. 
An LHC measurement of $\Mh$ could provide a diagnostic discriminating between models with light and heavy
spectra of third-generation squarks, and help forecast the likelihood of
discovering these sparticles in future LHC runs. Within the CMSSM
and NUHM1, measuring $\Mh$ could advise us whether to take \gmt\ at face
value or, conversely, hint towards extensions of those models. 

\begin{table*}[htb!]
\renewcommand{\arraystretch}{1.0}
\begin{center}
\small{
\begin{tabular}{|c|c|c|c|c|c|} \hline
Observable & Source & Constraint & $\Delta \chi^2$ & $\Delta \chi^2$ & $\Delta \chi^2$ \\
& Th./Ex. & & (CMSSM) & (NUHM1) & (\SM) \\
\hline \hline
$\mt$ [GeV] & \cite{mt1731} & $173.2 \pm 0.90$ & 0.05 & 0.06 & {-} \\
\hline
$\Delta\alpha_{\rm had}^{(5)}(\MZ)$ 
     &\cite{newDavier} &$0.02749 \pm 0.00010$ & 0.009 & 0.004 &  {-} \\
\hline
$\MZ$ [GeV]         
     &\cite{lepewwg} &$91.1875\pm0.0021$  & 2.7$\times10^{-5}$ & 0.26 &  {-} \\ 
\hline\hline
$ \Gamma_{Z}$ [GeV]    
     &\cite{Svenetal}   
     /\cite{lepewwg} &$2.4952\pm0.0023\pm0.001_{\rm SUSY}$   & 0.078  & 0.047 & 0.14 \\ 
\hline
$\sigma_{\rm had}^{0}$ [nb] 
     &\cite{Svenetal} 
     /\cite{lepewwg} &$41.540\pm0.037$    & 2.50 & 2.57 & 2.54 \\
\hline
$R_l$ &\cite{Svenetal}   
      /\cite{lepewwg} &$20.767\pm0.025$    & 1.05 & 1.08 & 1.08 \\ 
\hline
$ A_{\rm fb}(\ell)$ &\cite{Svenetal}   
                   /\cite{lepewwg} &$0.01714\pm0.00095$ & 0.72 & 0.69 & 0.81 \\ 
\hline
$ A_{\ell}(P_\tau)$ &\cite{Svenetal}   
                  /\cite{lepewwg} & 0.1465 $\pm$ 0.0032 & 0.11 & 0.13 & 0.07 \\ 
\hline
$ R_{\rm b}$ &\cite{Svenetal}   
            /\cite{lepewwg} & 0.21629 $\pm$ 0.00066 & 0.26 & 0.29 & 0.27 \\ 
\hline
$ R_{\rm c}$ &\cite{Svenetal}   
            /\cite{lepewwg} & 0.1721 $\pm$ 0.0030 & 0.002 & 0.002 & 0.002 \\ 
\hline
$ A_{\rm fb}({b})$ &\cite{Svenetal}   
                  /\cite{lepewwg} & 0.0992 $\pm$ 0.0016 & 7.17 & 7.37 & 6.63 \\ 
\hline
$ A_{\rm fb}({c})$ &\cite{Svenetal}   
                  /\cite{lepewwg} & 0.0707 $\pm$ 0.0035 & 0.86 & 0.88 & 0.80 \\ 
\hline
$ A_{b}$  &\cite{Svenetal}   
          /\cite{lepewwg} & 0.923 $\pm$ 0.020 & 0.36 & 0.36 & 0.35 \\ 
\hline
$ A_{c}$ &\cite{Svenetal}   
         /\cite{lepewwg} & 0.670 $\pm$ 0.027 & 0.005 & 0.005 & 0.005 \\ 
\hline
$ A_\ell({\rm SLD})$ &\cite{Svenetal}   
                     /\cite{lepewwg} & 0.1513 $\pm$ 0.0021 & 3.16 & 3.03 & 3.51 \\ 
\hline
$ \sin^2 \theta_{\rm w}^{\ell}(Q_{\rm fb})$ 
        &\cite{Svenetal}   
        /\cite{lepewwg} & 0.2324 $\pm$ 0.0012 & 0.63 & 0.64 & 0.59 \\ 
\hline
$\MW$ [GeV]
     &\cite{Svenetal}
     /\cite{lepewwg} & $80.399 \pm 0.023\pm0.010_{\rm SUSY}$  & 1.77 & 1.39 & 2.08 \\
\hline\hline
$ a_{\mu}^{\rm EXP} - a_{\mu}^{\rm SM}$
     &\cite{g-2}
     /\cite{newBNL,newDavier}
     &$(30.2 \pm 8.8 \pm 2.0_{\rm SUSY})\times10^{-10}$ & 4.35 & 1.82 & 11.19 (N/A) \\
\hline
$\Mh$ [GeV]
     & \cite{FeynHiggs}
     / \cite{Barate:2003sz,Schael:2006cr} & 
     $> 114.4 [\pm 1.5_{\rm SUSY}]$ & 0.0 & 0.0 & 0.0 \\
\hline\hline
BR$_{\rm b \to s \gamma}^{\rm EXP/SM}$
     &\cite{bsgth}
     /\cite{HFAG}
     &$1.117 \pm 0.076_{\rm EXP}$  & 1.83 & 1.09 & 0.94 \\
     & 
     &\phantom{1.117}$\pm 0.082_{\rm SM}\pm 0.050_{\rm SUSY}$  &  & & \\
\hline
BR$(B_{s} \to \mu^{+} \mu^{-})$
     &\cite{SuFla} /\cite{CMSLHCb}
     & CMS \& LHCb   & 0.04 & 0.44 & 0.01 \\
\hline
BR$_{\rm B \to \tau\nu}^{\rm EXP/SM}$
     &\cite{SuFla} 
     /\cite{HFAG}
     &  $1.43 \pm 0.43_{\rm EXP+TH}$  & 1.43 & 1.59 & 1.00 \\
\hline
${\rm BR}({B_d \to \mu^+ \mu^-})$
     & \cite{SuFla} /\cite{HFAG}
     & $ < 4.6 [\pm 0.01_{\rm SUSY}] \times 10^{-9}$ & 0.0 & 0.0 & 0.0 \\
\hline
$\br_{B \to X_s \ell \ell}^{\rm EXP/SM}$
     & \cite{Bobeth}/\cite{HFAG}
     & $0.99 \pm 0.32$ & 0.02 & $\ll 0.01$ & $\ll 0.01$\\
\hline
BR$_{K \to \mu \nu}^{\rm EXP/SM}$
     & \cite{SuFla} /\cite{Antonelli}
     & $1.008 \pm 0.014_{\rm EXP+TH}$   & 0.39 & 0.42 & 0.33 \\
\hline
BR$_{K \to \pi \nu \bar{\nu}}^{\rm EXP/SM}$
     & \cite{Buras:2000qz}/\cite{Artamonov:2008qb}
     & $ < 4.5 $ & 0.0 & 0.0 & 0.0 \\
\hline
$\Delta M_{B_s}^{\rm EXP/SM}$
     & \cite{Buras:2000qz} /\cite{Bona:2007vi,Lubicz:2008am}
     & $0.97 \pm 0.01_{\rm EXP} \pm 0.27_{\rm SM}$ & 0.02 & 0.02 & 0.01 \\
\hline
$\frac{\Delta M_{B_s}^{\rm EXP/SM}}
           {\Delta M_{B_d}^{\rm EXP/SM}}$
     & \cite{SuFla}
     /\cite{HFAG,Bona:2007vi,Lubicz:2008am}
     & $1.00 \pm 0.01_{\rm EXP} \pm 0.13_{\rm SM} $  & $\ll 0.01$ & 0.33 & $\ll 0.01$ \\
\hline
$\Delta \epsilon_K^{\rm EXP/SM}$
     & \cite{Buras:2000qz} /\cite{Bona:2007vi,Lubicz:2008am}
     & $1.08 \pm 0.14_{\rm EXP+TH}$ & 0.27 & 0.37 & 0.33 \\
\hline \hline
$\Omega_{\rm CDM} h^2$
     &\cite{MicroMegas} 
     /\cite{Komatsu:2010fb}
         &$0.1120 \pm 0.0056 \pm 0.012_{\rm SUSY}$ & 8.4$\times10^{-4}$ & 0.1 & N/A \\
\hline
$\ssi$ & \cite{XE100} & $(\mneu{1}, \ssi)$ plane & 0.13 & 0.13 & N/A \\
\hline\hline
jets + $\ETslash$ & \cite{CMSsusy,ATLASsusy} & $(m_0, m_{1/2})$ plane & 1.55 & 2.20 & N/A \\
\hline
$H/A, H^\pm$ & \cite{CMSHA} & $(\MA, \tb)$ plane & 0.0 & 0.0 & N/A \\
\hline \hline
Total $\chi^2$/d.o.f. & All & All & 28.8/22 & 27.3/21 & 32.7/{23}~(21.5/{22}) \\
$p$-values & & & 15\% & 16\% & {9\%} ({49\%}) \\
\hline
\end{tabular}
\caption{\it List of experimental constraints used in this work, including experimental and (where applicable)
theoretical errors: supersymmetric theory uncertainties in the interpretations of one-sided experimental limits
are indicated by [...]. Also shown are the contributions that these constraints
 make to the total $\chi^2$ functions at the best-fit points in the CMSSM and NUHM1, respectively, and
 (for comparison) in the SM limit of the CMSSM (called \SM) including (excluding) \gmt. 
 The total values of $\chi^2$, the numbers of degrees of freedom and the
 $p$-values at these points are shown in the two bottom rows.
  \label{tab:chi2}}  
  }
\end{center}

\end{table*}

Table~\ref{tab:chi2} displays the contributions to the
total $\chi^2$ of each of the observables at the best-fit points in the CMSSM and NUHM1, revealing where this tension
originates. We note, as already mentioned above, that there is an 
important contribution
to the $\chi^2$ function coming from the \gmt\ constraint, as
well as the LHC$_{\rm 1/fb}$ constraints. We return later to a more complete
discussion of the tension between these constraints, and also of the treatment
of the \bsg\ constraint.

Table~\ref{tab:chi2} also displays a similar $\chi^2$ breakdown evaluated assuming the central
values of $m_t, \Delta\alpha_{\rm had}^{(5)}(\MZ)$ and $\MZ$ for a
CMSSM point at very large $(m_0, m_{1/2}) = (15, 15)$~TeV, $A_0 = 100 \gev$ and
$\tb = 10$, near the limit in which its predictions coincide
with those of the SM for the same value of $\Mh$. This is very similar to the global
best-fit value of $\Mh$ in the SM obtained incorporating the limits from the direct 
Higgs searches at LEP, the Tevatron and the LHC~\cite{Gfitter}. 
Using the SM limit of the CMSSM within the {\tt MasterCode}
framework ensures that this
evaluation of $\chi^2$ in the \SM\ can be compared directly with 
those at the best-fit
points in the CMSSM and NUHM1.
In the \SM\ case we discard the constraints imposed by the cosmological dark matter density 
and XENON100,
since there is no way to explain dark matter with the SM. We also discard
the LHC missing-energy and $H/A$ constraints, but all other constraints are
  kept. Consequently, 
we {\it do} list the contribution to $\chi^2$ from \gmt. If this is included (omitted), the
global $\chi^2$ for the \SM\ is {32.7 (21.5)}. The number of degrees of 
freedom for the ``SM" is consequently 23 (22) and the $p$-value is {9\%} ({49\%}).
We observe that the $p$-value for the CMSSM is rather larger than that for the
``SM'' if \gmt\ is included, though similar if \gmt\ is not
included in the ``SM"  and MSSM analyses.

We also note that one of the big contributors to the global $\chi^2$
functions for all the models is $A_{\rm fb}(b)$, 
which contributes $\Delta \chi^2 \sim 7$ to
each of the fits, suppressing all their $p$-values.
Prior to the LHC results, the CMSSM yielded a significant improvement to $\chi^2$,
with the dominant contribution coming from \gmt\ ($\Delta \chi^2 = -10.8$).  Other contributing observables
were $\MW$ ($\Delta \chi^2 = -1.6$) and $A_\ell({\rm SLD})$  ($\Delta \chi^2 = -1.3$),
though $A_{\rm fb} (b)$ was somewhat worse in the CMSSM  ($\Delta \chi^2 = 2.1$).
The post-LHC comparison with the ``SM" is shown in Table~\ref{tab:chi2}. 
Looking at the entries for the electroweak precision observables,
{the only significant change is now that for \gmt ($\Delta \chi^2 = -6.8$),
with all other observables showing changes $|\Delta \chi^2| < 1$. Thus, when \gmt is dropped
as a constraint, the resulting $\chi^2$ for the best-fit points in the CMSSM and ``SM" 
are very similar. In the case of the post-LHC NUHM1, we also see a large drop
in $\chi^2$ relative to the ``SM" due to \gmt  ($\Delta \chi^2 = -9.4$) and again all
others give $|\Delta \chi^2| < 1$.
We also note that in both
the CMSSM and NUHM1 the best fits receive significant contributions from the LHC
$\ETslash$ + jets searches.

To illustrate further the impact of LHC$_{\rm 1/fb}$ experimental constraints relative to 
pre-LHC preferred regions, we display  in Fig.~\ref{fig:p} colour-coded 
contours of approximate%
\footnote{Strictly speaking, transforming a $\chi^2$ value to a $p$-value, 
using a specified number of degrees of freedom, is valid for 
Gaussian-behaved constraints. Because some of the experimental limits are one-sided and modelled in a non-Gaussian 
manner as previously described, the $p$-values reported here can therefore only be considered approximate.}%
~$p$-values from our global fits
for the CMSSM and NUHM1. Care is taken to count the effective number of degrees of 
freedom at each point, considering all constraints that contribute 
non-trivially to the $\chi^2$ functions.  Thus, for example, we drop the
contribution of the LHC$_{\rm 1/fb}$ missing-energy constraints where they contribute $\Delta \chi^2 < 0.1$, causing
the visible changes in shading along drooping diagonal lines in both panels of
Fig.~\ref{fig:p}. (This cut is applied {\em only} to the LHC$_{\rm 1/fb}$
missing-energy constraint.)
Substantial non-zero $p$-values are observed to extend to high 
$m_0$ and $m_{1/2}$, in both pre- (upper panels) and post-LHC$_{\rm 1/fb}$ (lower panels), and 
both the CMSSM (left panels) and NUHM1 (right panels) models.  
As also seen earlier, the primary effect of the LHC$_{\rm 1/fb}$ searches for jets + $\ETslash$ is most 
evident for $m_{1/2}$, preferring higher values than that predicted by the pre-LHC
global fits.
At even higher ($m_0$, $m_{1/2}$), beyond the drooping diagonal line, slight
increases in approximate $p$-values appear when comparing the pre-LHC results
with the post-LHC$_{\rm 1/fb}$ and post XENON100 results.  
This is due partly to the experimental
constraint on \bmm, which is nearing the SM prediction. Regions of the CMSSM and NUHM1
parameter spaces approaching the
high-mass decoupling limit receive a non-zero contribution from  \bmm\ in the
post-LHC$_{\rm 1/fb}$ era, resulting in $\Delta \chi^2 < 1$, which actaually improves the overall  $\chi^2$ per
effective degree of freedom.
Additionally, the XENON100 constraint slightly prefers high mass scales, so as to
  accomodate the small ``excess'' in events, also resulting in 
slightly better values of $\chi^2$ per effective degree of freedom.

\begin{figure*}[htb!]
\resizebox{8.5cm}{!}{\includegraphics{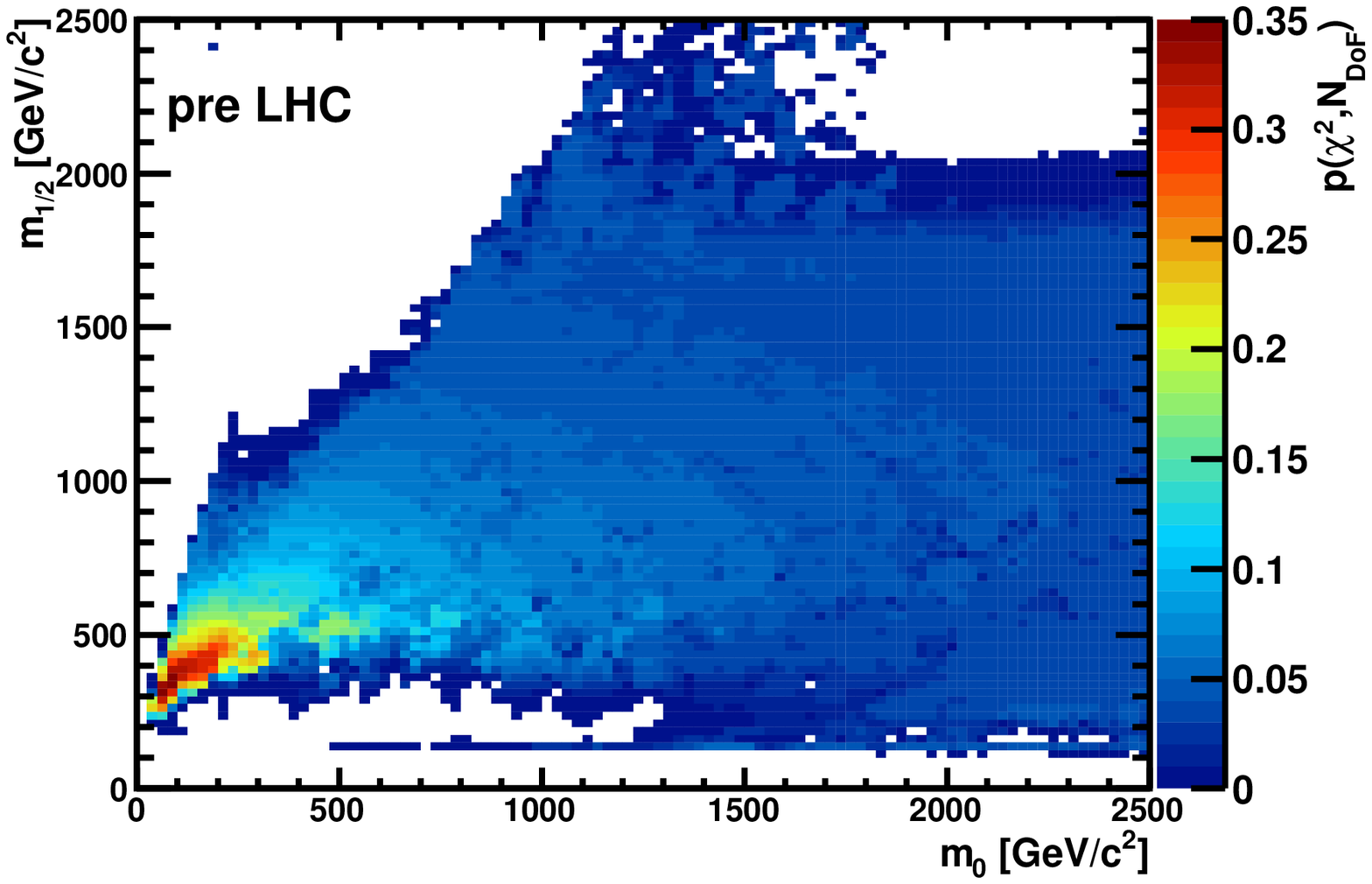}}
\resizebox{8.5cm}{!}{\includegraphics{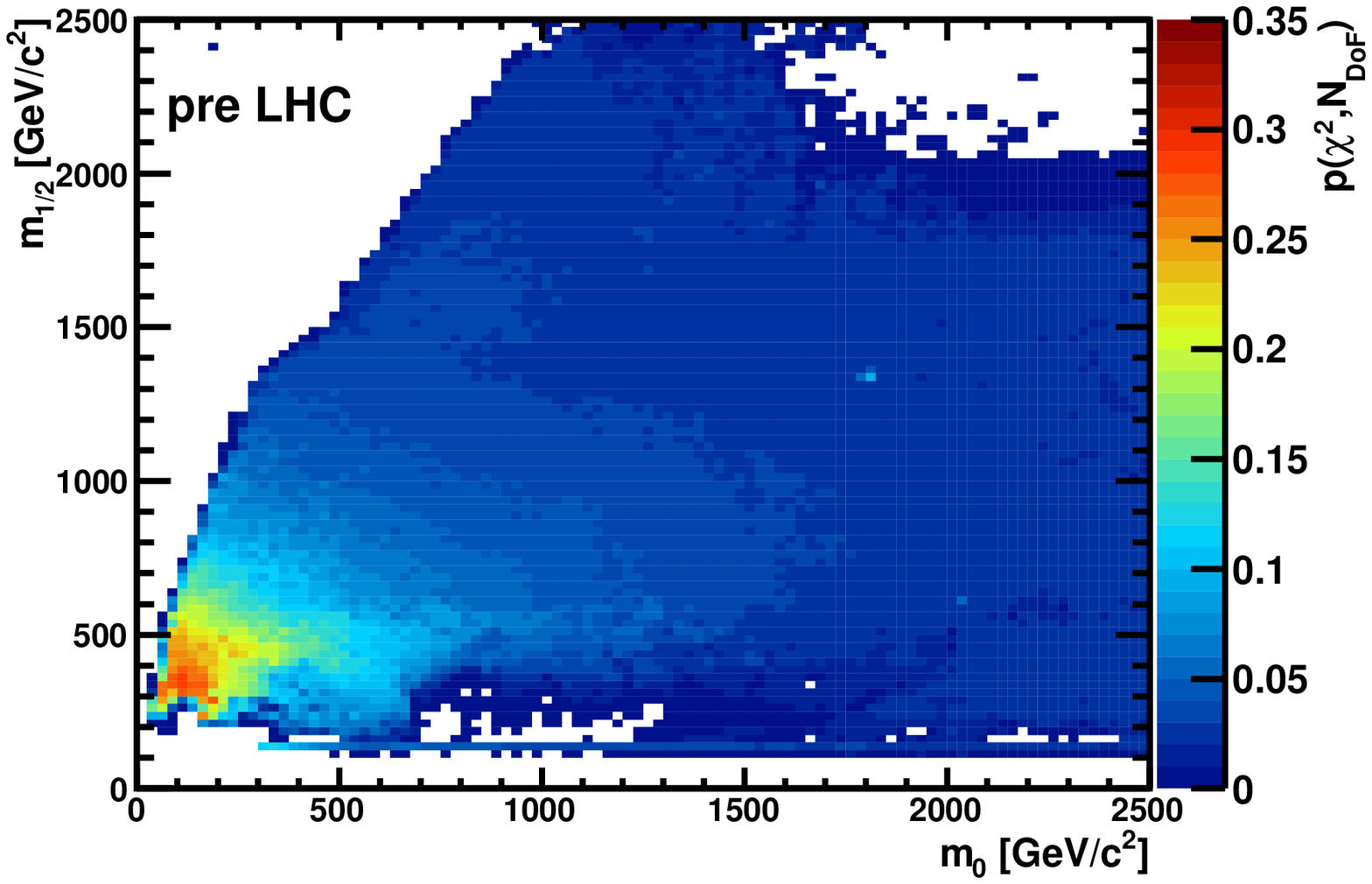}}
\resizebox{8.5cm}{!}{\includegraphics{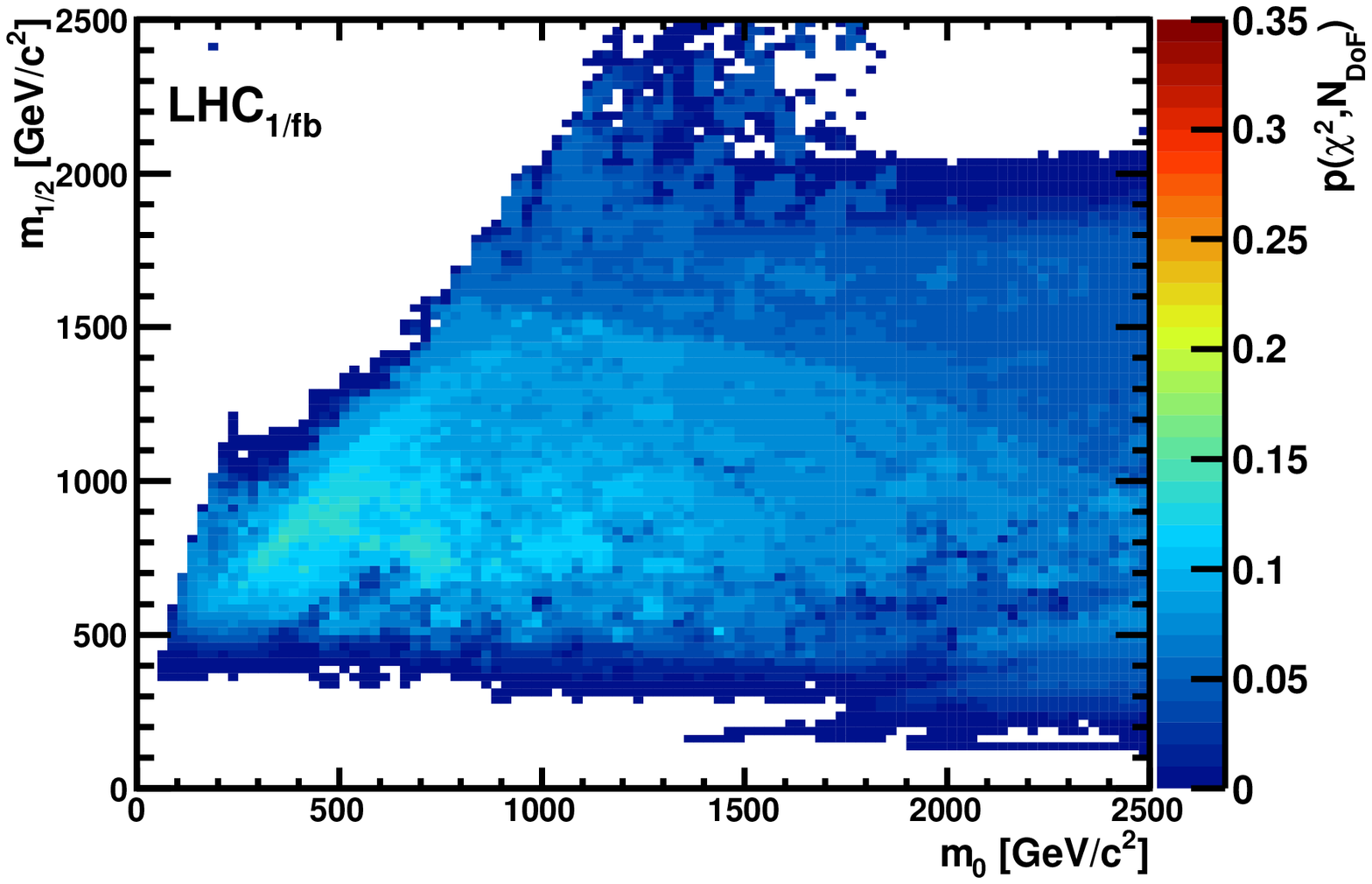}}
\resizebox{8.5cm}{!}{\includegraphics{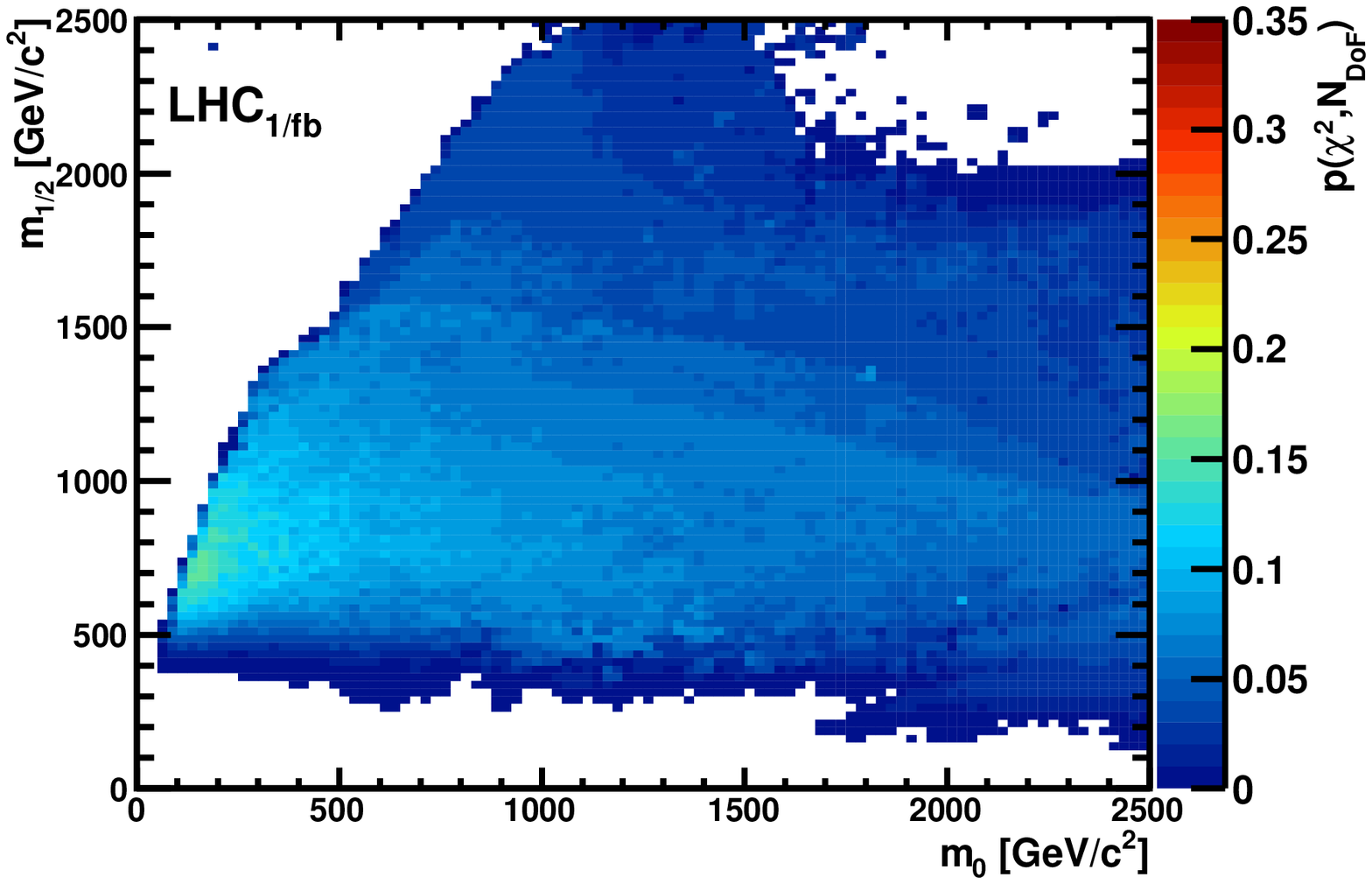}}
\vspace{-1cm}
\caption{\it The $(m_0, m_{1/2})$ planes in the CMSSM (left) and the
  NUHM1 (right), for the pre-LHC data set (upper) and LHC$_{\rm 1/fb}$ data set (lower). 
  In each plane, different regions are colour-coded
  according to the $p$-values found in our global fits. We note that in the LHC$_{\rm 1/fb}$ analysis
  the regions with $p > 0.05$ extend up to $m_{1/2} \sim 2000 \gev$
  in each case.
}
\label{fig:p}
\end{figure*}

Complementing the comparison of the $p$-values of the ``SM'', the CMSSM
and the NUHM1, we now use the standard F-test to test the utility of adding one
or several parameters to a model fit of data.
Given a set of data comprising $N$ observables
and a model using $m$ parameters, one may compute
$\chi^2(m)$  for $N-m$ degrees of freedom as done above. In general,
adding $r$ parameters produces a reduced value of $\chi^2(m+r)$, and
the difference between these two $\chi^2$ distributions is
itself a $\chi^2$ distribution for $r$ degrees of freedom.  
The F-statistic is defined by
\begin{equation}
F_\chi \; \equiv \; \frac{\chi^2(m) - \chi^2(m+r)}{\chi^2(m+r)/(N-m-r)} \; > \; 0.
\label{F}
\end{equation}
The probability that introducing the $r$ new parameters are warranted is 
found by integrating the F-distribution, $p_F(f,r,N-m-r)$, from $f=F_\chi$ to $\infty$.
We use the F-test to illustrate the relative preference for various models.

{In our case, for the \SM\ we have $\chi^2 = 32.7 (21.5)$ for 23 (22) 
degrees of freedom if \gmt\
is included in (omitted from) the fit. 
Using the CMSSM value of $\chi^2 = 28.8$ for 22 degrees of freedom,
we find $F_\chi = {2.98}$, and the probability that switching to the CMSSM is warranted is {$p_F = 90$}\%.
Correspondingly, using the NUHM1 value of $\chi^2 = 27.3$ for 21 degrees of freedom, we find
$F_\chi = {4.15}$, and the probability that switching to the NUHM1 is warranted is {$p_F = 97$}\%.
We can also compare the improvement in $\chi^2$ gained by moving from the 
CMSSM to the NUHM1. In this case the probability that the extra parameter
needed to define the NUHM1 model is preferred over the CMSSM case is 71\%.
Fig.~\ref{fig:F} uses shading to display values of $p_F$ in the $(m_0, m_{1/2})$ planes of the CMSSM and NUHM1.
On the basis of these plots, we find that present data may warrant
switching from the \SM\ to the CMSSM or NUHM1 
for values of $m_{1/2}$ out to  $\sim 1500 \gev$~\footnote{We note, however, that
these results may be too favourable to the CMSSM or NUHM1, since they do not include the impacts of the
many lower-sensitivity constraints from CMS and ATLAS. This problem could be avoided if the Collaborations publish
official combinations of the sensitivities of their searches.}. Beyond this range of $m_{1/2}$, the motivations for
these models would be significantly reduced. We also note that the F-test indicates that there would be no
advantage in switching to the CMSSM or the NUHM1 if \gmt\ were to be dropped.

\begin{figure*}[htb!]
\resizebox{8.5cm}{!}{\includegraphics{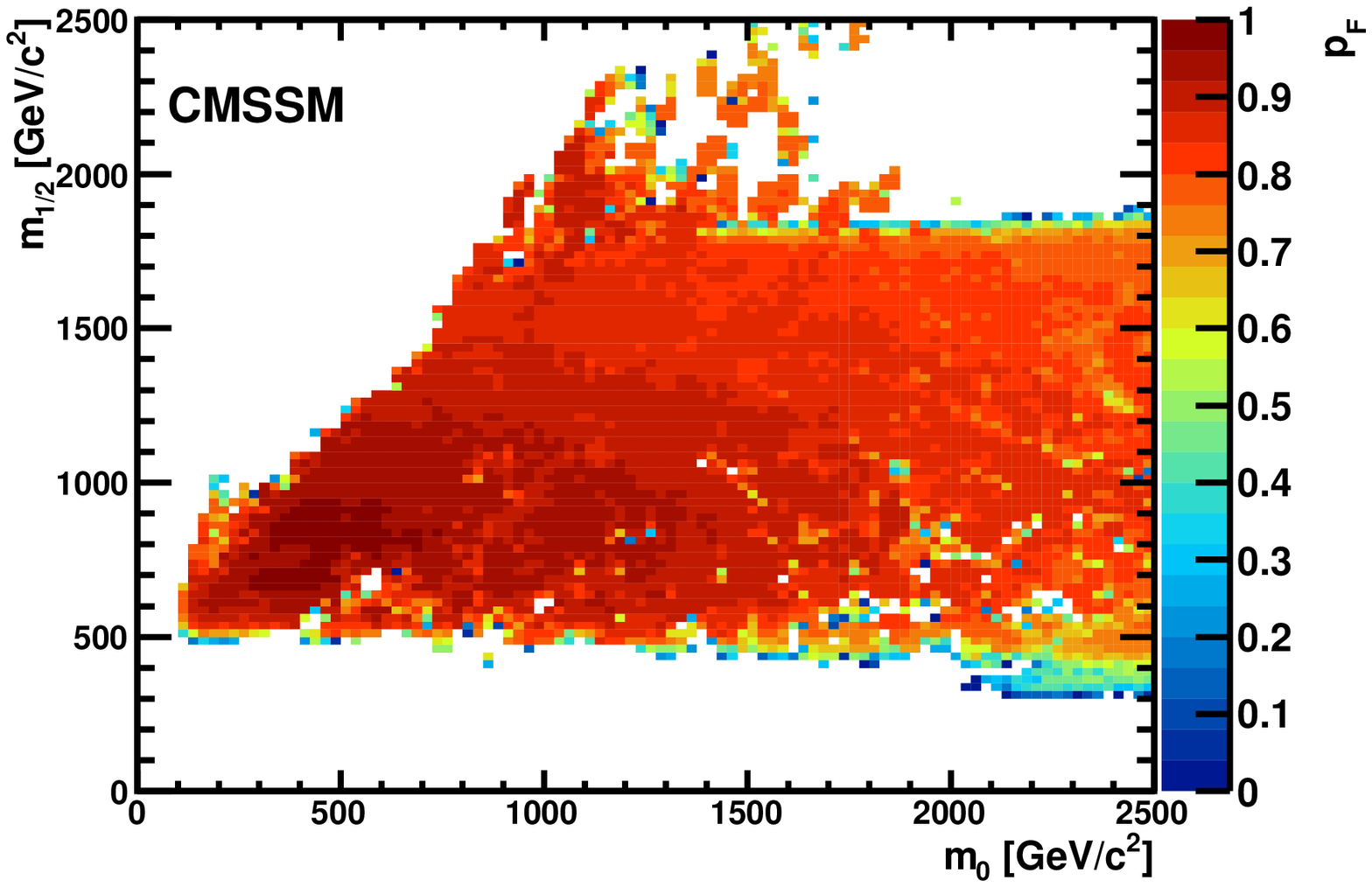}}
\resizebox{8.5cm}{!}{\includegraphics{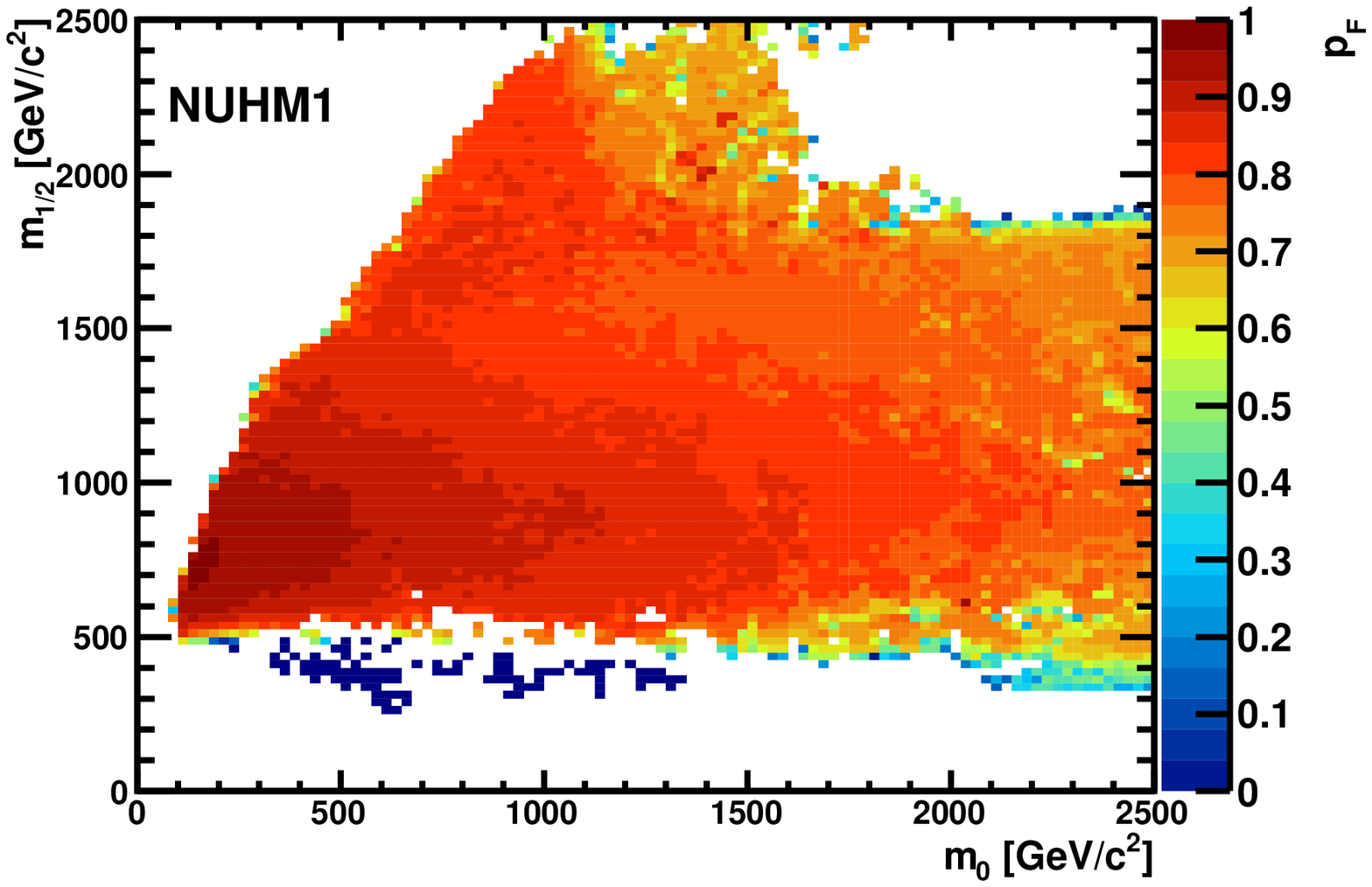}}
\vspace{-1cm}
\caption{\it The $(m_0, m_{1/2})$ planes in the CMSSM (left) and the
 NUHM1 (right). In each plane, different regions are colour-coded
 according to the values of $p_F$ found by applying the F-test to our global fits. We note that
 the regions with $p_F > 0.8$ extend up to $m_{1/2} \sim 1500 \gev$
 in each case.}
\label{fig:F}
\end{figure*}

An important part of the motivation for low-scale supersymmetry is to
alleviate the fine-tuning of the Higgs mass parameter in the Standard Model.
However, the problem of fine-tuning returns if the supersymmetric mass
scales become large~\cite{FT}. The required amount of fine-tuning is increased significantly 
in our LHC$_{\rm 1/fb}$ fits compared to our pre-LHC fits, principally because of
the increases in the best-fit values of $m_0$ and $m_{1/2}$. Specifically, in
the CMSSM our best pre-LHC fit required fine-tuning by a factor $\sim 100$, whereas
our best LHC$_{\rm 1/fb}$ fit requires fine-tuning by a factor $\sim 300$. The corresponding
numbers for the NUHM1 are $\sim 250$ pre-LHC and $\sim 600$ with the 
LHC$_{\rm 1/fb}$ data.

\subsection*{\it Uncertainties in the analysis}

In assessing the compatibility of the CMSSM and the
NUHM1 with the experimental data it is important also to
examine carefully the most important systematic
uncertainties in the constraints that drive the fit in this global likelihood analysis.
We saw in the Table~\ref{tab:chi2} that the most important contributions to the
global $\chi^2$ functions at the best-fit points in these models originate from the \gmt\ and
LHC$_{\rm 1/fb}$ constraints. 
Exploring this further, Fig.~\ref{fig:tension}
displays the CMSSM and NUHM1 $(m_0, m_{1/2})$ planes again, exhibiting the
contribution to $\chi^2$ from the \gmt\ constraint, evaluated for the
model parameter sets that minimize the total $\chi^2$ at each point in the plane.
Their shapes are different in the CMSSM and NUHM1, reflecting the
existence of the previously-mentioned `pit' in the latter model where $\chi^2$ may be reduced
by some judicious choice of the degree of Higgs non-universality.
{\it Prima facie}, there is tension between the \gmt\
constraint, which prefers small values of $(m_0, m_{1/2})$, and the LHC$_{\rm 1/fb}$
constraints, which prefer larger values of $(m_0, m_{1/2})$. This tension is
alleviated for larger values of $\tb$, which is why post-2010-LHC~\cite{mc6} and post-LHC$_{\rm 1/fb}$
fits have favoured larger values of $\tb$ than pre-LHC fits~\cite{mc4},
albeit with large uncertainties.

\begin{figure*}[htb!]
\resizebox{8.5cm}{!}{\includegraphics{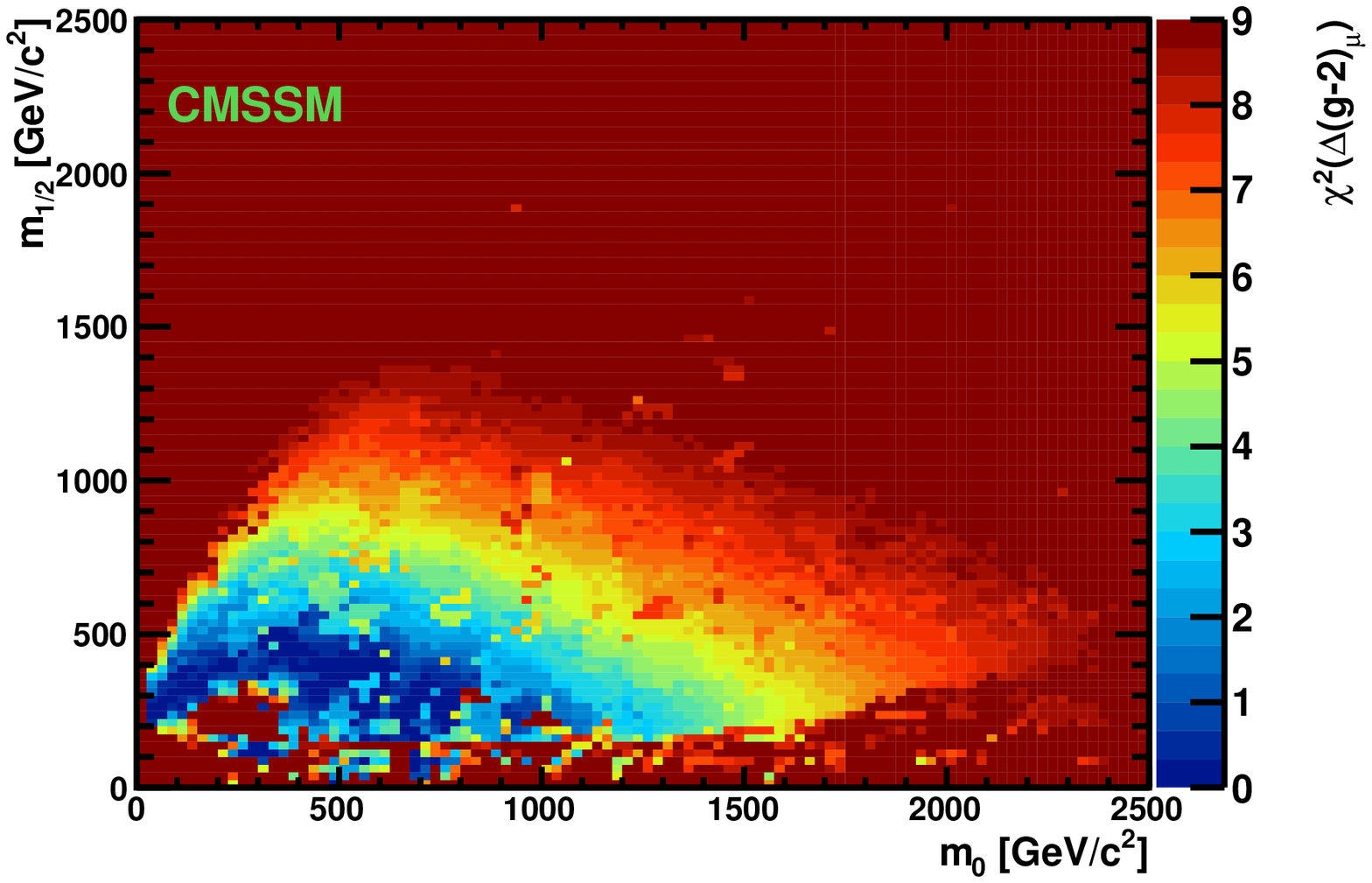}}
\resizebox{8.5cm}{!}{\includegraphics{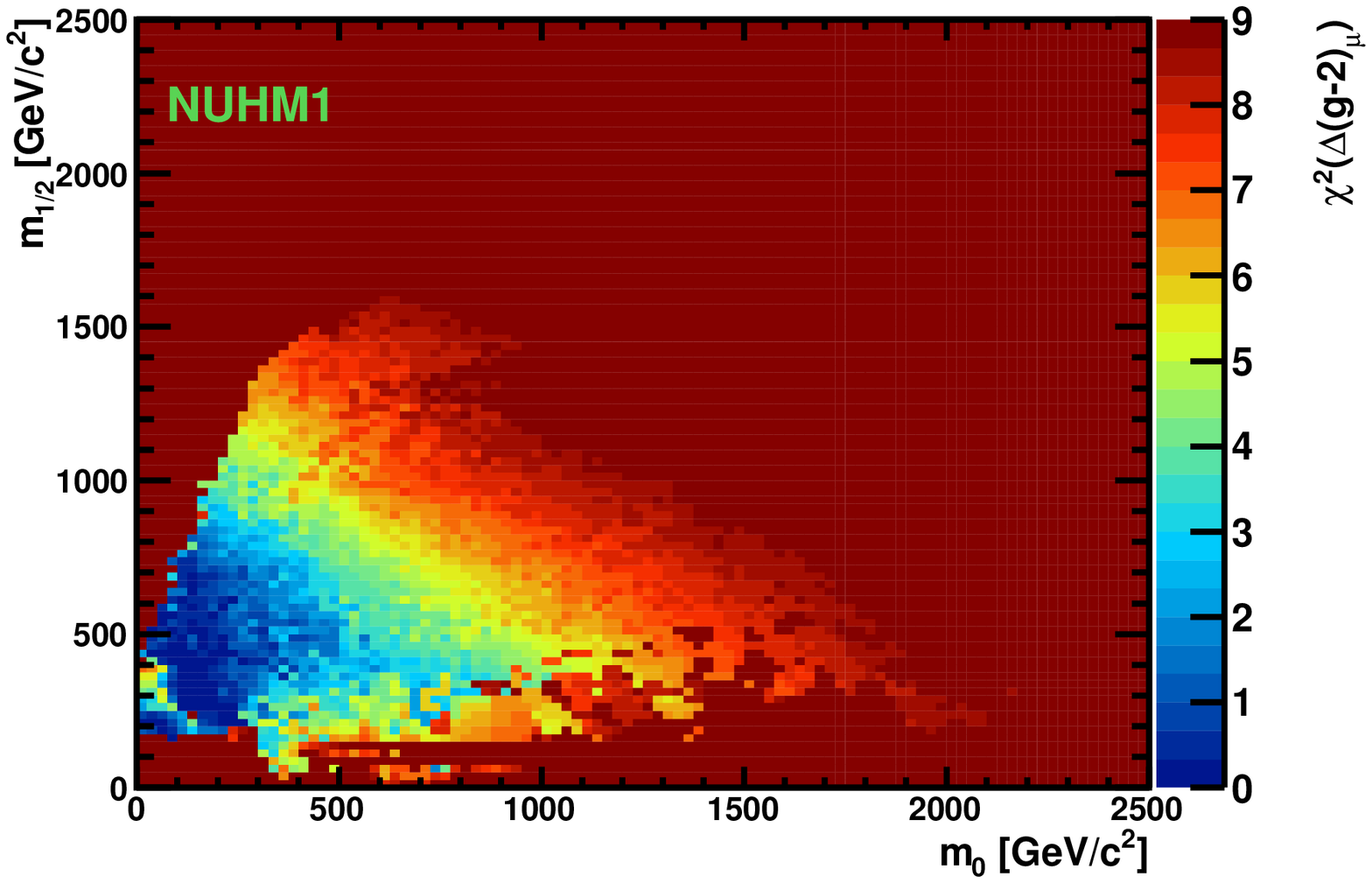}}\\
\vspace{-1cm}
\caption{\it The $(m_0, m_{1/2})$ planes in the CMSSM (left) and the
  NUHM1 (right), with shading displaying the contribution
  to the global $\chi^2$ function from \gmt\ as calculated using low-energy
  $e^+ e^-$ data to evaluate the SM contribution.
  These contributions are evaluated for the
  parameter sets that minimize $\chi^2$ at each point in the planes.
}
\label{fig:tension}
\end{figure*}

Fig.~\ref{fig:g-2} again displays the $(m_0, m_{1/2})$ planes in the CMSSM (left) and the
  NUHM1 (right), this time showing as solid lines the 68\% and 95\% CL contours
  obtained by dropping the \gmt\ constraint, the contours obtained applying the \gmt\ constraint as in
  Fig.~\protect\ref{fig:6895} being shown here for comparison as dotted lines.
  We see that, in the absence of the \gmt\ constraint, the outer parts of the 68\% CL
  contours are expanded outwards, close to the 95\% CL contours that are themselves
  close to the boundary set by the $\Omega_{\neu{1}} h^2$ constraint~\footnote{This constraint 
is not sacrosanct, but could be relaxed by postulating some amount of R-violation, 
or some other source of dark matter, or by modifying the expansion history of the Universe,
e.g., by altering the expansion rate during freeze-out, or by
postulating some subsequent injection of entropy.}. Within the overall
  range allowed by this constraint, the most important constraint is that provided by the LHC data.
  We note that the global likelihood functions in the CMSSM and NUHM1 are very flat, and that
  the best-fit points found dropping the \gmt\ constraints are correspondingly quite uncertain. However,
  it is clear that the amounts of fine-tuning at \gmt-less best-fit points are much higher than if \gmt\ is included.
  
\begin{figure*}[htb!]
\resizebox{8.5cm}{!}{\includegraphics{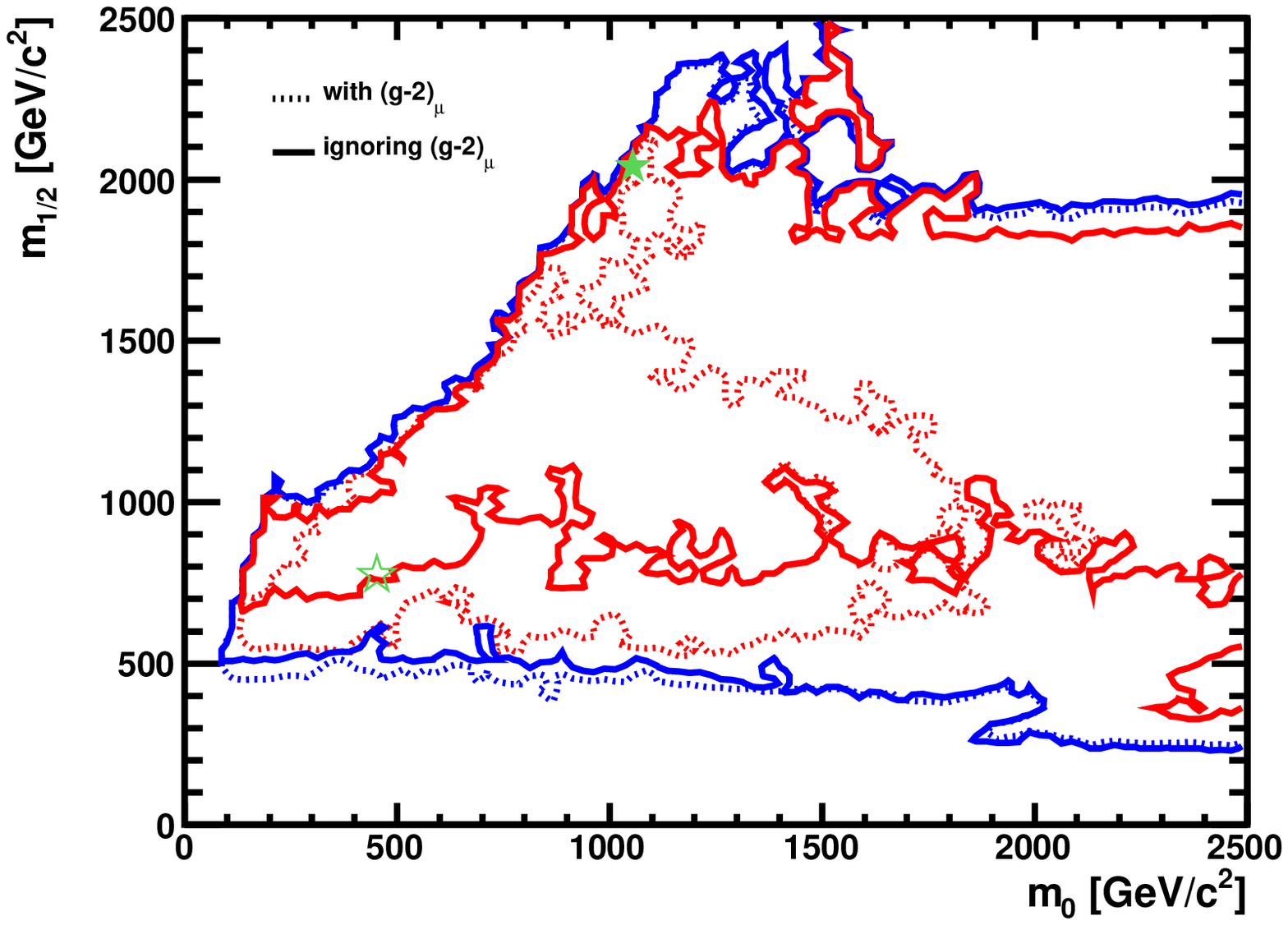}}
\resizebox{8.5cm}{!}{\includegraphics{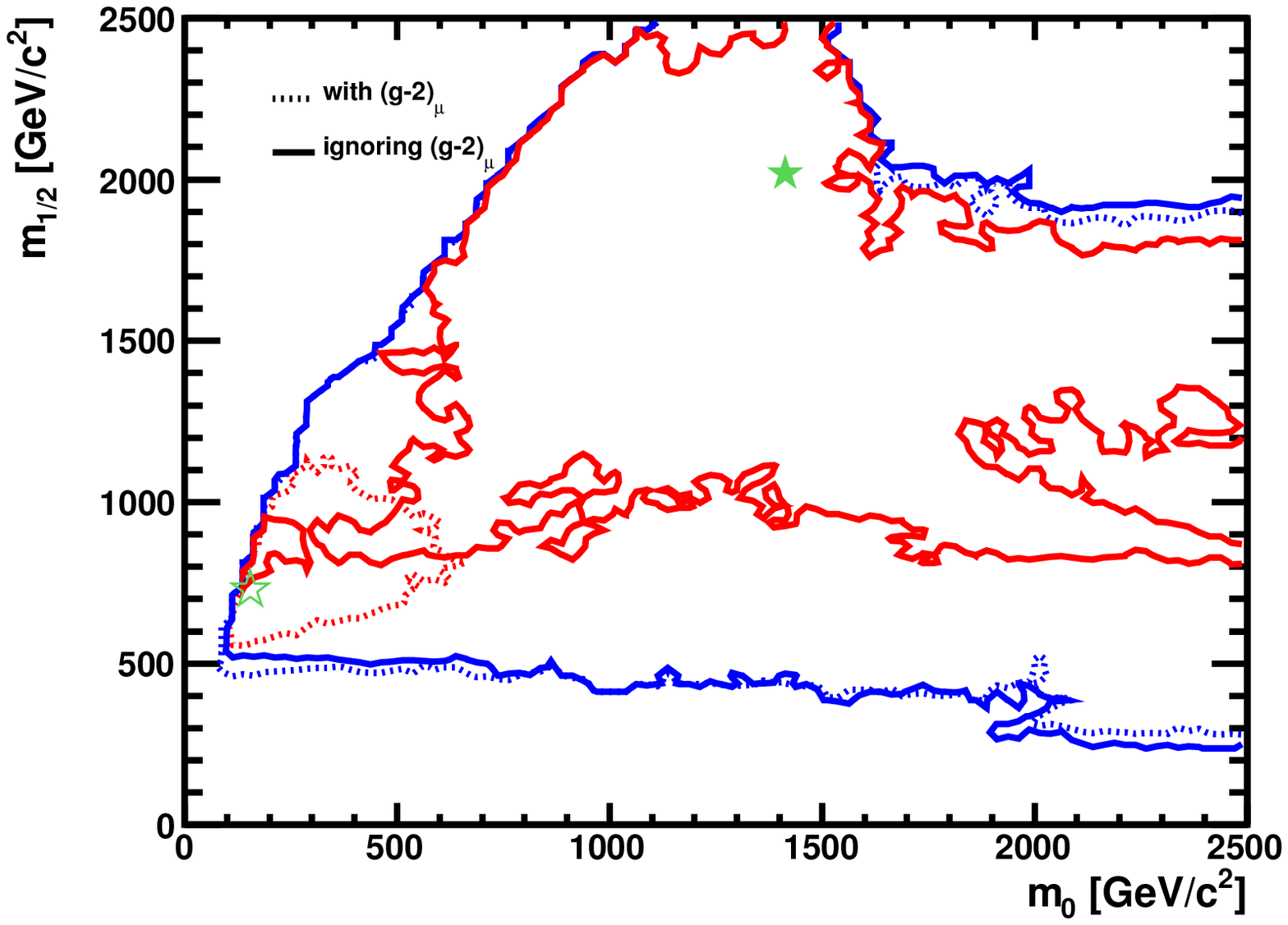}}
\vspace{-1cm}
\caption{\it The $(m_0, m_{1/2})$ planes in the CMSSM (left) and the
  NUHM1 (right). Here we show as solid lines the 68\% and 95\% CL contours
  obtained by dropping the \gmt\ constraint. The contours obtained applying the \gmt\ constraint as in
  Fig.~\protect\ref{fig:6895} are shown here for comparison as dotted lines.}
\label{fig:g-2}
\end{figure*}

We also comment on the treatment of \bsg, where different points of view have been taken concerning the
combination of the theoretical and experimental errors~\footnote{As in our previous analyses, 
we do not take into account constraints from exclusive $b\to s\gamma$ transitions. In particular, we do not
impose any constraint on the SUSY parameter space from the isospin asymmetry in $B\to K^*\gamma$, 
as included for instance in the {\tt SuperIso} package~\cite{SuperIso}. 
A conservative treatment of non-factorizable contributions to this observable
suggests a SM error exceeding $\pm 0.05$, i.e., a relative error exceeding 100\%
(see, e.g.,~\cite{Kim:2008rz}), and even larger errors are associated to the 
contributions of non-SM operators (see, e.g.,~\cite{Becher:2005fg}).
These uncertainties obscure possible SUSY contributions within the ranges 
currently of interest in the CMSSM and the NUHM1 models.}. 
The two uncertainties are of similar size, and
the issue of how they are combined is more severe than for other
observables. In our default 
implementation of \bsg\ we add the quoted errors in quadrature. However, it
might be argued that the theoretical error should be regarded as an overall
range of uncertainty that cannot be treated as an effective Gaussian error to
be added in quadrature to the experimental error. 
Therefore, we consider as
an alternative implementation of \bsg\ the possibility of adding linearly the
theoretical and experimental errors. As seen in
Fig.~\ref{fig:bsg}, the region of the CMSSM $(m_0, m_{1/2})$
plane that is favoured at the 68\% CL {\it contracts} significantly if the
errors in \bsg\ are added linearly, whereas there is little effect in the NUHM1.

This effect arises because the treatment of the \bsg\ errors does not change the
global $\chi^2$ function at large $(m_0, m_{1/2})$, where its value approaches that of
the \SM, since the experimental measurement of \bsg\ is in good agreement with the SM
prediction. On the other hand, adding the errors linearly relaxes the \bsg\ constraint at
smaller $(m_0, m_{1/2})$ in the CMSSM, reducing the tension with other observables and hence also
the minimum of $\chi^2$, as seen in Table~\ref{tab:bestfits}. The net result is to enhance
the rate of increase of $\chi^2$ for CMSSM parameters departing from the best fit, implying that the
68\% CL is reached more quickly. On the other hand, this effect is absent in the NUHM1 because
the freedom to adjust the degree of Higgs non-universality already mitigates the tension of \bsg\
with the other observables, leading to the `pit' in the global $\chi^2$ function mentioned above.

\begin{figure*}[htb!]
\resizebox{8.5cm}{!}{\includegraphics{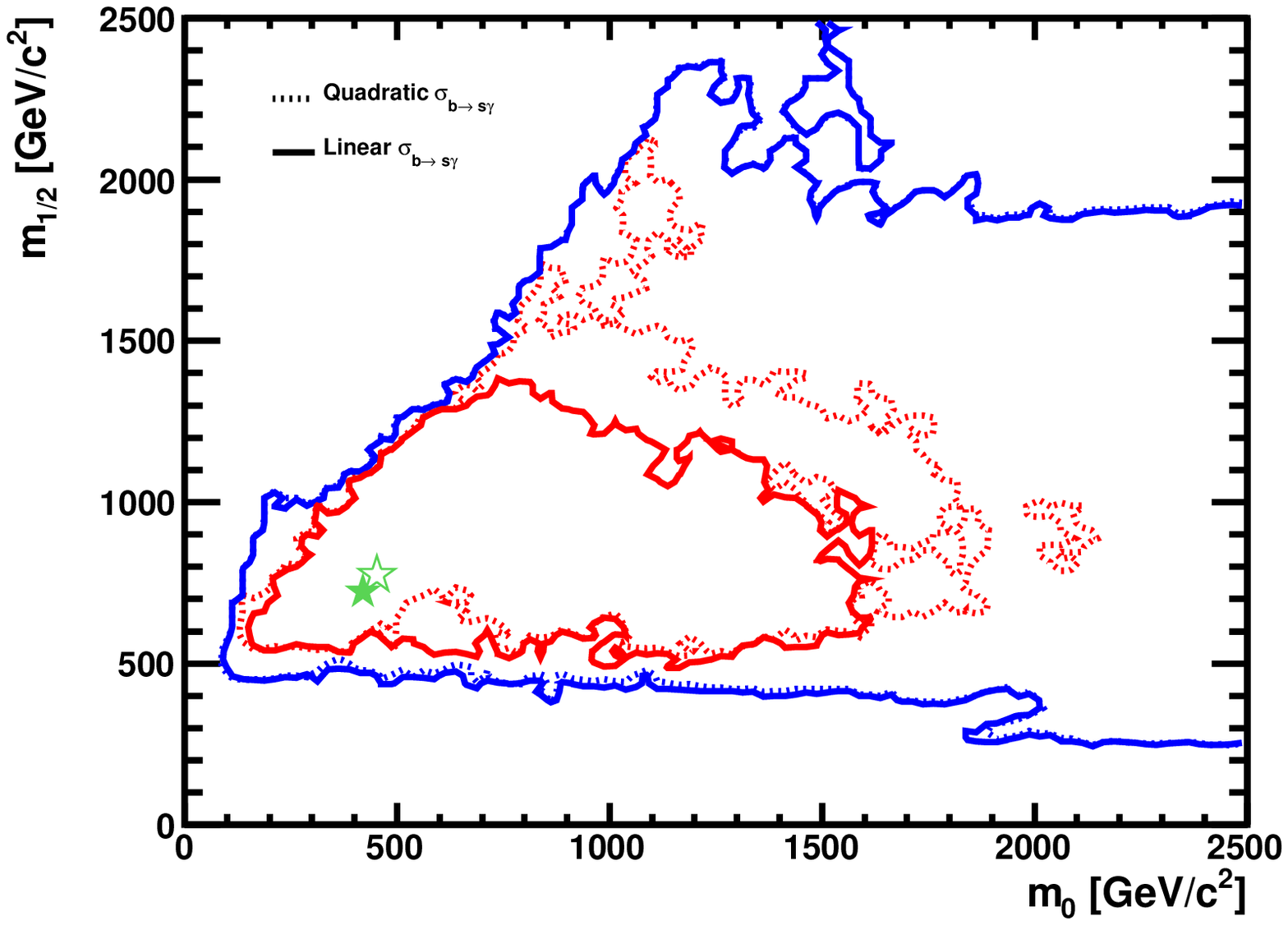}}
\resizebox{8.5cm}{!}{\includegraphics{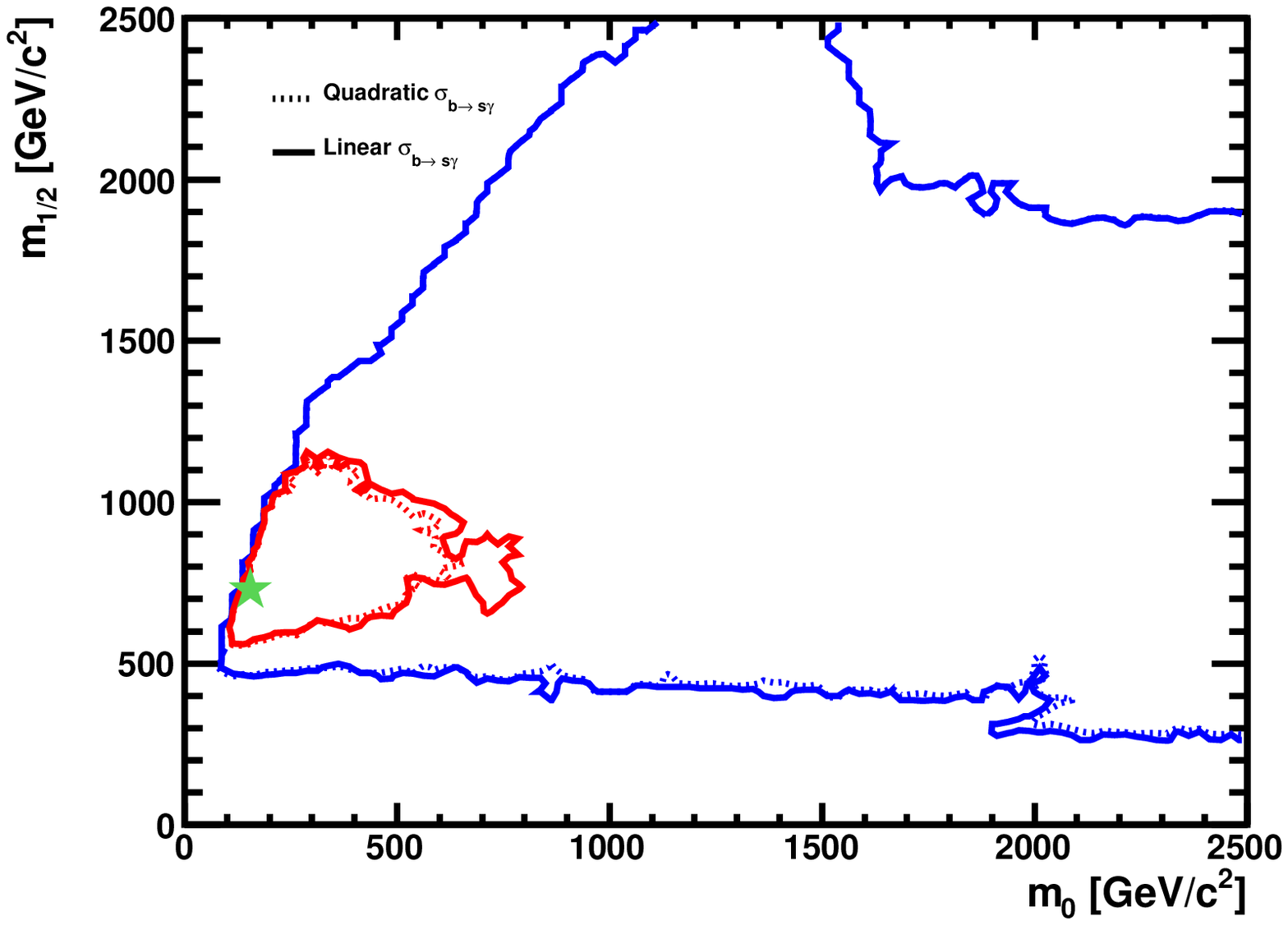}}
\vspace{-1cm}
\caption{\it The $(m_0, m_{1/2})$ planes in the CMSSM (left) and the
  NUHM1 (right). Here we show as solid lines the 68\% and 95\% CL contours
  obtained by adding linearly the experimental and theoretical errors
  in \bsg. The contours obtained by combining them quadratically as in
  Fig.~\protect\ref{fig:6895} are shown here for comparison as dotted lines.
}
\label{fig:bsg}
\end{figure*}

\subsection*{\it The $(\tb, m_{1/2})$ planes in the CMSSM and NUHM1}

Fig.~\ref{fig:tanbm12} displays the $(\tb, m_{1/2})$ planes in the CMSSM and NUHM1,
exhibiting clearly the movement of the best-fit points and 68\% and 95\%~CL
contours to larger $\tb$ that is driven by the tension between \gmt\ and the LHC push to larger $m_{1/2}$.
Comparing the dotted and solid contours, we see that the LHC $\ETslash$
constraints force the new best-fit points into what were previously the `tails'
of the 95\%~CL regions at large $m_{1/2}$ and hence $\tb$. However, it is
clear that the range of $\tb$ allowed at the 68\% CL is still very broad, extending from
$< 20$ to $> 50$ in both the CMSSM and the NUHM1. On the other hand, any future 
substantial increase in the LHC lower limit on $m_{1/2}$ would push $\tb$ in both 
models into a narrower range $\sim 50$, where it encounters pressure from \bmm\ as
discussed below.

\begin{figure*}[htb!]
\resizebox{8.5cm}{!}{\includegraphics{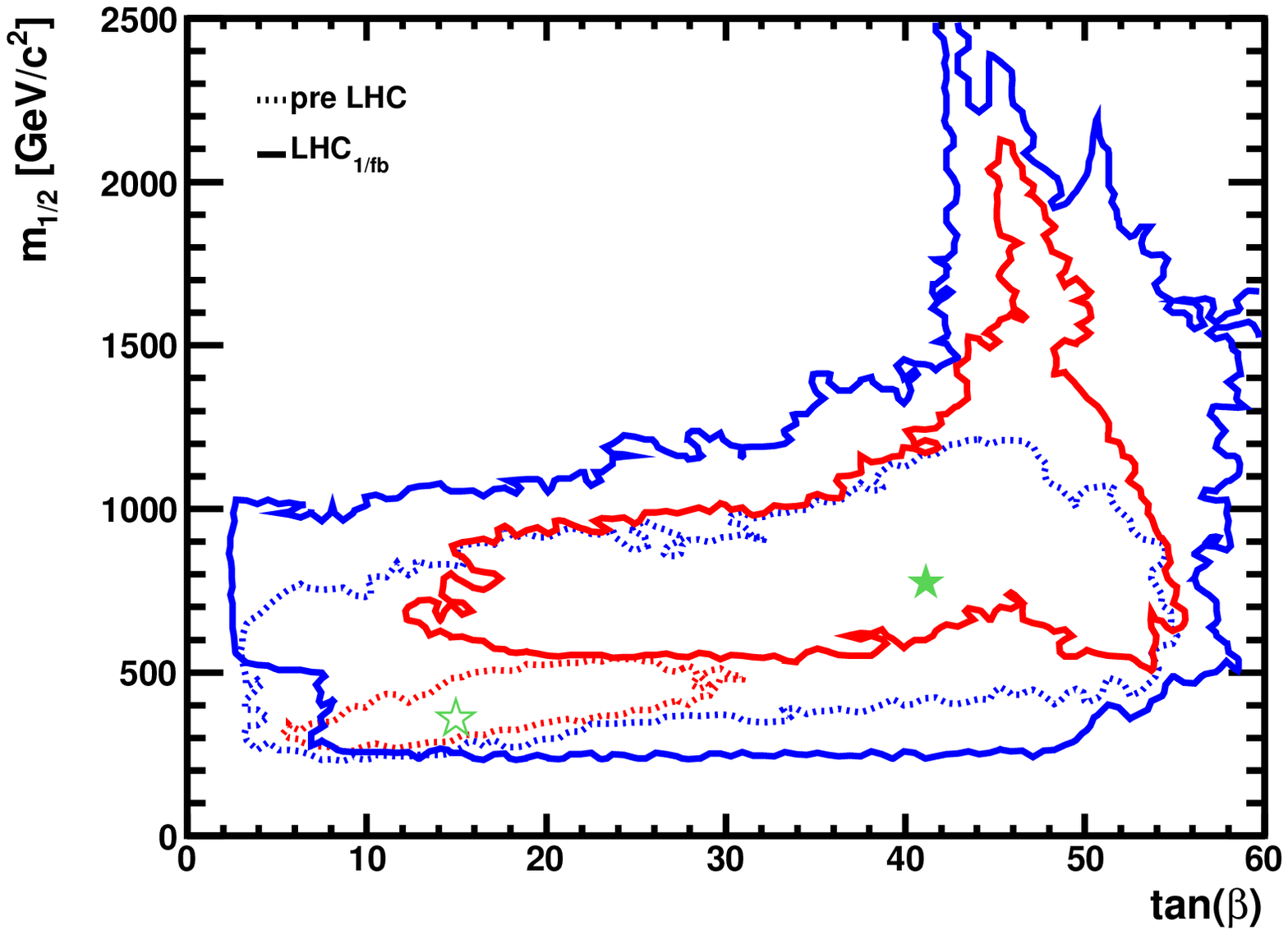}}
\resizebox{8.5cm}{!}{\includegraphics{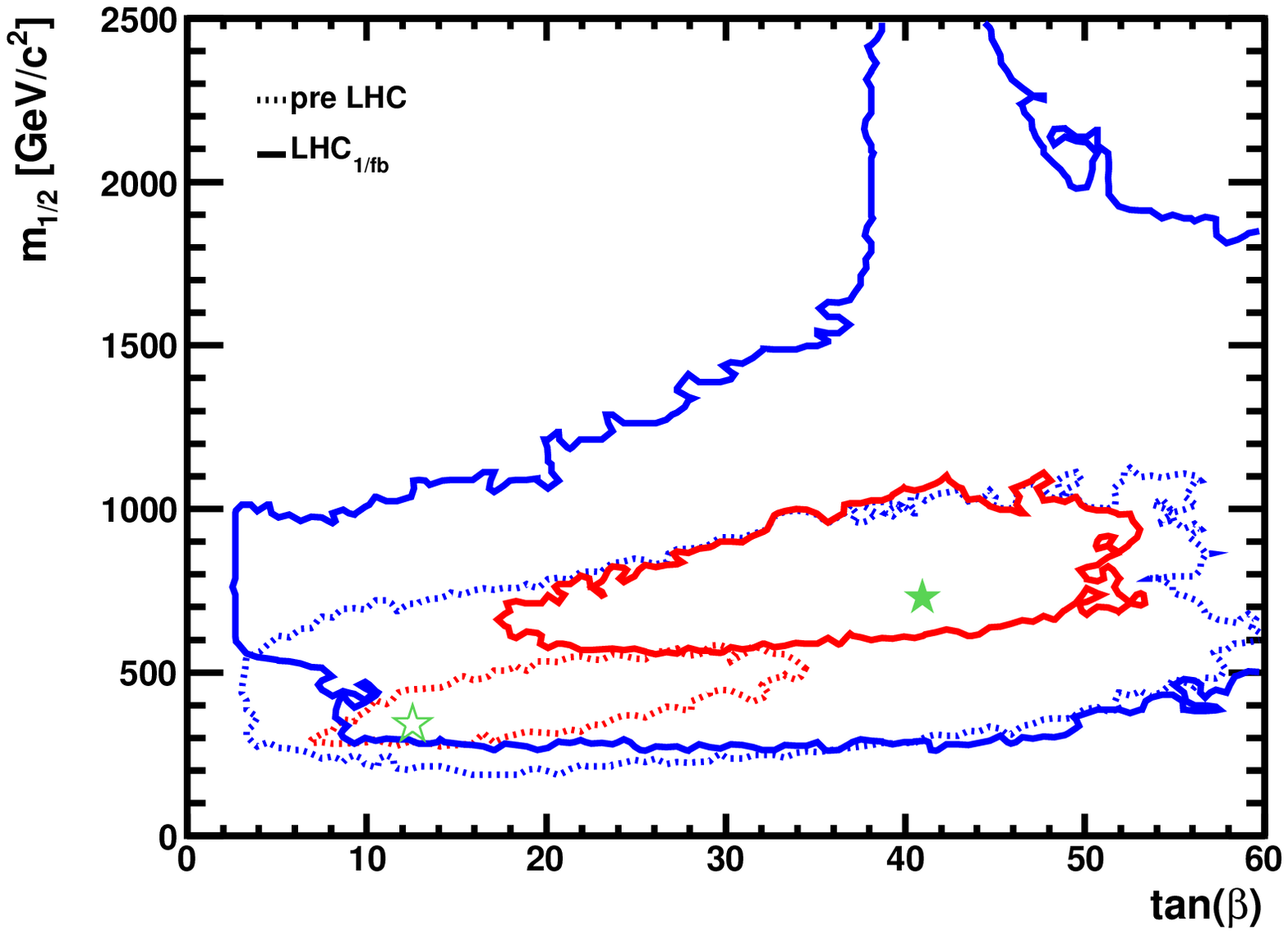}}
\vspace{-1cm}
\caption{\it The $(\tb, m_{1/2})$ planes in the CMSSM (left) and the
  NUHM1 (right). In each plane, the best-fit point after incorporation of the
  LHC$_{\rm 1/fb}$ constraints is indicated by a filled green star, and the
  pre-LHC fit~\protect\cite{mc4} by an open star. The $\Delta \chi^2= 2.30$ and 5.99
  contours, commonly interpreted as the boundaries of the 68\% and 95\%~CL regions, are indicated
  in red and blue, respectively, the solid lines including the
  LHC$_{\rm 1/fb}$ data and the dotted lines showing the pre-LHC fits.
}
\label{fig:tanbm12}
\end{figure*}

\subsection*{\it The $(\MA, \tb)$ planes in the CMSSM and NUHM1}

We now turn to the $(\MA, \tb)$ planes of the CMSSM and NUHM1, 
shown in Fig.~\ref{fig:MAtb},
which are affected directly by the new CMS constraints on the
heavy MSSM Higgs bosons $H/A, H^\pm$~\cite{ATLASHA,CMSHA,CMSHpm}, 
and by the CMS~\cite{CMSbmm} and LHCb~\cite{LHCbbmm}
constraints on \bmm. As already discussed, we include
the heavy Higgs constraints 
via the most sensitive CMS search for $H/A \to \tau^+ \tau^-$~\cite{CMSHA}.
In evaluating the constraint on the $(\MA, \tb)$ planes of 
the CMSSM and NUHM1 imposed by the new measurements of \bmm,
we use the official combination of the recent upper limits on \bmm\ from
the CMS and LHCb Collaborations~\cite{CMSLHCb}, which yields \bmm\ $< 1.08 \times 10^{-8}$
at the 95\%~CL. Our implementation actually includes the full likelihood
function arising from this combination. The measurement of \bmm\ by the
CDF Collaboration~\cite{CDFbmm} is in some tension with the CMS/LHCb combination,
but only at the $\Delta \chi^2 \sim 1$ level. However, since there is so far
no official combination of the CDF result with those of CMS and LHCb,
we limit ourselves to discussing later its compatibility with the predictions
for \bmm\ of our global fits. We see in Fig.~\ref{fig:MAtb} that the general effect
of the LHC$_{\rm 1/fb}$ data is to push the preferred range of $\MA$ to larger values,
as well as pushing $\tb$ towards larger values.

\begin{figure*}[htb!]
\resizebox{8.5cm}{!}{\includegraphics{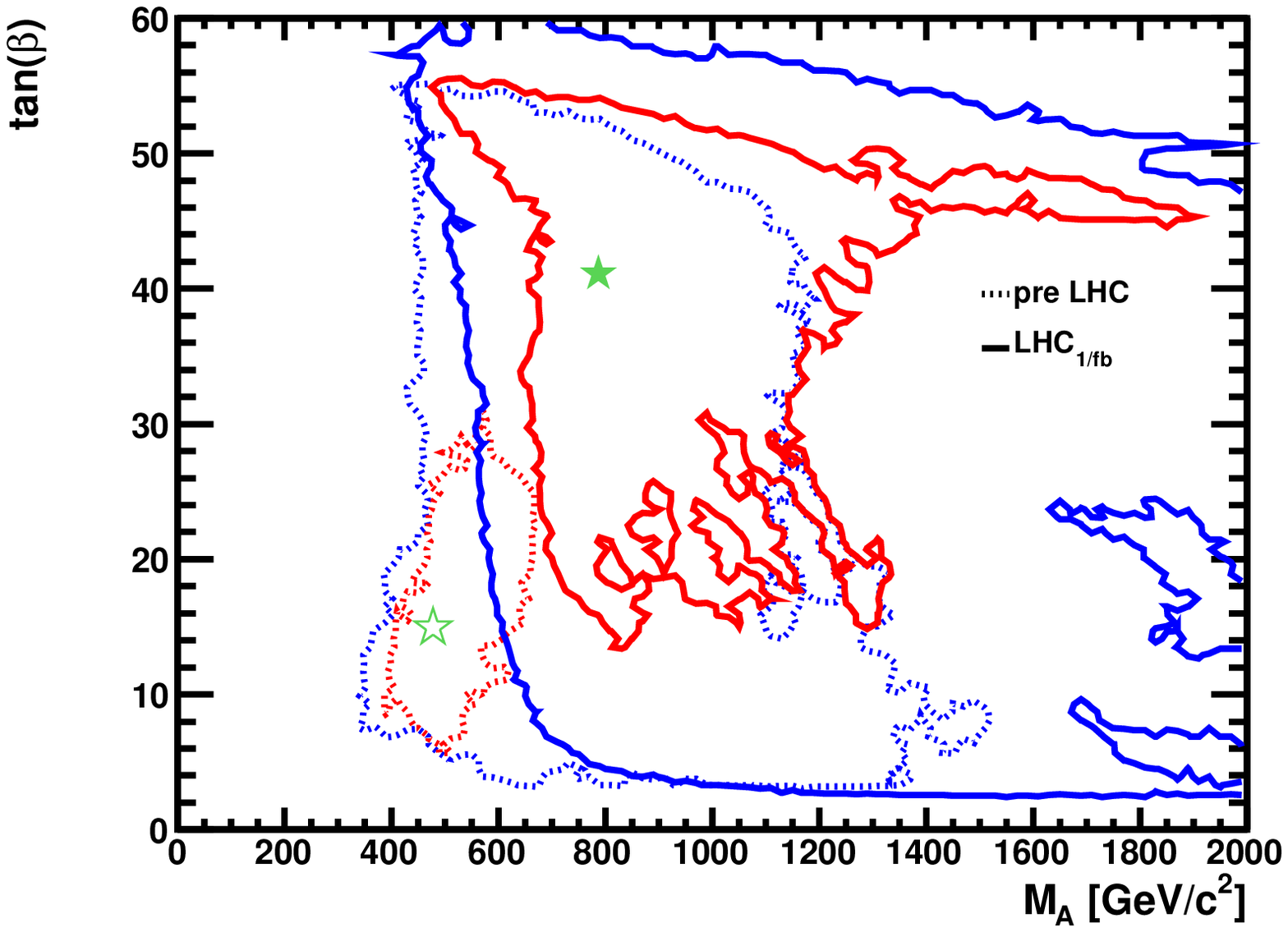}}
\resizebox{8.5cm}{!}{\includegraphics{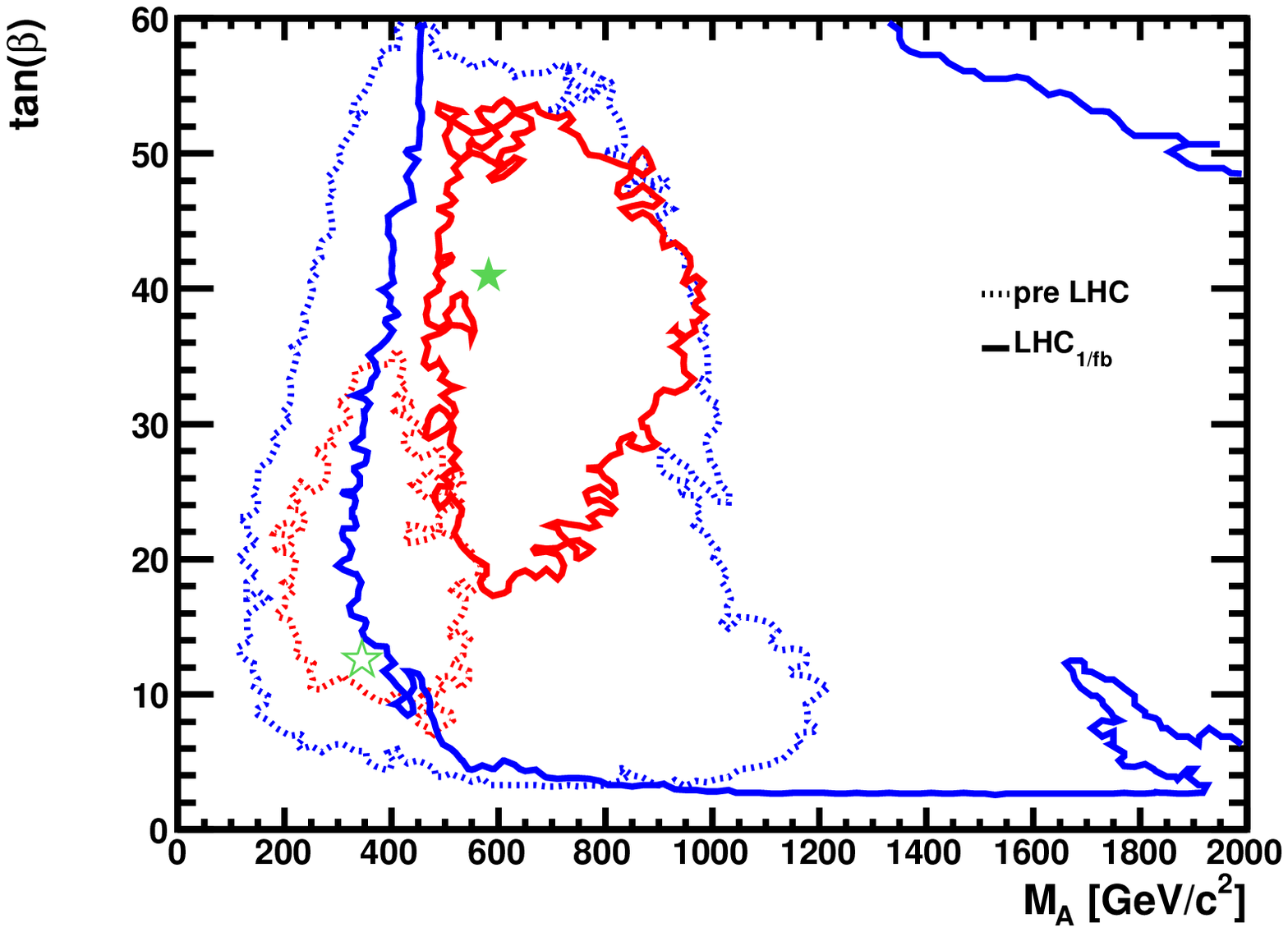}}
\vspace{-1cm}
\caption{\it The $(\MA, \tb)$ planes in the CMSSM (left) and the
  NUHM1 (right). In each plane, the best-fit point after incorporation of the
  LHC$_{\rm 1/fb}$ constraints is indicated by a filled green star, and the
  pre-LHC fit~\protect\cite{mc4} by an open star. The $\Delta \chi^2= 2.30$ and 5.99
  contours, commonly interpreted as the boundaries of the 68 and 95\%~CL regions, are indicated
  in red and blue, respectively, the solid lines including the
  LHC$_{\rm 1/fb}$ data and the dotted lines showing the pre-LHC fits.
}
\label{fig:MAtb}
\end{figure*}

The impacts of the $H/A$ and \bmm\ constraints are less important
in the CMSSM than in the NUHM1, so we discuss the latter in more detail.
Fig.~\ref{fig:MAtbNUHM1} shows four versions of the $(\MA, \tb)$ plane
in the NUHM1, with all the LHC$_{\rm 1/fb}$ constraints applied (upper left, equivalent to
the right panel of Fig.~\ref{fig:MAtb}),
dropping the $H/A$ constraint but keeping \bmm\ (upper
right), dropping \bmm\ but keeping the $H/A$ constraint
(lower left), and dropping both constraints (lower right). Comparing the
two upper panels, we see that the $H/A$ constraint is relevant for $\MA \lsim 450 \gev$,
and that applying it impacts the low-$\MA$ sides of the 68\% and 95\%~CL
contours, whereas the best-fit point is {unaffected}. Comparing the upper
and lower panels, we see that the \bmm\ constraint is relevant
{for low $\MA$ values.  In particular, the  68\%~CL 
contours extend to slightly lower $\tan \beta$ values
and the best-fit points (green stars) move to significantly lower $\tb$
when \bmm\ is included. 
as seen in Fig.~\ref{fig:MAtbNUHM1}.
Both the $H/A$ and \bmm\ constraints have the potential for
more significant impacts in the future.

\begin{figure*}[htb!]
\resizebox{8.5cm}{!}{\includegraphics{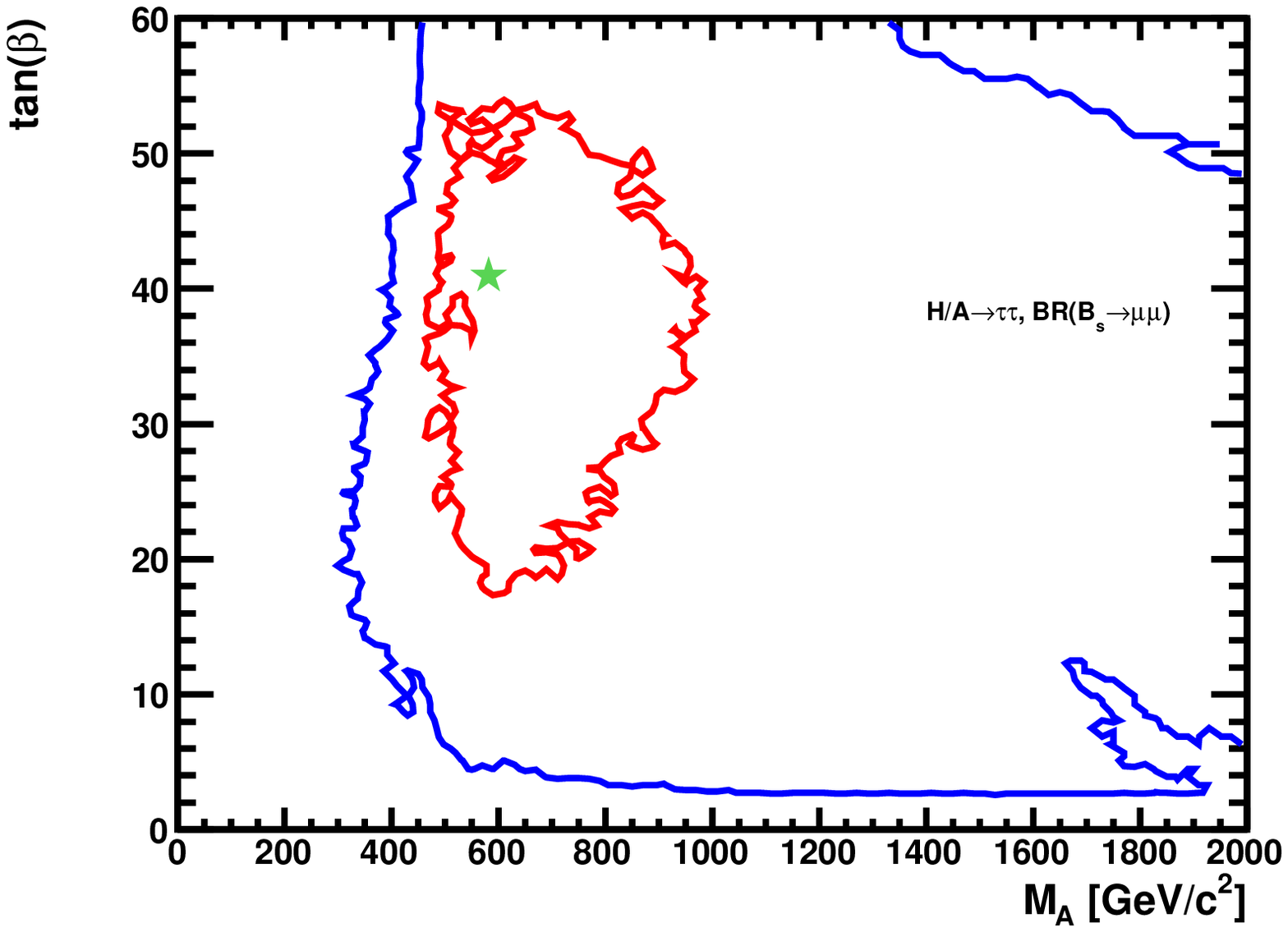}}
\resizebox{8.5cm}{!}{\includegraphics{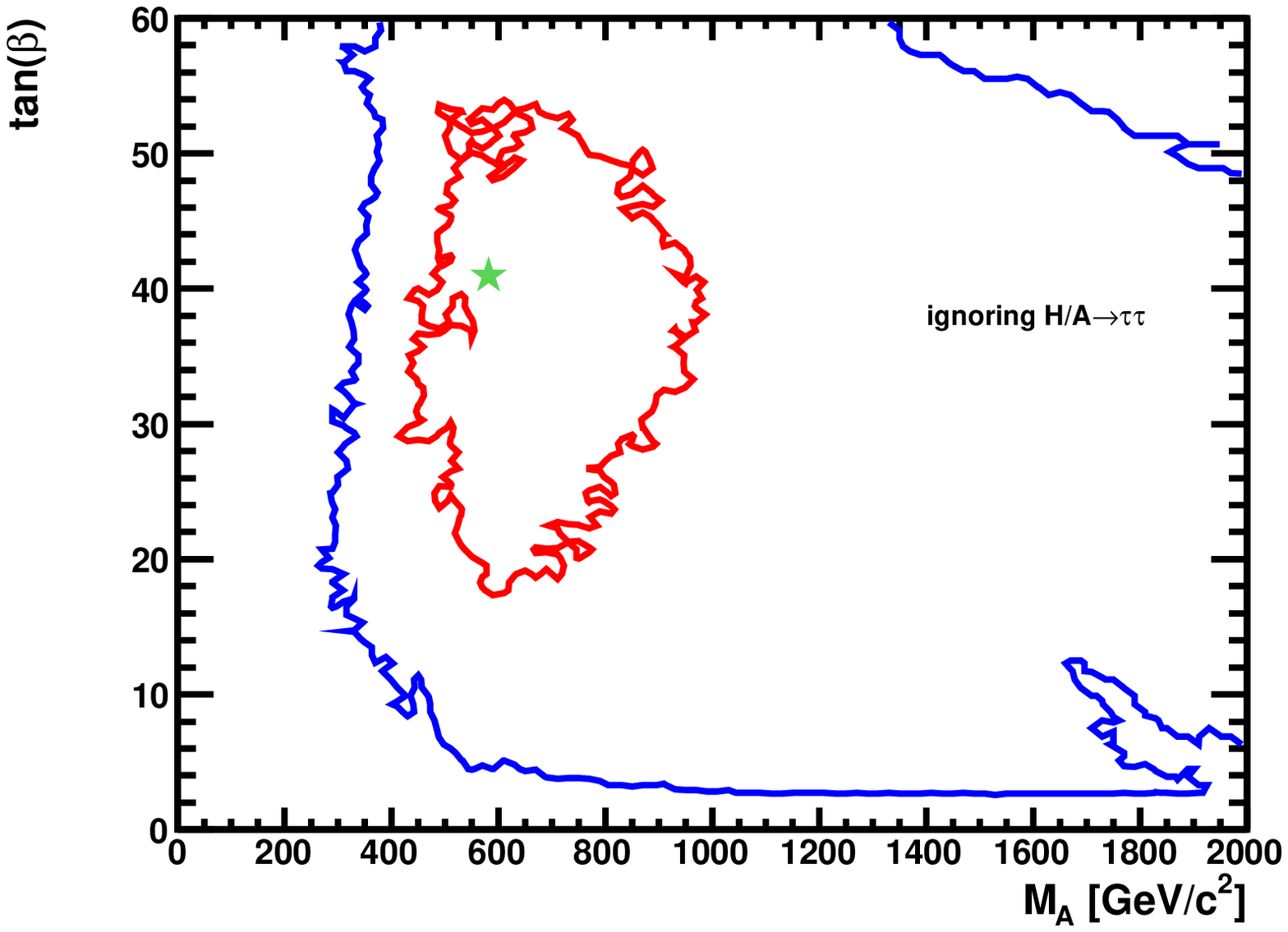}}
\resizebox{8.5cm}{!}{\includegraphics{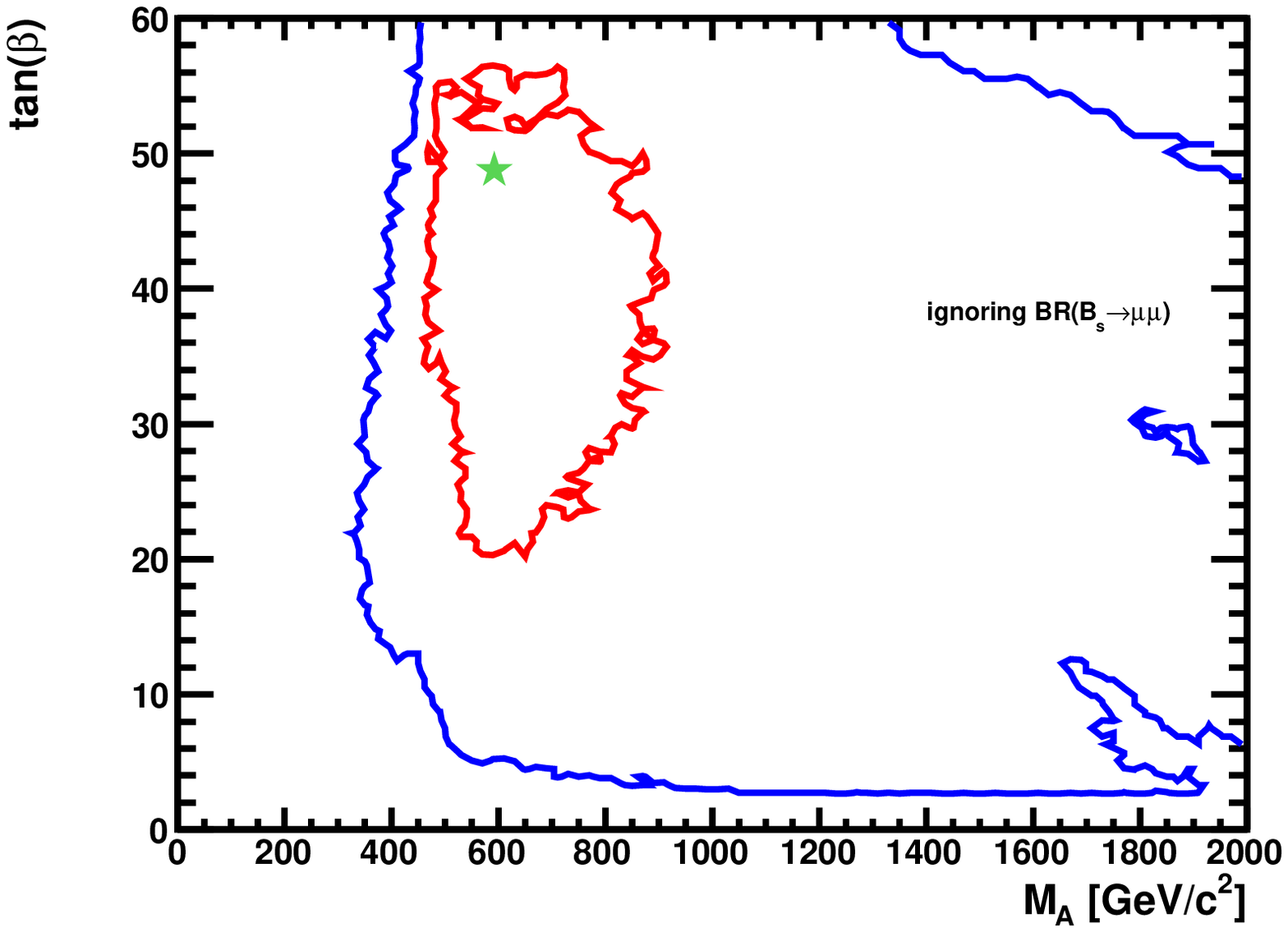}}
\resizebox{8.5cm}{!}{\includegraphics{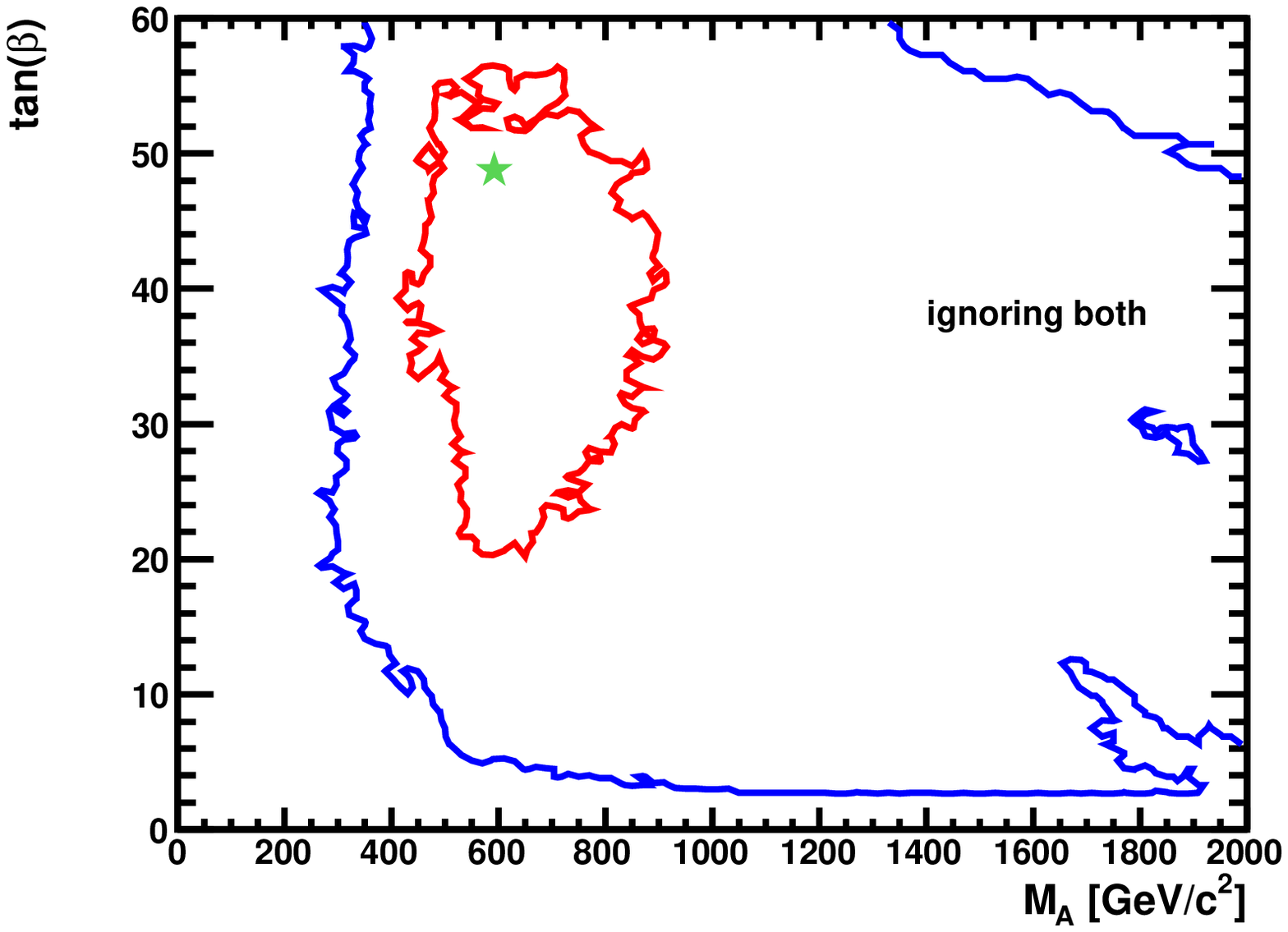}}
\vspace{-1cm}
\caption{\it The $(\MA, \tb)$ planes in the NUHM1 including both
the $H/A$~\protect\cite{CMSHA} and \bmm~\protect\cite{CMSLHCb} constraints (upper left), dropping the
$H/A$ constraint but keeping the \bmm\ constraint (upper right),
dropping \bmm\ but keeping $H/A$ (lower left), 
and dropping both constraints (lower right).
}
\label{fig:MAtbNUHM1}
\end{figure*}

\subsection*{\it Predictions for $\mgl$}

In Fig.~\ref{fig:mgl} we show the one-dimensional $\chi^2$
functions predicted by global fits for $\mgl$
in the CMSSM (left) and the NUHM1 (right). The solid lines are based on
our global fits including the LHC$_{\rm 1/fb}$ constraints,
whereas the dotted lines are based on our previous global fits
based on the pre-LHC constraints~\cite{mc4}~\footnote{The `stalactites' at
$\mgl \sim 400 \gev$ in the pre-LHC fits~\cite{mc4} were due to the light-Higgs funnel
that has now been excluded by the LHC$_{\rm 1/fb}$ data.
Likewise, the `stalactite' in the CMSSM LHC$_{\rm 1/fb}$ curve also originates in the light-Higgs funnel region,
and comes from points with large $m_0 > 3 \tev$ - which
is why they are not seen in Fig.~\ref{fig:6895} - and large $A_0$. These points might be
excluded by the ATLAS 1/fb 0-lepton search, whose published $(m_0, m_{1/2})$ exclusion for $\tb = 10$ and
$A_0 = 0$ extends only to $m_0 = 3 \tev$.}. We see that the
best-fit estimates of $\mgl$ have increased substantially to $\sim 1600 \gev$
as a result of the LHC$_{\rm 1/fb}$ data, but we also see that there is
considerable uncertainty in this estimate, with $\mgl > 2500 \gev$
subject to a penalty $\Delta \chi^2 \sim 2$ only.

\begin{figure*}[htb!]
\resizebox{8.5cm}{!}{\includegraphics{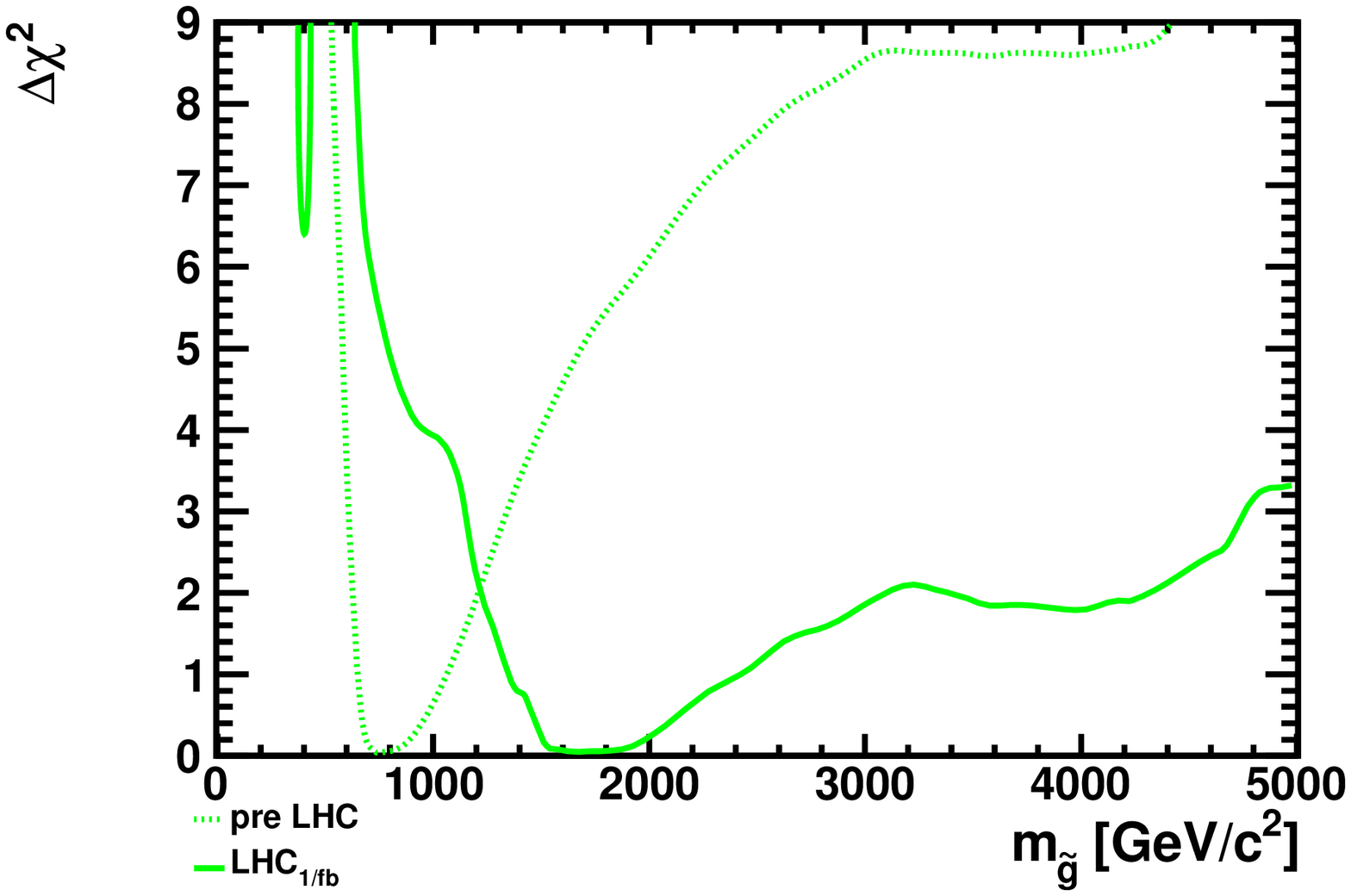}}
\resizebox{8.5cm}{!}{\includegraphics{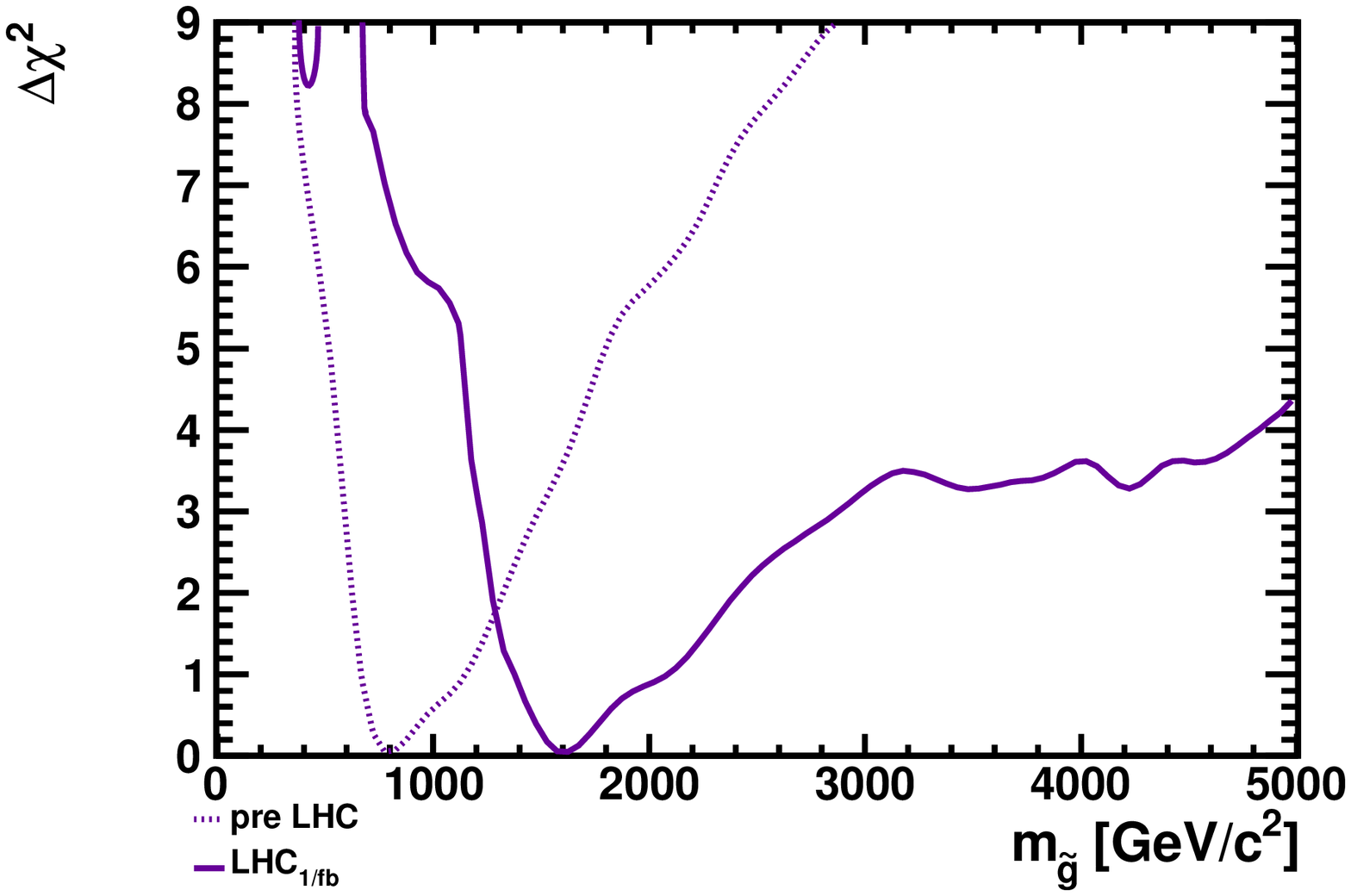}}
\vspace{-1cm}
\caption{\it The one-dimensional $\chi^2$ functions for $\mgl$ in
the CMSSM (left) and the NUHM1 (right). The solid lines are for
fits including the LHC$_{\rm 1/fb}$ data, and the dotted lines are for fits
based on the pre-LHC data~\protect\cite{mc6}.
}
\label{fig:mgl}
\end{figure*}

\subsection*{\it Predictions for \bmm}

In Fig.~\ref{fig:bmm} we show the one-dimensional $\chi^2$
functions predicted by our global fits for \bmm\
in the CMSSM (left) and the NUHM1 (right). The solid lines are based on
the official combination of the CMS and LHCb constraints on this decay~\cite{CMSLHCb},
whereas the dashed lines show results using an unofficial combination of
these constraints with the CDF measurement~\cite{CDFbmm}, and the
dotted lines represent pre-LHC predictions~\cite{mc4}. We see that the
best-fit estimates of \bmm\ are somewhat above the SM value, as a result
of the push towards larger $\tb$ required to accommodate the LHC data while
reconciling them with \gmt. In both the CMSSM and the NUHM1, the estimates of
\bmm\ are quite compatible with an unofficial combined fit to CDF, CMS and
LHCb data, where the main effect is a reduction of $\chi^2$ in a somewhat
  broader range of \bmm.

\begin{figure*}[htb!]
\resizebox{8.5cm}{!}{\includegraphics{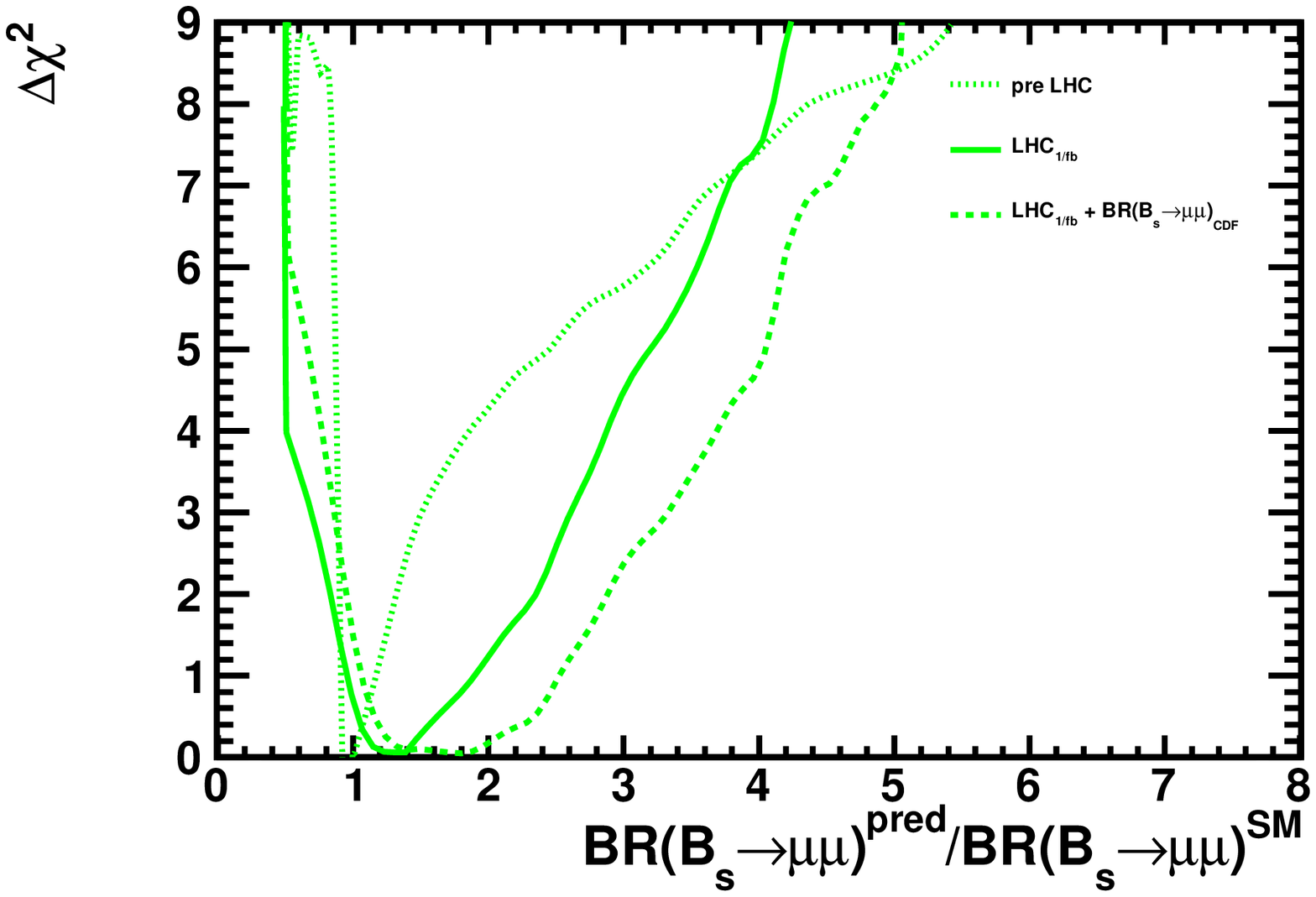}}
\resizebox{8.5cm}{!}{\includegraphics{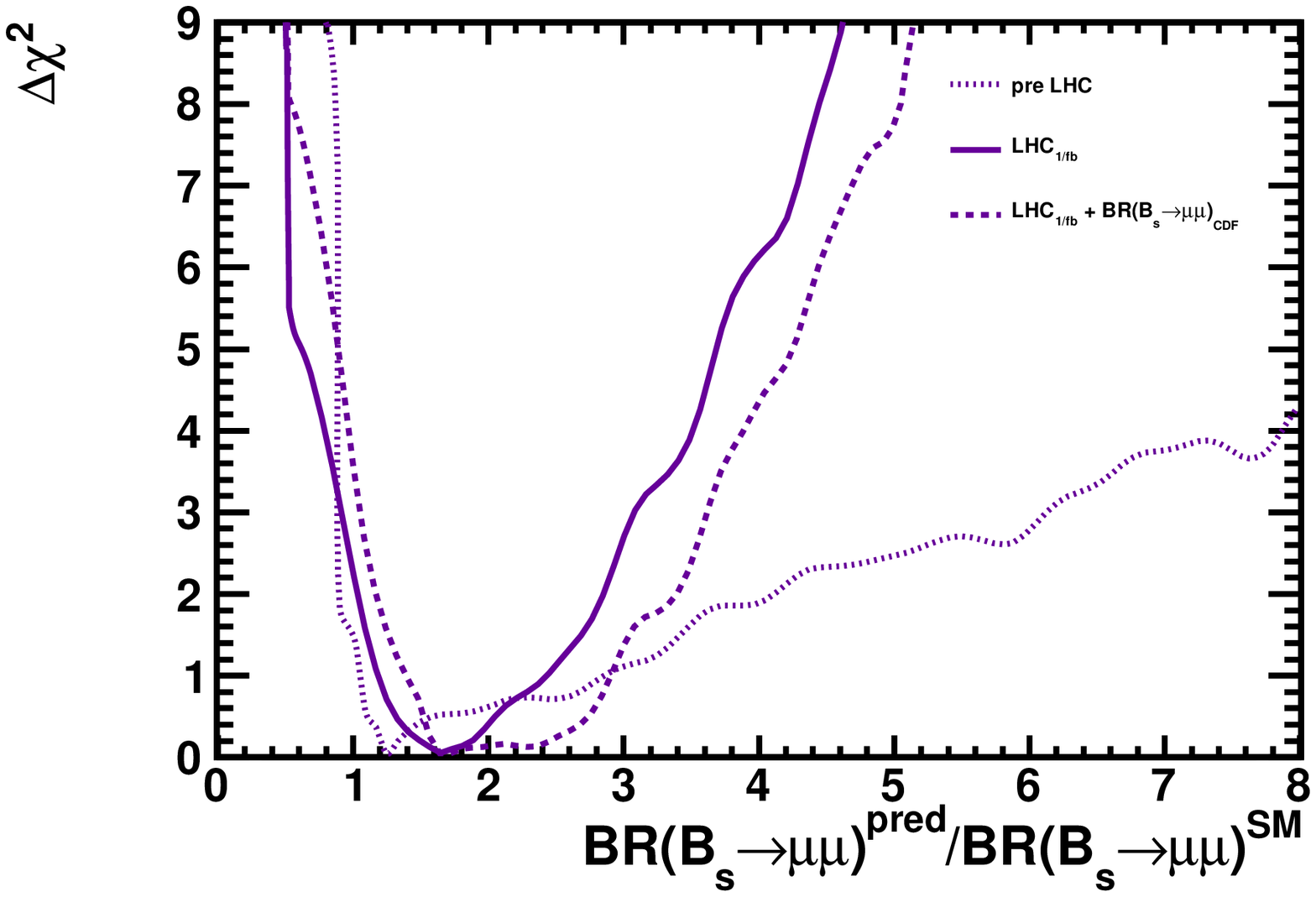}}
\vspace{-1cm}
\caption{\it The one-dimensional $\chi^2$ functions for \bmm\ in
the CMSSM (left) and the NUHM1 (right). The solid lines are for
fits including the official combination of the results from the CMS
and LHCb Collaborations~\protect\cite{CMSLHCb}, the dashed lines are for fits using
an unofficial combination of these results with the CDF result~\protect\cite{CDFbmm},
and the dotted lines represent pre-LHC predictions~\protect\cite{mc4}.
}
\label{fig:bmm}
\end{figure*}

\subsection*{\it Predictions for $\Mh$}

In Fig.~\ref{fig:Mh} we show the one-dimensional $\chi^2$
functions predicted by our global fits for $\Mh$
in the CMSSM (left) and the NUHM1 (right). 
In this figure we {\it do not} include the
direct limits from 
LEP~\cite{Barate:2003sz,Schael:2006cr} or the Tevatron, so as to 
illustrate whether there is a conflict between these 
limits and the predictions of supersymmetric models.
For each model we display the new likelihood functions corresponding to
the LHC$_{\rm 1/fb}$ data set, indicating the
theoretical uncertainty in the calculation of $\Mh$ of $\sim 1.5 \gev$
by red bands. We also show, as dashed lines without red bands, our
previous predictions based on the pre-LHC results (also discarding the LEP
constraint). We see that the LHC data improve the consistency of the
model predictions with the LEP exclusion, removing whatever tension
existed previously. We cannot resist pointing out that the best-fit value
for $\Mh$ found recently in a SM fit including LEP, Tevatron and LHC
exclusions as well as precision electroweak data 
$\sim 120 \gev$~\cite{Gfitter}, and
that this is also the value of the SM Higgs mass that is most compatible
with the ongoing LHC searches~\cite{LHCH}.

\begin{figure*}[htb!]
\resizebox{8.25cm}{!}{\includegraphics{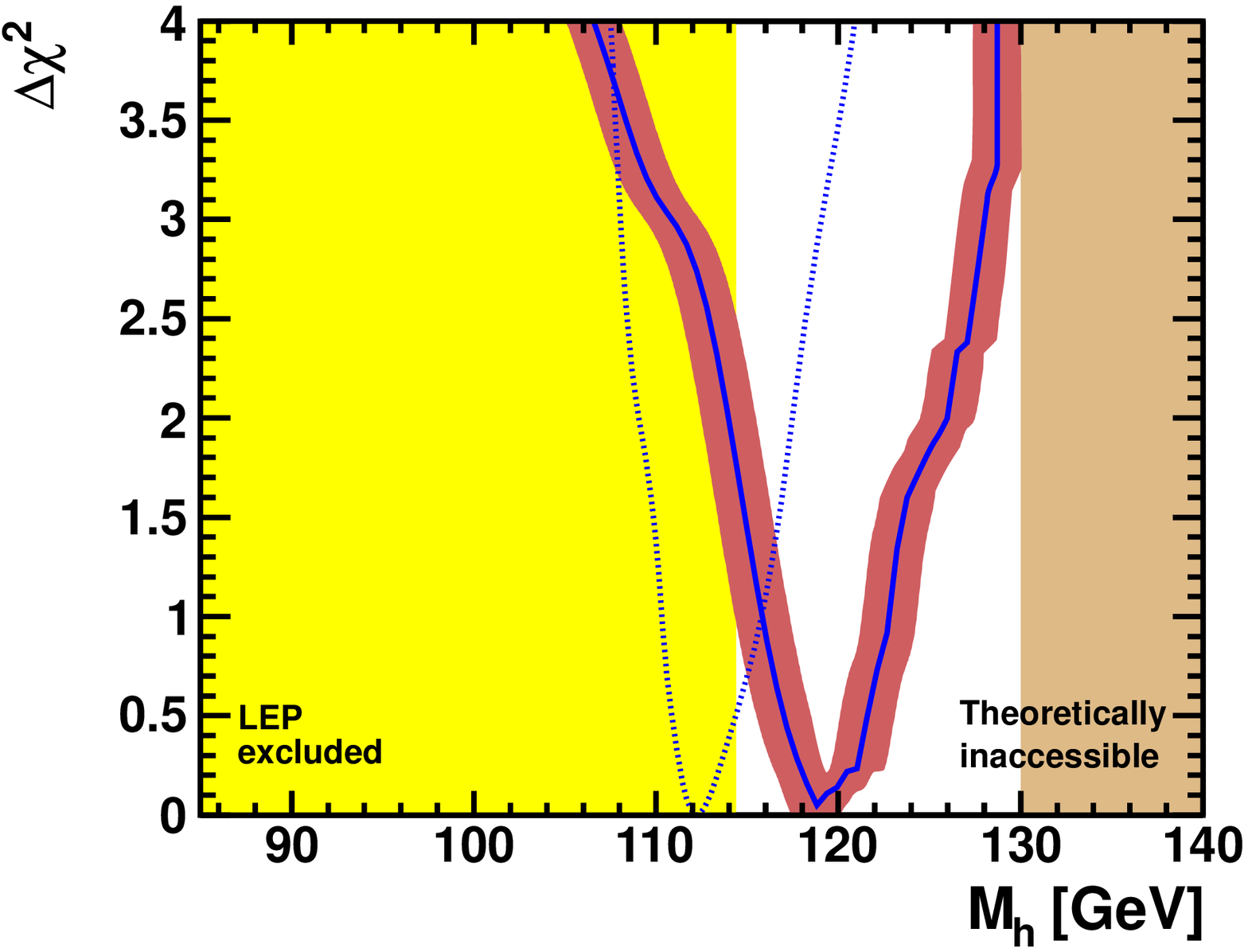}}
\resizebox{8.25cm}{!}{\includegraphics{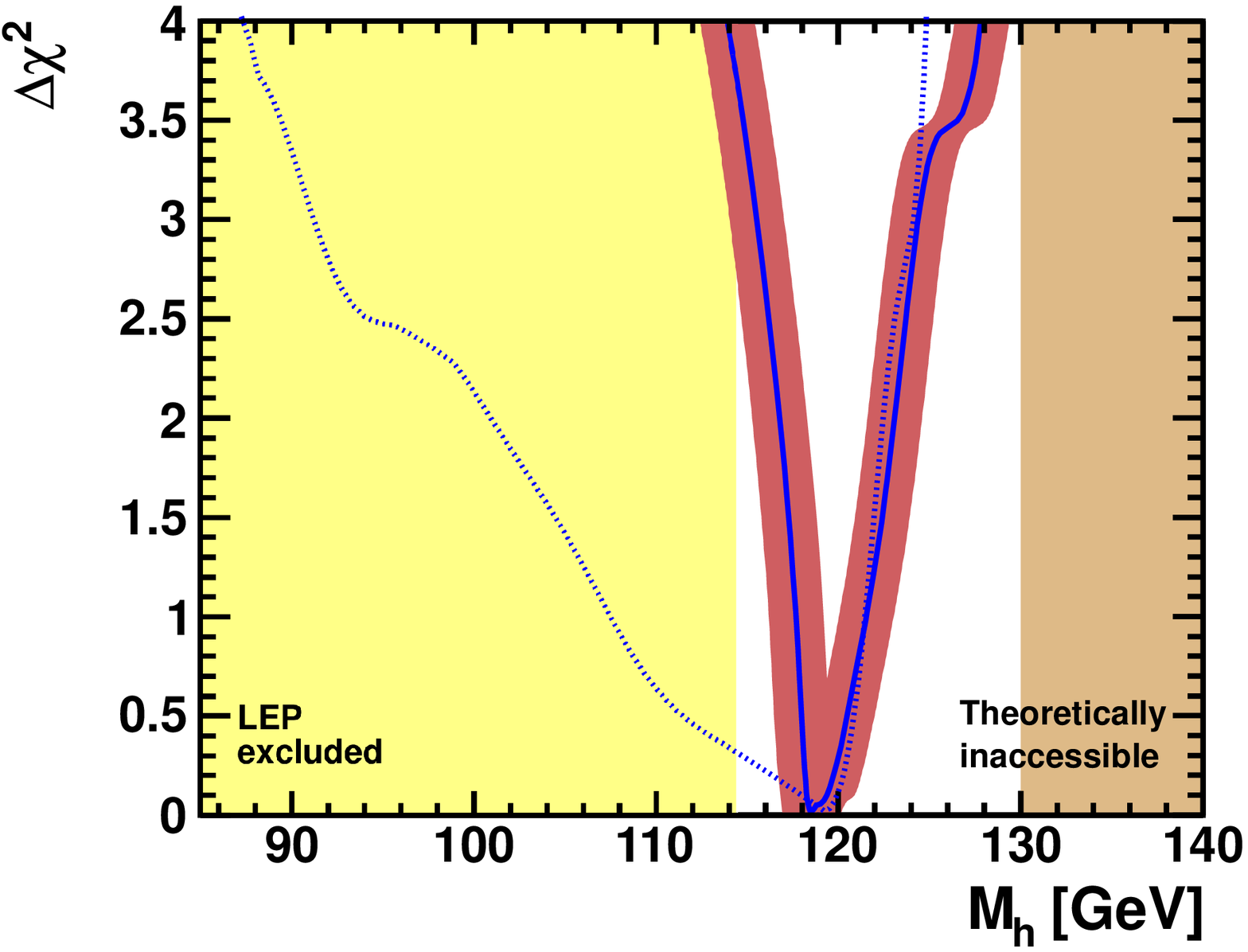}}
\vspace{-1cm}
\caption{\it The one-dimensional $\chi^2$ functions for $\Mh$ in
the CMSSM (left) and the NUHM1 (right). The solid lines are for
fits including all the available data, except the
LEP\cite{Barate:2003sz,Schael:2006cr}
constraints, with a red band indicating the estimated theoretical
  uncertainty in the calculation of $\Mh$
of $\sim 1.5 \gev$. The pre-LHC results~\protect\cite{mc4} are shown as dotted lines.
The yellow shading in the left panel shows the LEP exclusion of an
SM Higgs boson, which applies also to the lightest CMSSM Higgs boson~\cite{Ellis:2001qv,Ambrosanio:2001xb}.
The lighter yellow shading in the right panel reflects the fact that this mass range is
not completely excluded in the NUHM1 due to a possible suppression of the
  $ZZh$ coupling. The beige shading in both panels
indicates values of $\Mh$ inaccessible in the supersymmetric
models studied with GUT-scale unification.
}
\label{fig:Mh}
\end{figure*}

\subsection*{\it Predictions for $\MA$}

In Fig.~\ref{fig:MA} we show the one-dimensional $\chi^2$
functions predicted by our global fits for $\MA$
in the CMSSM (left) and the NUHM1 (right). We see that the
best-fit values of $\MA$ have increased in both models, by
$\sim 350 \gev$ and $\sim 250 \gev$, respectively.

\begin{figure*}[htb!]
\resizebox{8.5cm}{!}{\includegraphics{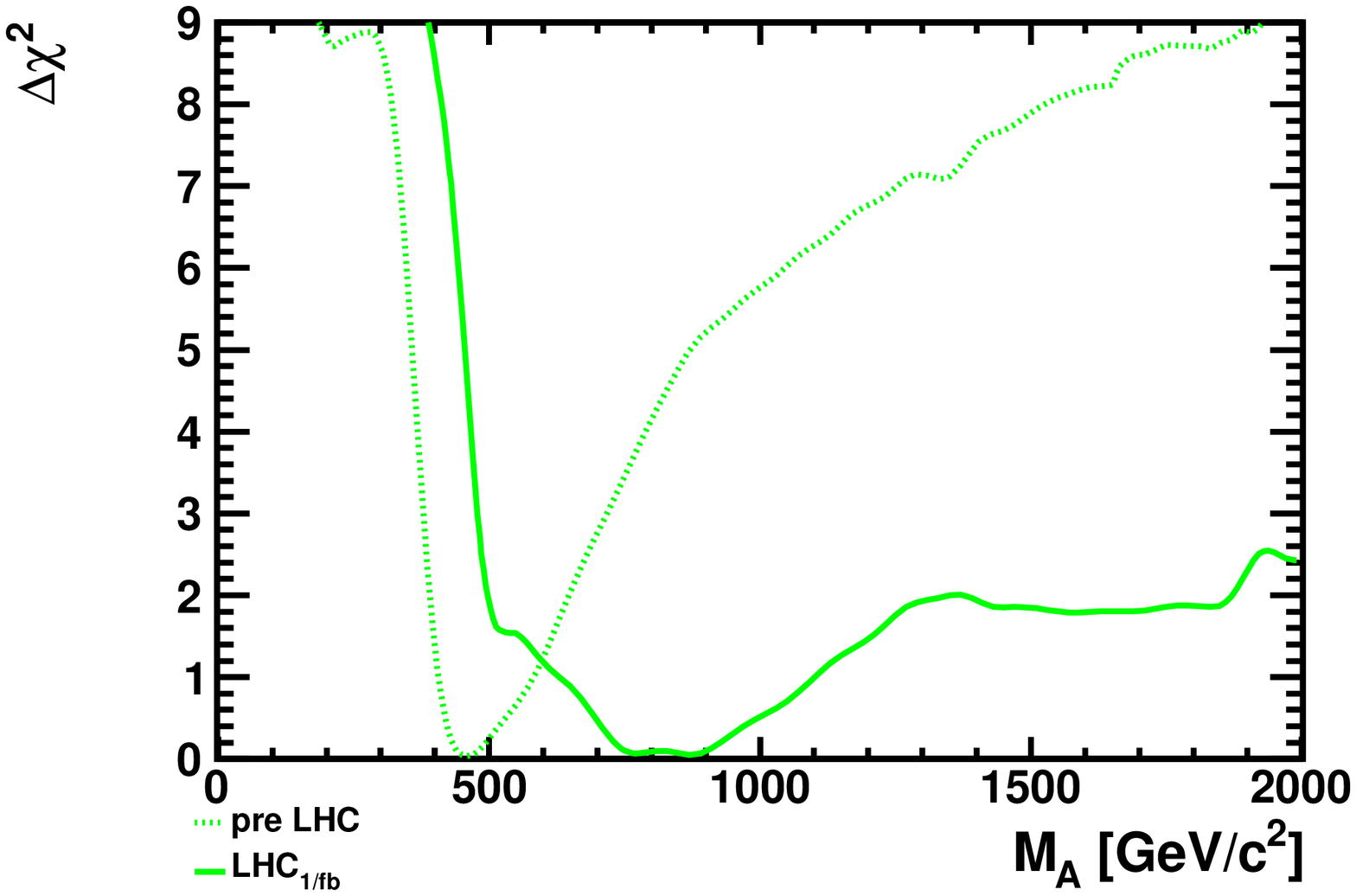}}
\resizebox{8.5cm}{!}{\includegraphics{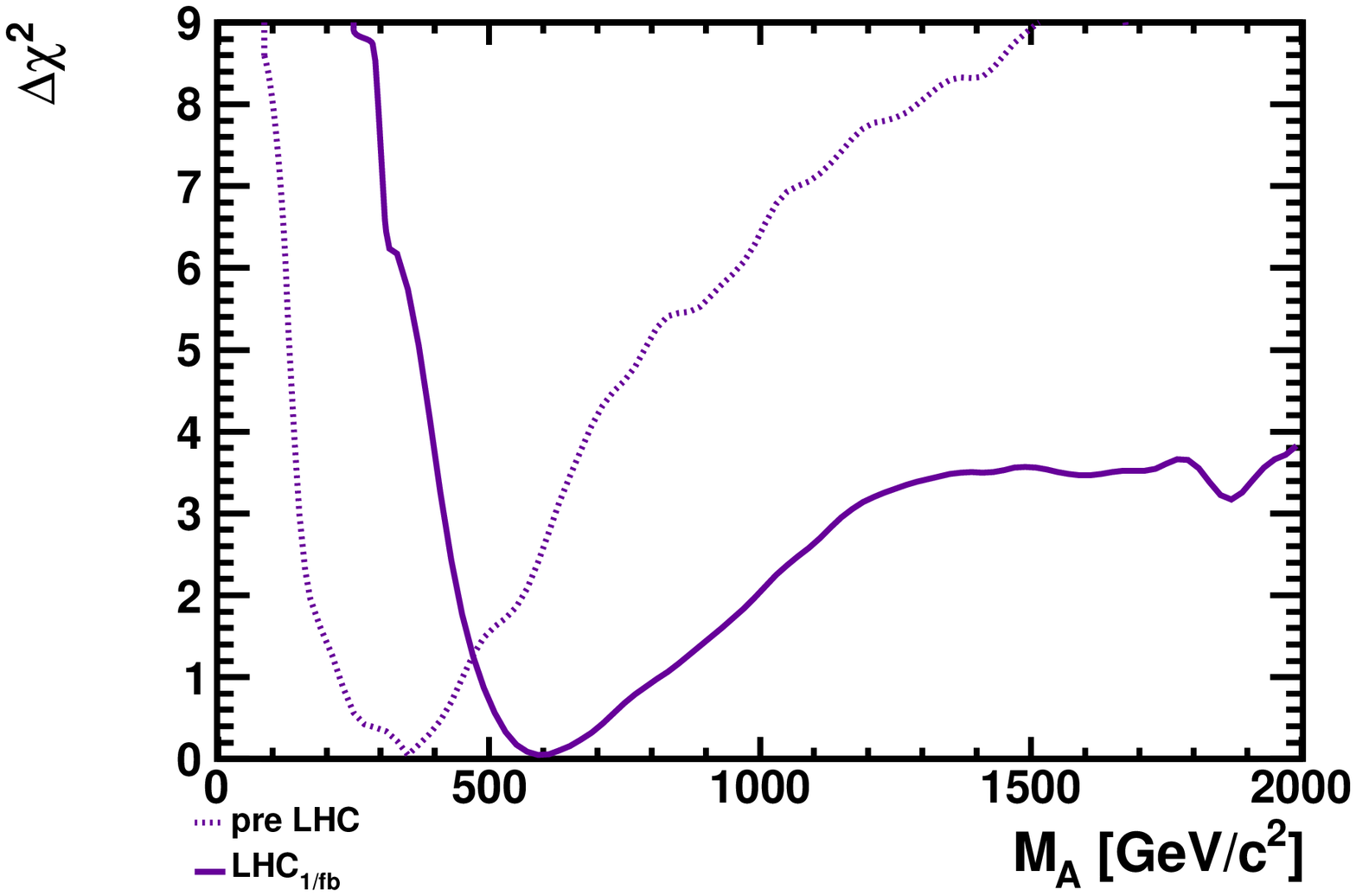}}
\vspace{-1cm}
\caption{\it The one-dimensional $\chi^2$ functions for $\MA$ in
the CMSSM (left) and the NUHM1 (right). The solid lines are for
fits including the LHC$_{\rm 1/fb}$ data, and the dotted lines are for fits
based on the pre-LHC data~\protect\cite{mc4}.}
\label{fig:MA}
\end{figure*}

\subsection*{\it Dark matter scattering cross sections}

In Fig.~\ref{fig:ssi} we show the 68\% and 95\%~CL contours in the $(\mneu{1}, \ssi)$ planes
for the CMSSM (left) and the NUHM1 (right). The solid lines are based on
our global fits including the LHC$_{\rm 1/fb}$ constraints, whereas the dotted lines
correspond to our previous fits using the pre-LHC constraints. In both
cases, we assume $\Sigma_{\pi N} = 50 \pm 14 \mev$~\cite{sigma}~\footnote{We recall
the sensitivity of predictions for \ssi\ to the uncertainty in $\Sigma_{\pi N}$~\cite{mc6}.}, 
and we include with the LHC$_{\rm 1/fb}$ data the
XENON100 constraint on $\ssi$~\cite{XE100}. We see that the LHC$_{\rm 1/fb}$ data
tend to push $\mneu{1}$ to larger values~\footnote{The slivers of points at
$\mneu{1} \sim 60 \gev$ originate in the light-Higgs funnel region with large $m_0 > 3 \tev$
mentioned earlier, which might be
excluded by the ATLAS 1/fb 0-lepton search.}, and that these are correlated with lower values
of \ssi, though with best-fit values still $\sim 10^{-45}$~cm$^2$. We do not present
here predictions for spin-dependent scattering or signatures of astrophysical
dark matter annihilations, which are further removed from the prospective
experimental sensitivities in the near future.

\begin{figure*}[htb!]
\resizebox{9cm}{!}{\includegraphics{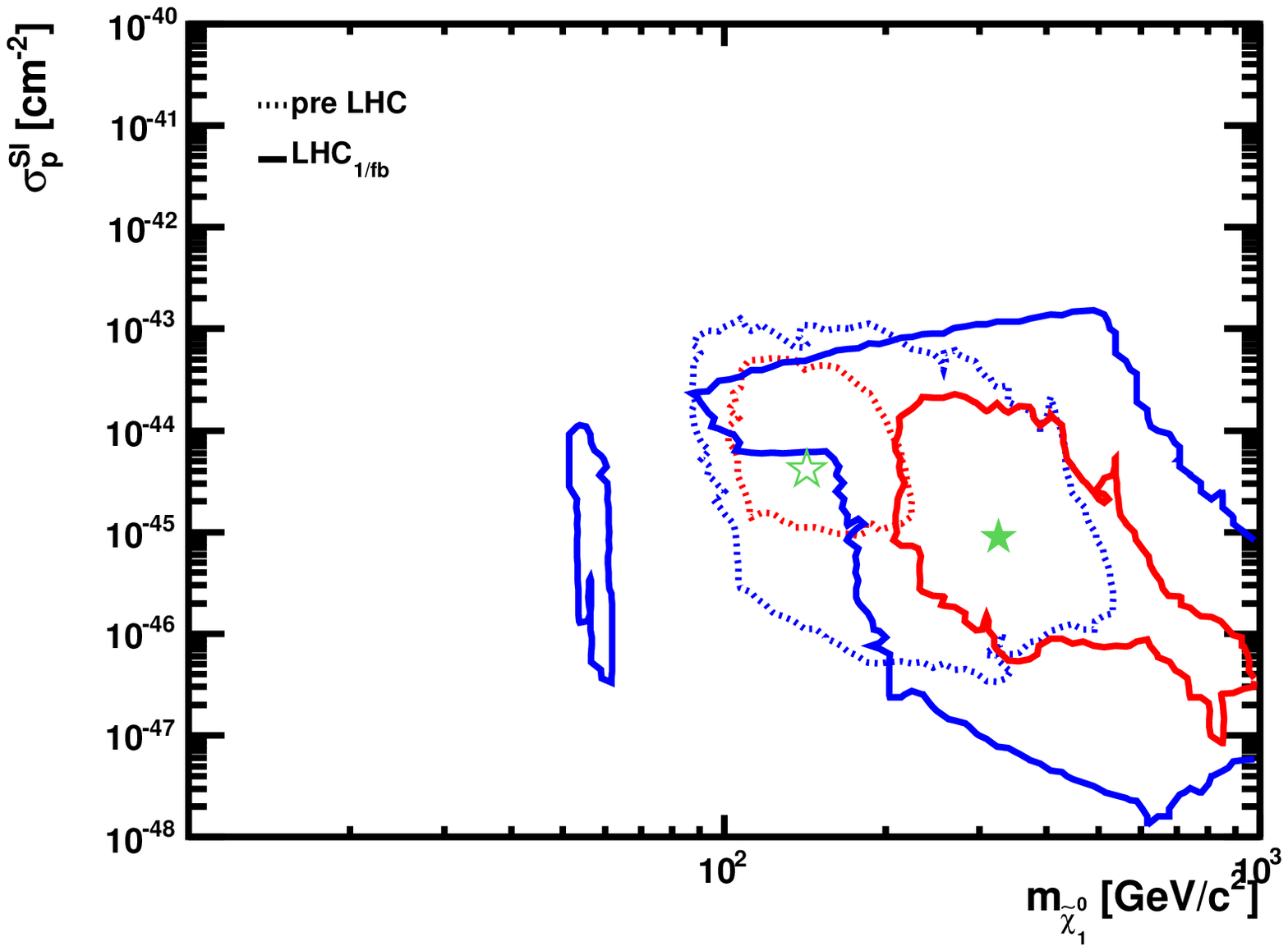}}
\resizebox{9cm}{!}{\includegraphics{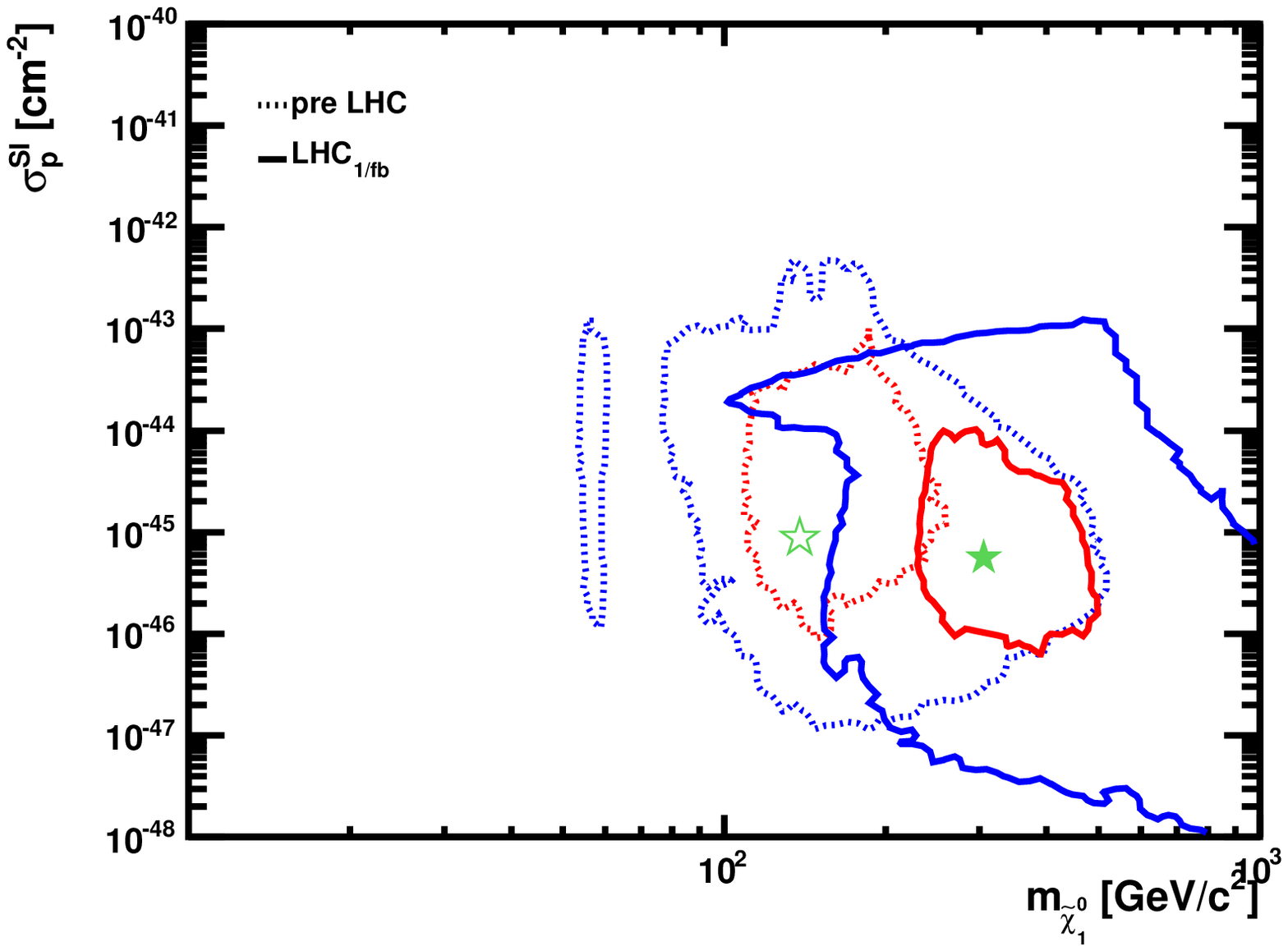}}
\vspace{-1cm}
\caption{\it The 68\% and 95\%~CL contours (red and blue, respectively)
in the CMSSM (left) and the NUHM1 (right). The solid lines are for
fits including the XENON100~\protect\cite{XE100} and LHC$_{\rm 1/fb}$ data, whereas the dotted lines 
include only the pre-LHC data~\protect\cite{mc4}.
}
\label{fig:ssi}
\end{figure*}

\subsection*{\it Sparticle thresholds in $e^+ e^-$ annihilation}

In view of the interest in building an $e^+ e^-$ collider as the next major
project at the energy frontier, we now analyze the implications of the LHC$_{\rm 1/fb}$ 
and XENON100 data for expectations for sparticle production in $e^+e^-$ annihilation 
within the CMSSM and NUHM1.
In this respect it has to be kept
in mind that the LHC searches are mainly sensitive to the 
production of coloured particles, whereas
lepton colliders will have a high sensitivity in particular for the 
production of colour-neutral states, such as
sleptons, charginos and neutralinos, as well as yielding high-precision measurements
that will provide indirect sensitivity to quantum effects of new
states. Anything inferred
from the coloured sector concerning the uncoloured sector depends
on the underlying model assumptions, and in particular on 
assumptions about the possible universality of soft supersymmetry breaking at the
GUT scale. Non-universal models, e.g., low-energy supersymmetric
models, or models with different GUT assumptions,
could present very different possibilities.

Fig.~\ref{fig:thresholds} compares the likelihood functions for
various thresholds in the CMSSM (upper panel) and the NUHM1 (lower panel),
based on the global fits made using the LHC$_{\rm 1/fb}$ and XENON100 constraints.
The lowest thresholds are those for
$e^+ e^- \to \neu{1} \neu{1}$, $\astaue \staue$, $\asel{R}\sel{R}$
 and $\asmu{R}\smu{R}$ (the latter is not shown, it is similar
to that for $\asel{R}\sel{R}$). We see that, within the CMSSM and 
NUHM1, it now seems that these thresholds may well lie above 500~GeV, though in the CMSSM significant fractions of their
likelihood functions still lie below 500~GeV.
The thresholds for $\neu{1} \neu{2}$ and
$\asel{R}\sel{L} + \asel{L}\sel{R}$ are expected to be somewhat higher,
possibly a bit below 1~TeV.
The preferred value for the threshold for $\cha{1}\champ{1}$ 
lies at about 1700 GeV in both the CMSSM and NUHM1 scenarios, 
that for the $HA$ threshold lies above 1~TeV, and that
for first- and second-generation squark-antisquark pair production lies beyond 2.5~TeV in both models.
It should be kept in mind that these high thresholds are linked with
the reduced $p$-value of the model. Further increases in the
excluded regions would yield even higher thresholds, but would also
make the  CMSSM or NUHM1 seem even less likely.

\begin{figure*}[htb!]
\begin{center}
\resizebox{7.5cm}{!}{\includegraphics{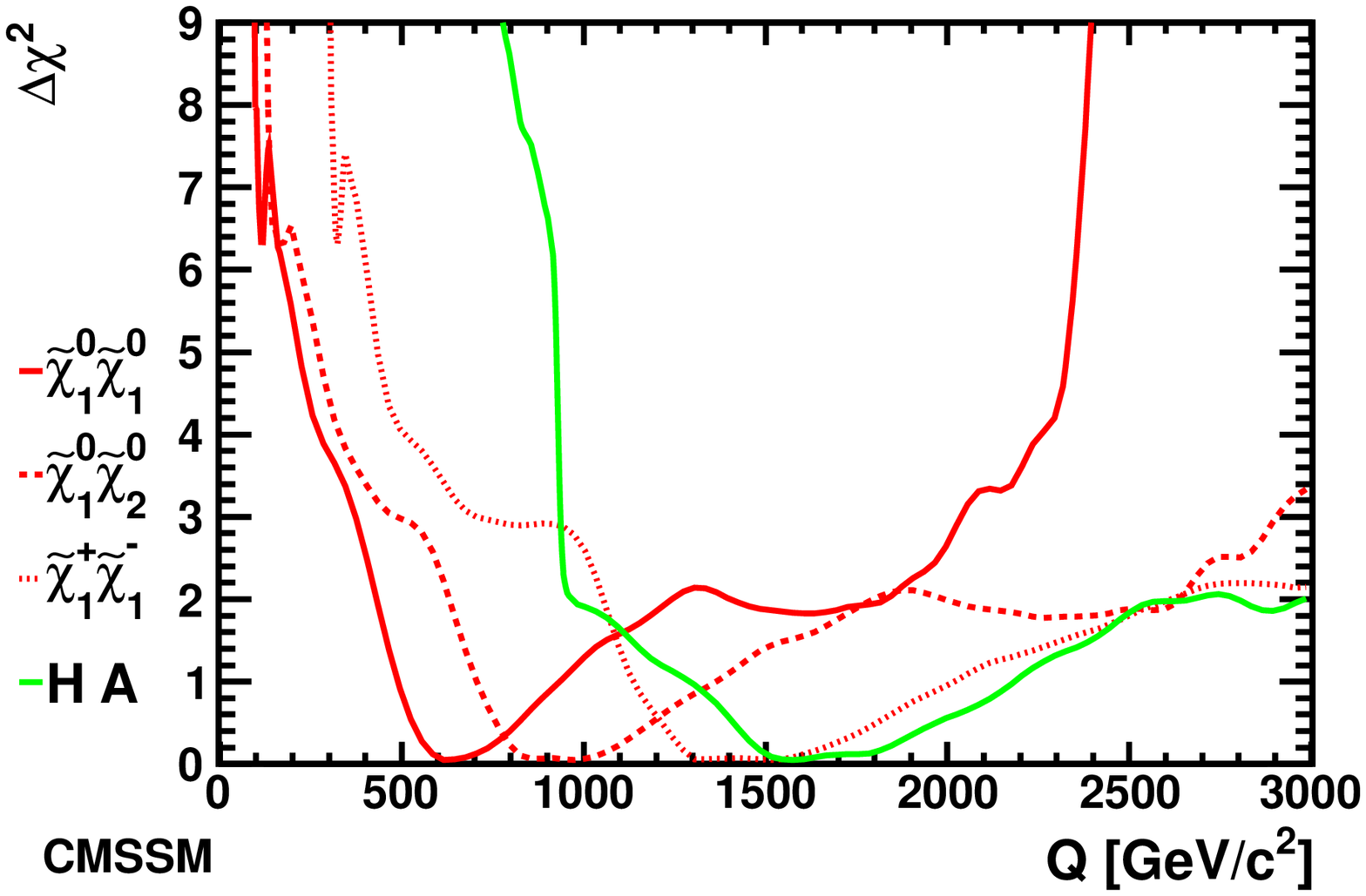}}
\resizebox{7.5cm}{!}{\includegraphics{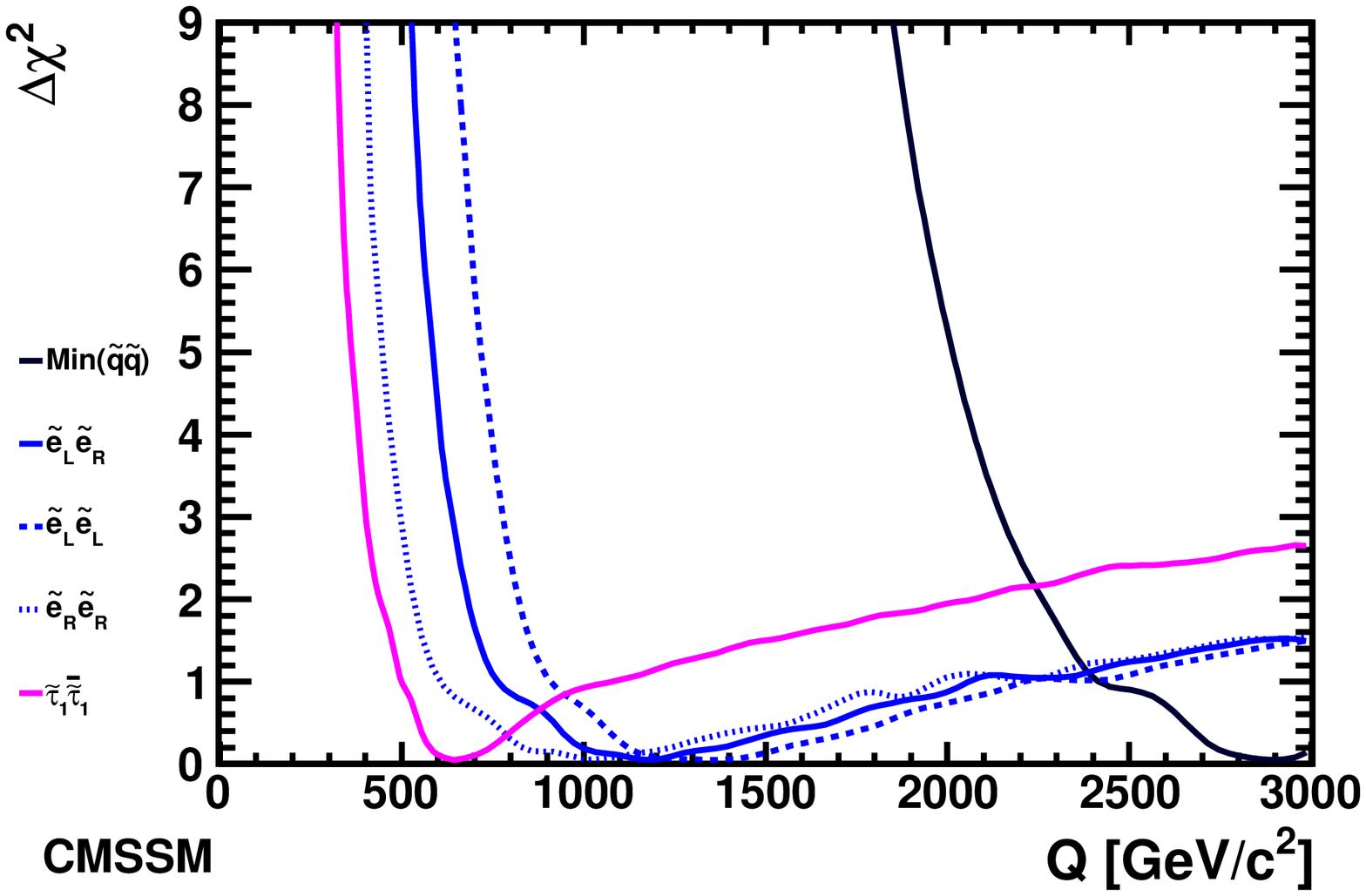}}\\
\resizebox{7.5cm}{!}{\includegraphics{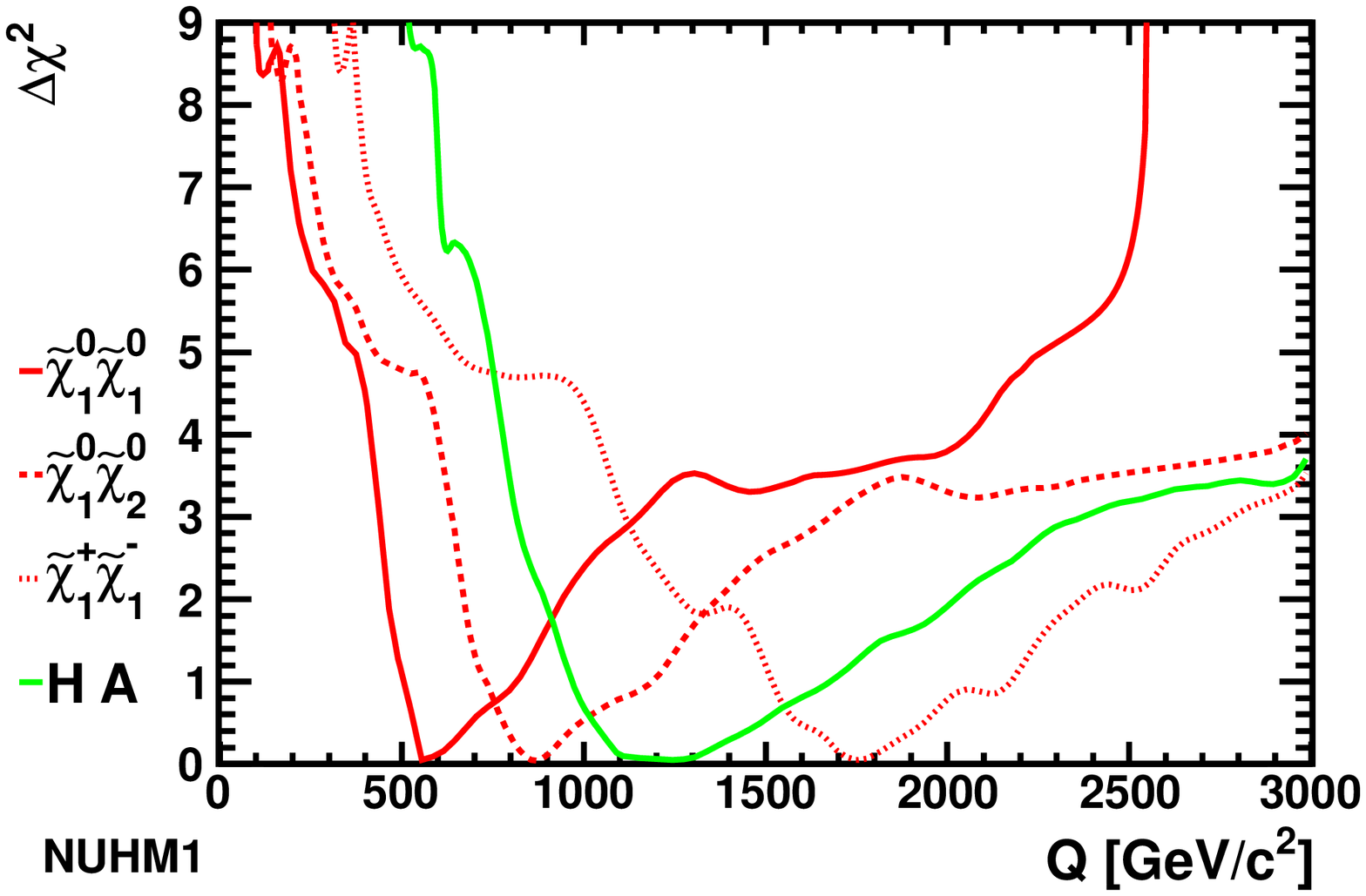}}
\resizebox{7.5cm}{!}{\includegraphics{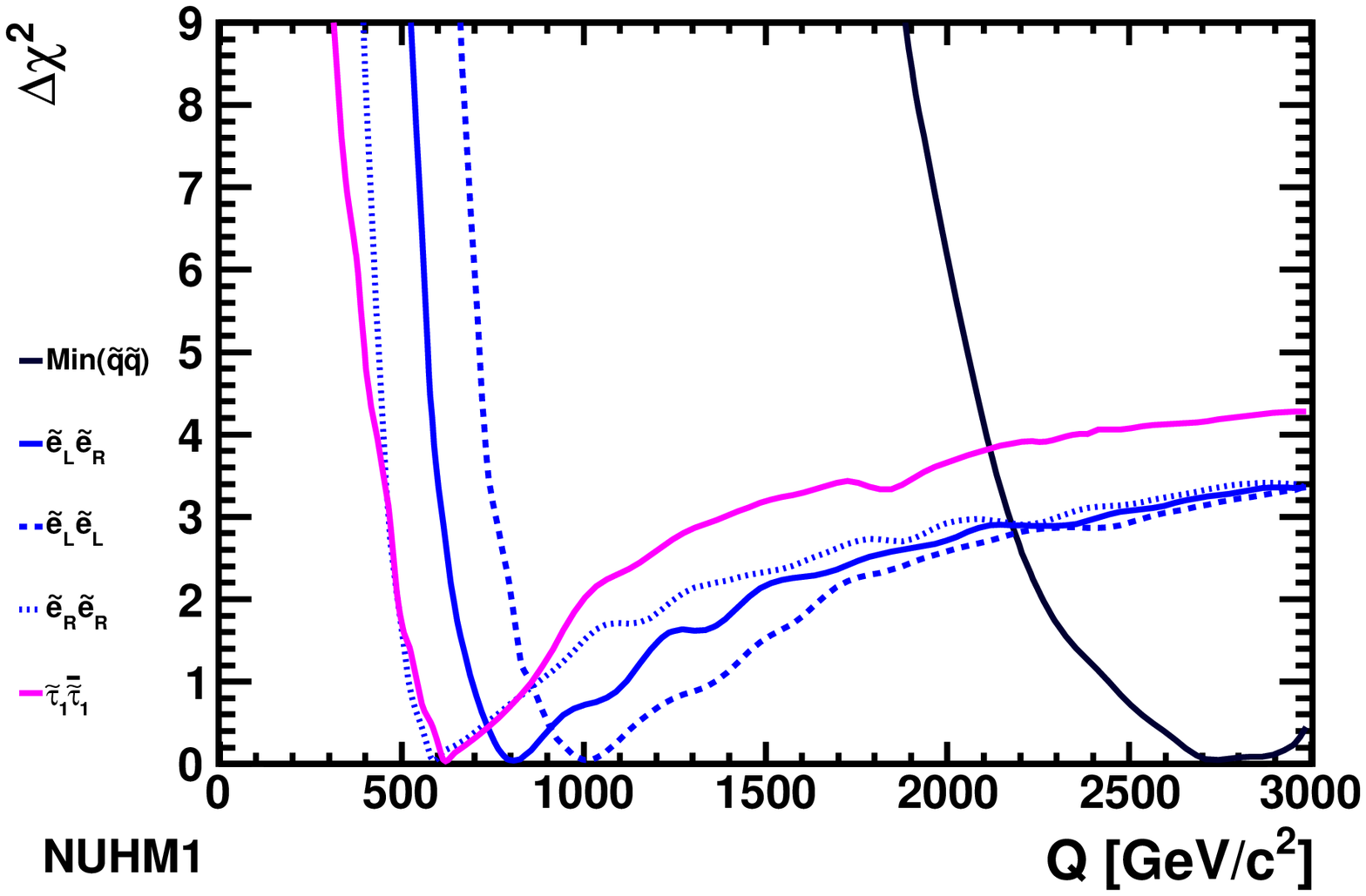}}\\
\vspace{-1cm}
\end{center}
\caption{\it The $\chi^2$ likelihood functions for various pair-production
thresholds in $e^+ e^-$, as estimated in the CMSSM (upper panel) and the NUHM1 (lower panel)
after incorporating the XENON100~\protect\cite{XE100} and LHC$_{\rm 1/fb}$ constraints.
The likelihood function for the $\asmu{R}\smu{R}$ threshold (not shown)
is very similar to that for $\asel{R}\sel{R}$.
}
\label{fig:thresholds}
\end{figure*}

\section{Summary and Conclusions}

There is some disappointment in the air that the LHC has found no signs of
supersymmetry in its first $\sim 1$/fb of data. However, it should
be kept in mind that the searches performed at the LHC so far have essentially
only been able to set limits on the production of the gluino and the
squarks of the first two generations, and the resulting limits depend
sensitively on the mass assumed for the lightest supersymmetric
particle~\cite{stealth}.
On the other hand, the sensitivities of direct searches for stops 
and sbottoms and colour-neutral superpartners are very limited up
to now. This situation will improve, as several
times more data can be expected by the end of 2012, there is the prospect
subsequently of an increase in the energy by a factor up to two,
and the LHC is expected eventually to accumulate orders of magnitude
more data.

The initial optimistic prospects for SUSY searches at the LHC were
largely driven by two indications that the supersymmetric mass
scale might not be very high: \gmt\ and the need for dark matter that should
not be overdense. Neither of these indications has weakened recently. Indeed, the
\gmt\ hint has even strengthened with the convergence of the previously
discrepant SM calculations using low-energy $e^+ e^-$ and $\tau$ decay
data~\cite{newDavier,Jegerlehner}. However, as we have discussed in this
paper, significant tension is 
now emerging between the \gmt\ constraint and LHC data within the
specific context of the CMSSM and NUHM1. {\it A priori}, in a general SUSY
model there is not necessarily a tension between a heavy gluino and
heavy squarks of the first two generations on the one hand, as favoured
by the LHC limits, and light colour-neutral states on the other hand, 
as favoured by \gmt.

The tension within the CMSSM and NUHM1 can be reduced to some extent
by adopting a larger value of $\tan \beta$, but
this may eventually lead to subsidiary tension with the LHC $H/A$ constraints and the
tightening experimental vise on \bmm. 
In any case, it will be important to subject the \gmt\ constraint to
closer scrutiny,
and the upcoming Fermilab and J-PARC
experiments on \gmt~\cite{FNALg-2} are most welcome and timely in this regard. In parallel,
refinements of the experimental inputs for the prediction of \gmt\ from 
both low-energy $e^+ e^-$ and $\tau$ decay data would also be welcome.
It will be also necessary to subject the theoretical 
calculations within the SM and the corresponding estimates of the
remaining theoretical uncertainties to further scrutiny.

The dark matter upper limit on the sparticle mass scale remains unchanged,
and is responsible for the disfavoured region above $m_{1/2} \sim 2500$~GeV visible in
our figures for the CMSSM and the NUHM1. On the other hand, the dark matter constraint on $m_0$ is not so
strong, as also seen in the figures, extending well beyond the range displayed.
Considering the impact of direct jets + $\ETslash$ searches only, 
the regions of the CMSSM and NUHM1 $(m_0, m_{1/2})$ planes in Fig.~\ref{fig:p} with $p$-values 
significantly non-zero 
extend beyond the likely reach even of the full-energy LHC in its high-luminosity 
incarnation. {\it A fortiori}, the same is true for the regions of these planes allowed
at the current 95\%~CL ($\Delta \chi^2 = 5.99$ relative to the global minima, 
bounded by the blue contours in Fig.~\ref{fig:6895}). This is even more true of the full regions of
the CMSSM and NUHM1 $(m_0, m_{1/2})$ planes that are allowed by the dark matter
constraint.

In light of this discussion, under what circumstances could one conclude that
the CMSSM or NUHM1 is excluded? Currently, our best fits in both these models 
have $p$-values above 10\%, comparable to that of SM fits to precision electroweak
data from LEP and SLD, and the F-test shows that both the CMSSM and NUHM1
are warranted extensions of the SM, in the sense that introducing their parameters provides an improvement
in $\chi^2$ that is valuable in both cases. Moreover, it seems unlikely that the LHC will
soon be able to explore all the region of the $(m_0, m_{1/2})$ planes in
Fig.~\ref{fig:p} where the models' $p$-values exceed 5\%, nor does the LHC seem
likely soon to push $F_\chi$ (see Fig.~\ref{fig:F}) to uninterestingly
low levels. This is not surprising, as in the high-mass
limit the superpartners decouple and one is left essentially with the SM
with a light Higgs.

One way for the LHC 
to invalidate the models studied here would be to discover an SM-like 
Higgs boson weighing substantially more
than the range $\sim 120 \gev$ predicted in Fig~\ref{fig:Mh}. 
A value of $\Mh \sim 125 \gev$ or more would be in some tension with \gmt, and perhaps hint towards models beyond the CMSSM and NUHM1, whereas a value of $\sim 130 \gev$ or more would cast severe doubt on most simple GUT-based models.
As already mentioned,
range $\Mh \sim 120$ to $130 \gev$ is precisely that currently favoured independently by precision electroweak
data and by LEP, Tevatron and LHC searches. If a Higgs-like signal
were to be discovered in the lower part of this range, supersymmetry might not be far away,
whereas if $\Mh$ is in the upper part of this range, indicating that at
  least the third-generation squarks could be heavy, one might for some time be in the frustrating situation of acquiring ever
more circumstantial hints for supersymmetry, but with no direct evidence. On the other hand, if the LHC
discovers that $\Mh > 130 \gev$, the time might come to take another look at
non-minimal supersymmetric models.

\clearpage
\pagebreak
\newpage

\subsubsection*{Acknowledgements}

The work of O.B., J.E. and K.A.O. is supported partly by the London
Centre for Terauniverse Studies (LCTS), using funding from the European
Research Council 
via the Advanced Investigator Grant 267352. 
The work of S.H. was supported 
in part by CICYT (grant FPA 2010--22163-C02-01) and by the
Spanish MICINN's Consolider-Ingenio 2010 Program under grant MultiDark
CSD2009-00064. The work of K.A.O. is supported in part by DOE grant DE-FG02-94ER-40823 at the University of Minnesota.


\end{document}